\documentclass[aps,pra,showpacs, twocolumn,amsmath,amssymb,tightenlines,epsfig,floatfix]{revtex4-1}
\usepackage{graphicx}
\usepackage{dcolumn}	
\usepackage{bm}			
\usepackage{amsfonts}
\usepackage{xspace}
\usepackage{color}
\usepackage{epstopdf}

\usepackage{multirow}
\usepackage{array}

\begin{document}

\newcommand{\psihat}{\ensuremath{\hat{\psi}}\xspace}
\newcommand{\psihatd}{\ensuremath{\hat{\psi}^{\dagger}}\xspace}
\newcommand{\ahat}{\ensuremath{\hat{a}}\xspace}
\newcommand{\Ham}{\ensuremath{\mathcal{H}}\xspace}
\newcommand{\ahatd}{\ensuremath{\hat{a}^{\dagger}}\xspace}
\newcommand{\bhat}{\ensuremath{\hat{b}}\xspace}
\newcommand{\bhatd}{\ensuremath{\hat{b}^{\dagger}}\xspace}
\newcommand{\boldr}{\ensuremath{\mathbf{r}}\xspace}
\newcommand{\dr}{\ensuremath{\,d\mathbf{r}}\xspace}
\newcommand{\dk}{\ensuremath{\,d^3\mathbf{k}}\xspace}
\newcommand{\etal}{\emph{et al.\/}\xspace}
\newcommand{\ie}{i.e.\:}
\newcommand{\eq}[1]{Eq.\,(\ref{#1})\xspace}
\newcommand{\fig}[1]{Figure\,(\ref{#1})\xspace}
\newcommand{\abs}[1]{\left| #1 \right|} 
\newcommand{\proj}[2]{\left| #1 \rangle\langle #2\right| \xspace} 
\newcommand{\Qhat}{\ensuremath{\hat{Q}}\xspace}
\newcommand{\Qhatd}{\ensuremath{\hat{Q}^\dag}\xspace}
\newcommand{\phihatd}{\ensuremath{\hat{\phi}^{\dagger}}\xspace}
\newcommand{\phihat}{\ensuremath{\hat{\phi}}\xspace}
\newcommand{\boldk}{\ensuremath{\mathbf{k}}\xspace}
\newcommand{\boldp}{\ensuremath{\mathbf{p}}\xspace}
\newcommand{\boldsigma}{\ensuremath{\boldsymbol\sigma}\xspace}
\newcommand{\boldalpha}{\ensuremath{\boldsymbol\alpha}\xspace}
\newcommand{\grad}{\ensuremath{\boldsymbol\nabla}\xspace}
\newcommand{\parti}[2]{\frac{ \partial #1}{\partial #2} \xspace}
 \newcommand{\vs}[1]{\ensuremath{\boldsymbol{#1}}\xspace}
\renewcommand{\v}[1]{\ensuremath{\mathbf{#1}}\xspace}
\newcommand{\Psihat}{\ensuremath{\hat{\Psi}}\xspace}
\newcommand{\Psihatd}{\ensuremath{\hat{\Psi}^{\dagger}}\xspace}
\newcommand{\Vhatd}{\ensuremath{\hat{V}^{\dagger}}\xspace}
\newcommand{\Xhat}{\ensuremath{\hat{X}}\xspace}
\newcommand{\Xhatd}{\ensuremath{\hat{X}^{\dag}}\xspace}
\newcommand{\Yhat}{\ensuremath{\hat{Y}}\xspace}
\newcommand{\Yhatd}{\ensuremath{\hat{Y}^{\dag}}
\xspace}
\newcommand{\ddt}{\ensuremath{\frac{d}{dt}}
\xspace}
\newcommand{\nset}{\ensuremath{n_1, n_2,\dots, n_k}
\xspace}
\newcommand{\sah}[1]{{\textcolor{red}{#1}}}
\newcommand{\jhat}{\ensuremath{\hat{J}}
\xspace}

\title{Squeezed-light-enhanced atom interferometry below the standard quantum limit}
\author{Stuart S. Szigeti$^{1,2,*}$, Behnam Tonekaboni$^1$, Wing Yung S. Lau$^1$, Samantha N. Hood$^1$, and Simon A. Haine$^{1}$}
\affiliation{$^1$School of Mathematics and Physics,  University of Queensland, Brisbane, QLD, 4072, Australia}
\affiliation{$^2$ARC Centre for Engineered Quantum Systems, The University of Queensland, Brisbane, QLD 4072, Australia}
\email{s.szigeti@uq.edu.au}

\begin{abstract}
We investigate the prospect of enhancing the phase sensitivity of atom interferometers in the Mach-Zehnder configuration with squeezed light. Ultimately, this enhancement is achieved by transferring the quantum state of squeezed light to one or more of the atomic input beams, thereby allowing operation below the standard quantum limit. We analyze in detail three specific schemes that utilize (1) single-mode squeezed optical vacuum (i.e. low frequency squeezing), (2) two-mode squeezed optical vacuum (i.e. high frequency squeezing) transferred to both atomic inputs, and (3) two-mode squeezed optical vacuum transferred to a single atomic input. Crucially, our analysis considers incomplete quantum state transfer (QST) between the optical and atomic modes, and the effects of depleting the initially-prepared atomic source. Unsurprisingly, incomplete QST degrades the sensitivity in all three schemes. We show that by measuring the transmitted photons and using information recycling [Phys.~Rev.~Lett.~\textbf{110}, 053002 (2013)], the degrading effects of incomplete QST on the sensitivity can be substantially reduced. In particular, information recycling allows scheme (2) to operate at the Heisenberg limit irrespective of the QST efficiency, even when depletion is significant. Although we concentrate on Bose-condensed atomic systems, our scheme is equally applicable to ultracold thermal vapors.   
\end{abstract}

\pacs{03.75.Dg, 42.50.Gy, 42.50.Dv, 03.75.Be, 42.50.-p}

\maketitle

\section{Introduction}
Atom interferometry is a leading precision measurement technology, having demonstrated state-of-the-art measurements of accelerations and rotations \cite{Peters:1999, Muller:2008c, Altin:2013, Canuel:2006, Lenef:1997, Gustavson:1997}, gravity gradients \cite{Snadden:1998, McGuirk:2002}, magnetic fields \cite{Vengalattore:2007}, the fine structure constant ($\alpha$) \cite{Gupta:2002, Bouchendira:2011}, and Newton's gravitational constant ($G$) \cite{Fixler:2007, Lamporesi:2008, Andia:2013, Rosi:2014}. Further increases to the sensitivity of atom interferometers would allow for some exciting science, such as improved tests of the weak equivalence principle \cite{Fray:2004, Dimopoulos:2007, Schlippert:2014}, searches for quantum gravitational effects \cite{Amelino-Camelia:2009}, and the measurement of gravitational waves \cite{Tino:2007, Dimopoulos:2008}. Current state-of-the-art atom interferometers utilize uncorrelated sources, which can operate no better than the \emph{standard quantum limit} (SQL) - i.e. the sensitivity scales as $1/\sqrt{N}$ where $N$ is the number of detected atoms. Unfortunately, current atomic sources have low fluxes (in comparison with photon sources), and there exist considerable technical barriers to increasing this flux \cite{Robins:2013}. Hence, developing atom interferometers that operate below the SQL, which have a better `per atom sensitivity' than current devices, is of great interest. 

Sub-SQL atom interferometers necessarily exploit entanglement, and a number of proposals exist for generating the required entanglement between the atomic degrees of freedom. These are based on phenomena as diverse as molecular dissociation \cite{Kheruntsyan:2005b}, spin-exchange collisions \cite{Pu:2000, Lucke:2011, Gross:2011}, atomic four-wave mixing \cite{Jaskula:2010, Bucker:2011, Haine:2011, Lewis-Swan:2014}, and atomic Kerr squeezing \cite{Kitagawa:1993, Sorensen:2002, Gross:2010, Riedel:2010, Johnsson:2007a, Haine:2009, Haine:2014}. However, all these schemes require large inter-atomic interactions, small atom number, and give little control over the motional atomic state. These are the \emph{opposite} conditions required for precision atom interferometry. Alternatively, squeezed atomic states can be generated by mapping the quantum state of squeezed light to an atomic field \cite{Jing:2000, Fleischhauer:2002b, Haine:2005, Haine:2005b, Haine:2006b, Hammerer:2010}. Given that squeezed light is known to give sub-SQL sensitivities in optical interferometers \cite{Caves:1981}, transferring the entanglement to atomic degrees of freedom should similarly allow sub-SQL atom interferometry. Importantly, since the squeezing is generated independently of the atomic source, in principle this technique gives high flux (relative to state-of-the-art atomic sources), weakly-interacting squeezed atomic states in targeted motional states - ideal for atom interferometry. 

In this paper, we present three squeezed-light-enhanced atom interferometry schemes, and show that these are all capable of sub-SQL phase sensitivities. We explicitly consider Bose-condensed atomic sources, as the narrow velocity distribution and large coherence length of a Bose-Einstein condensate (BEC) offers considerable advantages over thermal sources, including more precise manipulation of the motional state, increased visibility, and the prospect of feedback-stabilization under high flux outcoupling \cite{Debs:2011, Szigeti:2012, Haine:2004, Szigeti:2009, Szigeti:2010, Szigeti:2013, Hush:2013}. However, many of the results in this paper are equally applicable to ultracold thermal sources. Our analysis considers the effects of incomplete quantum state transfer (QST) between the optical and atomic modes, and the effect of depleting the initial condensate mode. We also incorporate the technique of \emph{information recycling} \cite{Haine:2013} into our schemes, and demonstrate that this can be used to combat the negative effects of incomplete QST. Given the maturity of squeezed light generation technology, and the high efficiency of photon detection, it would be relatively straightforward to incorporate our schemes into existing state-of-the-art atom interferometers. Consequently, squeezed-light enhancement and information recycling offer a promising path to improved sensitivity in atom-interferometer-based technologies. 

The structure of our paper is as follows. In Sec.~\ref{sec_theoretical_model} we derive a simplified quantum model of atom-light coupling, based on two-photon Raman transitions, and show how this atom-light coupling can be used to achieve QST between the optical and atomic modes (Sec.~\ref{sec_QST}), and coherent atomic beamsplitting (Sec.~\ref{sec_coherent_atom_BS}). Sec.~\ref{sec_review_MZ} briefly reviews atom interferometry in the Mach-Zehnder configuration. Our first interferometry scheme, where enhancement is achieved with a single-mode squeezed optical vacuum (i.e. low frequency squeezing), is presented in Sec.~\ref{sec_single_mode_atom_int}. The sensitivity of this scheme is derived for complete QST and an undepleted initial atomic source (Sec.~\ref{sec_single_mode_CQST}), and compared to truncated Wigner simulations that include depletion. The effects of incomplete QST on this scheme, and how such effects can be reduced with information recycling, are considered in Sec.~\ref{sec_single_mode_incomplete}. The effects of losses are briefly explored in Sec.~\ref{sec_losses}. In Sec.~\ref{sec_high_freq_sq}, we present our second scheme, which utilizes a two-mode squeezed optical vacuum (i.e. high frequency squeezing) transferred to both atomic inputs. Once again, we quantitatively examine the phase sensitivity when QST is both complete (Sec.~\ref{sec_two_mode_CQST}) and incomplete (Sec.~\ref{sec_two_mode_incomplete}), and numerically consider the effects of depletion (Sec.~\ref{sec_loss_and_depletion}). Section~\ref{sec_high_to_low_sq} presents and analyses our final atom interferometry scheme, where only a single atomic input is enhanced with a two-mode squeezed optical vacuum. These three schemes are then compared and summarized in Sec.~\ref{sec_summary}.

\section{Theoretical model for atom-light coupling} \label{sec_theoretical_model}
\begin{figure}[!tb]
\centering
\includegraphics[width=0.8\columnwidth]{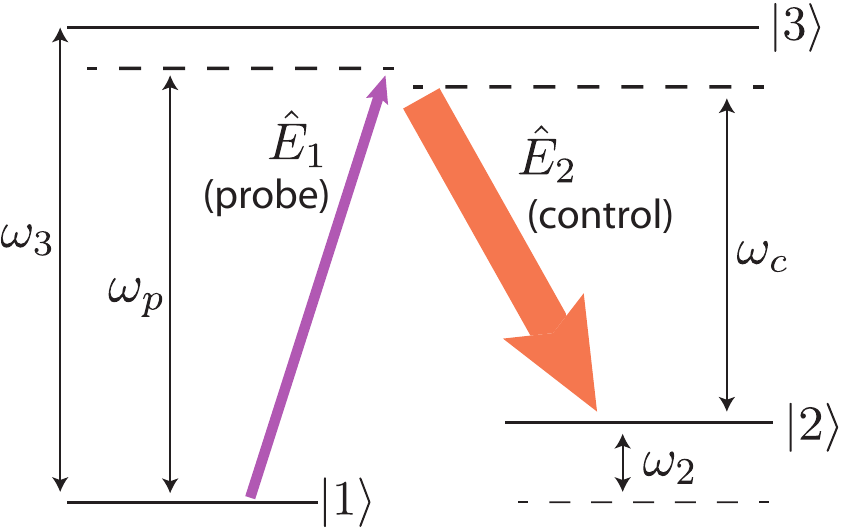}
\caption{ Energy level scheme for a three-level Raman transition comprising two non-degenerate hyperfine ground states, $|1\rangle$ and $|2\rangle$. The BEC is initially formed in the state $|1\rangle$, and population is transferred to $|2\rangle$ via the absorption of a photon from $E_1$ (the probe beam) and the emission of a photon into $E_2$ (the control beam). The probe and control beams are detuned from the excited state $|3\rangle$ by an amount $\Delta_p = \omega_3-\omega_p$ and $\Delta_c = \omega_3-\omega_2-\omega_c$, respectively.}
\label{levelsfig}
\end{figure}

Our system has been previously described in \cite{Haine:2005, Haine:2005b, Haine:2006b}. We begin with a BEC consisting of atoms with two hyperfine states $|1\rangle$ and $|2\rangle$, separated in energy by an amount $\hbar \omega_2$. These two levels are coupled with a Raman transition via two optical fields, $\hat{E}_1$ (probe beam, wavevector $\mathbf{k}_1$) and $\hat{E}_2$ (control beam, wavevector $\mathbf{k}_2$), which are both detuned from an excited state $|3\rangle$ (see Fig.~\ref{levelsfig}). We assume that the control field $\hat{E}_2$ is much more intense than the probe field $\hat{E}_1$, allowing us to ignore depletion and quantum fluctuations and approximate the control field as a classical plane wave - i.e. $\hat{E}_2(\textbf{r}, t) \approx \mathcal{E}_2 \exp[i (\textbf{k}_2 \cdot \textbf{r} - \omega_c t)]$. Furthermore, by design $\Delta_p \equiv \omega_3-\omega_p$ and $\Delta_c \equiv \omega_3-\omega_2-\omega_c$ are large compared with the Rabi frequencies of the $|1\rangle \rightarrow |3\rangle$ and $|2\rangle \rightarrow |3\rangle$ transitions. Therefore, the excited state can be adiabatically eliminated \cite{Sinatra:1995, Brion:2007, Paulisch:2014}, giving an effective coupling between atomic states $| 1 \rangle$ and $|2 \rangle$. Finally, we assume a dilute atomic sample, since it is generally optimal for atom interferometry to operate in a regime where the inter-atomic interactions are negligible \cite{Jamison:2011, Debs:2011, Altin:2011, Altin:2011b}. Under these approximations and the rotating-wave approximation \cite{Fewell:2005}, the Hamiltonian for the system becomes
\begin{align}
	\hat{\mathcal{H}} 	&= \sum_{j=1,2} \int  \dr \, \psihatd_j(\boldr) H_j(\textbf{r}) \psihat_j(\boldr)  \nonumber \\  
					&+\hbar g  \int \dr \left( \psihat_1(\boldr) \psihatd_2(\boldr) \hat{E}_1(\boldr) e^{- i (\boldk_2\cdot \boldr -\omega_c t)} + h.c.\right) \nonumber \\
	&+ \hat{\mathcal{H}}_{\mathrm{light}} \, ,			 \label{full_Ham}
\end{align} 
where $\hat{\mathcal{H}}_{\text{light}}$ is the Hamiltonian for the free photon field. Here, $H_1(\textbf{r}) = H(\textbf{r})$ and $H_2(\textbf{r}) = H(\textbf{r}) + \hbar \omega_{2}$, where $H(\textbf{r})$ is the single-atom Hamiltonian common to both hyperfine states. $\psihat_1(\boldr)$ and $\psihat_2(\boldr)$ are the usual bosonic field operators for atoms in hyperfine states $|1\rangle$ and $|2\rangle$, respectively, satisfying the commutation relation $[\hat{\psi}_i(\boldr),\hat{\psi}^\dag_j(\boldr^\prime)]=\delta_{ij} \delta(\boldr-\boldr^\prime)$. Similarly, $\hat{E}_1(\boldr)$ is the annihilation operator for the probe field satisfying $[\hat{E}_1(\boldr),\hat{E}_1^\dag(\boldr^\prime)]= \delta(\boldr-\boldr^\prime)$. We assume that the probe field $\hat{E}_1(\boldr)$ has a small spread of frequencies around $\omega_p = c |\mathbf{k}_1|$, in which case the effective coupling strength is given by
\begin{equation}
	g = d_{12} \sqrt{\frac{ \omega_p}{2 \hbar \epsilon_0}}\frac{\Omega}{ \Delta_p}, \label{effective_coupling}
\end{equation}
where $d_{12}$ is the dipole moment of the $|1\rangle \rightarrow |3\rangle$ transition, $\Omega$ is the Rabi frequency for the $|2\rangle \rightarrow |3\rangle$ transition, effected by the classical control field, and $\epsilon_0$ is the permittivity of free space.

As is typical in atom interferometry, we assume that all atoms are initially in a single motional mode $u_0(\boldr)$ of state $|1\rangle$. Furthermore, we will assume that our probe field is vacuum, except for occupation in a pulse characterized by a wave packet propagating in the $z$-direction: 
\begin{equation}
	u_p(\boldr,t)e^{i\boldk_1\cdot \boldr} = u_{\text{trans}}(x,y)u_{\text{prop}}(z - ct)e^{i\boldk_1\cdot \boldr}, \label{eq_pulse_envelope}
\end{equation}
satisfying $L |\textbf{k}_1|\gg1$, where $L$ is the characteristic length scale of $u_{\text{prop}}(z)$.  Provided the timescale for population transfer is fast compared with the timescale for atomic motion, the Hamiltonian~(\ref{full_Ham}) can be simplified considerably by expanding the field operators in the appropriate mode basis, and keeping only those modes that are highly occupied. Explicitly, we approximate 
\begin{subequations}
\begin{eqnarray}
\psihat_1(\boldr)  &\approx& u_0(\textbf{r}) \hat{a}_{1}, \\
\psihat_2(\boldr) &\approx& u_0(\boldr) e^{i(\boldk_1-\boldk_2)\cdot\boldr} \hat{a}_{2}, \\
\hat{E}_1(\boldr) &\approx& u_p(\boldr,t) e^{i\boldk_1\cdot \boldr} \bhat.
\end{eqnarray}
\end{subequations}
The simplified Hamiltonian is therefore $\hat{\mathcal{H}} \approx \hat{\mathcal{H}}_0 + \hat{\mathcal{H}}_\text{int}'$, where 
\begin{subequations}
\label{simple_ham_3mode}
\begin{align}
	\hat{\mathcal{H}}_0 		&= \left(\hbar\omega_2 + \frac{\hbar^2}{2m}|\boldk_1-\boldk_2|^2 \right)\ahatd_2\ahat_2 + \hbar \omega_p \bhatd \bhat \\
	\hat{\mathcal{H}}_\text{int}	' &=  \hbar gf(t) \left( \ahat_1 \ahatd_2 \bhat \, e^{i\omega_c t} + h.c. \right), 
\end{align}
\end{subequations}
and 
\begin{equation}
	f(t) = \int \dr \, |u_0(\boldr)|^2 u_p(\boldr,t).
\end{equation}

Note that we have neglected the single-atom energy contribution due to $H_j(\textbf{r})$, which is approximately a constant energy offset on the timescale of the population transfer. The time dependence in $\hat{\mathcal{H}}$ is due to the propagation of the probe wave packet $u_p(\boldr, t)$. Interestingly, an identical Hamiltonian is obtained if the probe beam is continuous-wave and the classical control field is shaped by the temporal function $f(t)$. This would couple the mode of the probe field defined by the mode function $u_p(\boldr,t)$ to the condensate.

Finally, after moving to the interaction picture $\hat{\mathcal{H}} \to \hat{\mathcal{H}}_\text{int} = \hat{U}^\dag \hat{\mathcal{H}}_\text{int}' \hat{U}$ with $\hat{U} = \exp(i \hat{\mathcal{H}}_0 t / \hbar)$,
we obtain the following Heisenberg equations of motion:
\begin{subequations}
\label{3modeEOM}
\begin{eqnarray}
i\dot{\ahat}_1 &=& g f(t)\ahat_2 \bhatd e^{i\delta t} \label{ahat1dot} \\
i\dot{\ahat}_2 &=& g f(t) \ahat_1\bhat e^{-i\delta t} \label{ahat2dot} \\
i\dot{\bhat} &=& g f(t) \ahat_2\ahatd_1 e^{i\delta t} \label{bhatdot} \, ,
\end{eqnarray}
\end{subequations}
where $\delta = \omega_p - \omega_c - \omega_2 - \hbar |\boldk_1-\boldk_2|^2/(2m)$ is the two-photon detuning, which is an experimental parameter freely adjustable by tuning the frequency offset between $E_1$ and $E_2$. We only consider the optimal case $\delta=0$, when the system is on-resonance such that the energy transferred by the two-photon transition perfectly matches the change in electronic and kinetic energies of the atom. Equations~(\ref{3modeEOM}) form the basis for the rest of this paper, as they allow us to describe the process of QST (of the light state to the atomic state) and our conventional coherent atomic beamsplitters. Each of these processes is discussed in more detail below. 

\subsection{Quantum state transfer (QST)} \label{sec_QST}
The goal of this process is to take a pulse of light, described by annihilation operator $\bhat$, and coherently map its quantum state onto the atomic mode $\ahat_2$. Thus, if $\bhat$ was initially in some interesting quantum state, such as a squeezed state, or was entangled with another mode, after QST $\ahat_2$ will also be in this state and/or be entangled with this other mode. To achieve perfect QST, the number of photons in mode $\bhat$ must be much less than the initial number of condensate atoms in mode $\ahat_1$. Then, for a sufficiently short atom-light coupling time, a small number of atoms can be transferred to mode $\hat{a}_2$, which now have the initial quantum state of the light in mode $\hat{b}$, while leaving the number of atoms in mode $\ahat_1$ essentially unchanged. We can therefore make the \emph{undepleted reservoir approximation}  $\ahat_1\rightarrow\sqrt{N_{a_1}}$, where $N_{a_1}$ is the mean number of atoms in mode $\ahat_1$. Under this approximation, the dynamics of QST are described wholly by Eqs~(\ref{ahat2dot}) and (\ref{bhatdot}), which can be solved exactly:
\begin{subequations}
\label{analytic_QST}
\begin{align}
	\ahat_2(t) 		&= \ahat_2(t_0) \cos (\theta_\text{QST}/2) - i\bhat(t_0)\sin (\theta_\text{QST}/2), \label{asol} \\ 
	\bhat(t) 		&= \bhat(t_0) \cos (\theta_\text{QST}/2) - i\ahat_2(t_0)\sin (\theta_\text{QST}/2) \, \label{bsol} ,
\end{align}
\end{subequations}
where $\theta_\text{QST}(t) \equiv 2 g \sqrt{N_{a_1}} \int_0^{t} dt^\prime \, f(t^\prime)$. As was shown in Jing \etal \cite{Jing:2000}, when $\theta_\text{QST} = \pi$ we have \emph{complete QST}, such that the quantum state of $\bhat(t_0)$ is perfectly mapped to the quantum state of $\ahat_2(t_1)$, up to a phase factor. However, achieving the required coupling strength for complete QST is likely to be challenging in practice. Thus \emph{incomplete QST}, with $0< \theta_\text{QST} < \pi$, is the likely experimental regime, and consequently is considered in detail throughout this paper. Of course, incomplete QST also occurs when the undepleted reservoir approximation breaks down, which occurs when the initial number of photons in $\bhat$ becomes comparable to the number of atoms in $\ahat_1$. In this regime the solution (\ref{analytic_QST}) is invalid, and we require a numeric solution to Eqs~(\ref{3modeEOM}). Moreover, as $\theta_\text{QST}$ is no longer well-defined, the QST process must be described by an alternative metric. The natural choice is the \emph{QST efficiency}:
\begin{equation}
	\mathcal{Q}(t) \equiv \frac{\langle \hat{a}_2^\dag(t)\hat{a}_2(t) \rangle}{\langle \hat{b}^\dag(t_0)\hat{b}(t_0) \rangle}, \label{defn_F} 
\end{equation}
which is a measure of the percentage of atoms outcoupled compared with the total number of input photons. Although this is a somewhat cruder metric than the fidelity, it is operationally more convenient, and certainly more than adequate for our purposes. When $\hat{a}_1$ is treated as an undepleted reservoir, and Eqs~(\ref{analytic_QST}) apply, then $\mathcal{Q} = \sin^2 (\theta_\text{QST}/2)$. 

\begin{figure}[t]
\includegraphics[width=1.0\columnwidth]{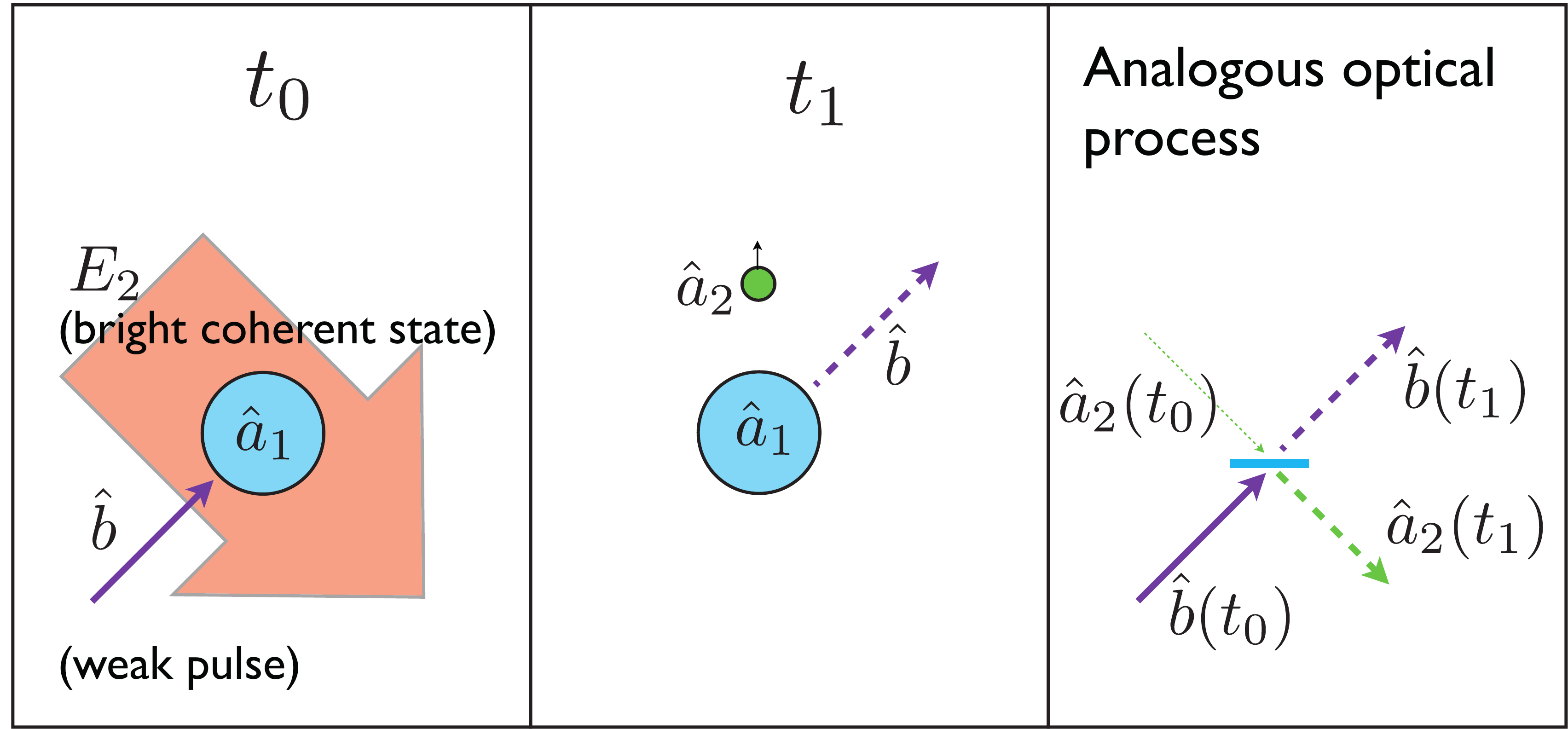}
\caption{Analogy between atom-light QST process, and a beamsplitter. A photon from mode $\bhat$ can either implement a Raman transition, resulting in an atom being outcoupled into mode $\ahat_2$, or the photon can be transmitted, remaining in mode $\bhat$. In the regime where the number of photons in mode $\bhat$ is much less than the number of atoms in the condensate $\ahat_1$, we can treat $\ahat_1$ as an undepletable reservoir. }
\label{fig:QST}
\end{figure}

It is instructive to conceptualize the process of QST as an \emph{atom-light beamsplitter}. That is, a type of beamsplitter that distributes an initial quantum state of light $\bhat(t_0)$ amongst an atomic mode $\ahat_2(t_1)$ and an outgoing mode of light $\bhat(t_1)$, in much the same way as a conventional beamsplitter distributes a quantum state of light amongst two outgoing modes of light. This helpful analogy is illustrated diagrammatically in Fig.~\ref{fig:QST}. As a concrete example, note that when $\theta_\text{QST} = \pi/2$, we have a $50/50$ atom-light beamsplitter, corresponding to a QST efficiency of $50\%$. We invoke the atom-light beamsplitter analogy throughout this paper, as it allows quite complicated atom-interferometric schemes to be conceptualized as simple linear-optical setups.  

\subsection{Coherent atomic beamsplitters} \label{sec_coherent_atom_BS}
The coherent beamsplitting and reflection of two atomic modes via two-photon Raman transitions is a mature experimental technique that has been used with much success in atom interferometry \cite{Kasevich:1992}. In these experiments, \emph{both} light pulses have a mean photon number much larger than the number of atoms in modes $\ahat_1$ and $\ahat_2$, and are therefore well-approximated as undepletable coherent states. In this regime, the atom-light coupling is a conventional Raman transition, in the sense that the coupling coherently transfers population between the two atomic modes without significantly affecting the state of the optical modes. This can be explicitly seen by making the replacement $\hat{b} \to \sqrt{N_b}$, for mean photon number $N_b$, and subsequently solving Eqs (\ref{ahat1dot}) and (\ref{ahat2dot}), yielding
\begin{subequations}
\label{coherent_BS_soln}
\begin{align}
	\ahat_1(t) &= \ahat_1(t_0) \cos (\theta_\text{BS}/2) - i\ahat_2(t_0)\sin (\theta_\text{BS}/2),  \\ 
	\ahat_2(t) &= \ahat_2(t_0) \cos (\theta_\text{BS}/2) - i\ahat_1(t_0)\sin (\theta_\text{BS}/2), \, 
\end{align}
\end{subequations}
with $\theta_\text{BS}(t) \equiv 2 g \sqrt{N_b}  \int_0^t dt^\prime \, f(t^\prime)$. When $\theta_\text{BS} = \pi/2$ and $\theta_\text{BS} = \pi$, we have a $50/50$ atomic beamsplitter and mirror, respectively. 

It is worth noting that Eqs~(\ref{coherent_BS_soln}) apply to other coherent beamsplitting techniques, such as Bragg pulses \cite{Giltner:1995b, Muller:2008} and Bloch oscillations \cite{Clade:2010}. Indeed, given that inertial sensors based on Bose-condensed sources and large momentum transfer beamsplitters offer a promising alternative route to improved sensitivity \cite{Chiow:2011, Szigeti:2012, Hardman:2014}, incorporating such interferometers within the squeezed-light enhanced schemes outlined below is a most attractive prospect.

\begin{figure}[t]
\includegraphics[width=1.0\columnwidth]{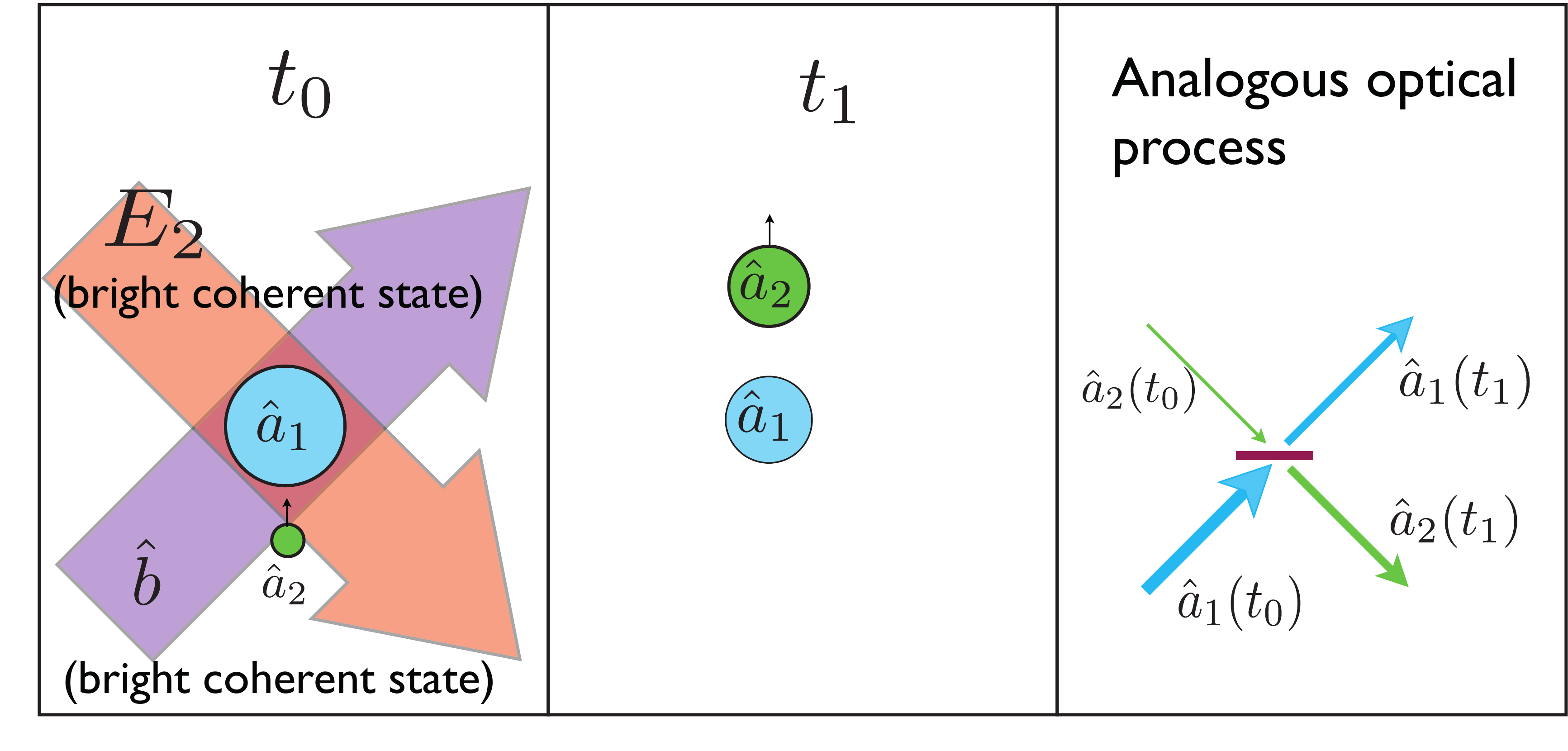}
\caption{Analogy between a conventional two-photon Raman transition and a beamsplitter. An atom in mode $\ahat_1$ can absorb a photon from mode $\bhat$, and emit it into $E_2$ via stimulated emission, producing one atom in mode $\ahat_2$. An atom can be transferred from $\ahat_2$ to $\ahat_1$ via the reverse process. In the regime where the number of photons in both $\bhat$ and $E_2$ is much greater than the number of photons in mode $\ahat_1$, we can treat the optical modes as undepletable reservoirs, and the process behaves as an atomic beamsplitter. }
\label{fig:BS}
\end{figure}

\section{Review: the Mach-Zehnder atom interferometer} \label{sec_review_MZ}
A standard atom interferometer in the Mach-Zehnder (MZ) configuration operates by first splitting a single-mode population of atoms (in mode $\ahat_1$, say) with a $50/50$ beamsplitter, and then letting a phase difference between the two modes accumulate due to the physical process of interest (e.g. linear acceleration). The two modes are redirected together via an atomic mirror, and subsequently mixed with a second $50/50$ beamsplitter. This converts the phase shift $\phi$ into a population difference between the two modes, $\hat{N}_{a_1} - \hat{N}_{a_2}$, where $\hat{N}_{a_i} = \hat{a}_i^\dag \hat{a}_i$. The population difference at the output can be measured, allowing for an accurate estimate of the phase shift and hence the relevant physical quantity of interest. This \emph{output signal} can be expressed in terms of the two input modes (i.e. the atomic modes prior to the first beamsplitter) by repeated application of Eqs~(\ref{coherent_BS_soln}):
\begin{align}
	\hat{S}_a 	&= \hat{N}_{a_1}(t_f) - \hat{N}_{a_2}(t_f) \notag \\ 
        			&= \cos\phi \left(\hat{N}_{a_2}(t_1) - \hat{N}_{a_1}(t_1)\right) \nonumber \\
        			&+ \sin\phi \left( \hat{a}_{1}^\dagger (t_1)\hat{a}_{2}(t_1) + \hat{a}_{2}^\dagger(t_1) \hat{a}_{1}(t_1)\right) \, , \label{sig1}
\end{align}
where $t_1$ and $t_f$ are the times immediately before the first beamsplitter and immediately after the second beamsplitter, respectively. The \emph{phase sensitivity} of the interferometer can be determined from this output signal via the expression \cite{Scully:1997}
\begin{equation}
	\Delta \phi = \sqrt{\frac{V(S_a)}{ (d\langle \hat{S}_a \rangle/d\phi)^2}} \, , \label{deltaphidef}
\end{equation}
where $V(Q) \equiv \langle \hat{Q}^2\rangle - \langle \hat{Q}\rangle^2$ is the variance. If one input is either a traditionally prepared Bose-condensed source (modelled as a coherent state or Fock state) or a laser-cooled thermal source, and the other input is vacuum, then the standard MZ interferometer can achieve a sensitivity no better than the SQL $\Delta \phi = 1 / \sqrt{N_t}$, where $N_t$ is the total number of atoms measured at the output. Sub-SQL sensitivities require a more exotic initial state \cite{Pezze:2009}. As shown below, QST is a neat and practical method of generating just such a state.

\section{Enhanced atom interferometry with single-mode squeezed light} \label{sec_single_mode_atom_int}
We first consider using a single-mode squeezed optical vacuum to enhance the sensitivity of atom interferometry. Although generating the single-mode squeezed vacuum state considered here is feasible, it is likely to be a technically challenging procedure. Ultimately, the difficulty stems from the frequencies of the light field where squeezing can be observed. Conceptually, it is impossible to squeeze only a single frequency of light; naturally squeezing occurs across a range of frequencies. More precisely, an optically squeezed state has quantum correlations between sidebands symmetrically distributed around a central carrier frequency, $\omega_{p} = c |\textbf{k}_2|$ [see Fig.~\ref{sidebands}(a)]. Below some critical frequency $\omega_\text{crit}$, technical considerations usually ensure that these correlations are masked by uncorrelated classical noise [the red frequencies in Fig.~\ref{sidebands}(a)]. Hence, in order for the optical mode taking part in the QST process to display quantum correlations, we require $\Delta \omega \gg 2 \omega_\text{crit}$, where $\Delta \omega$ is the characteristic width of $F(\omega)$, the Fourier transform of $f(t)$. Although this is technically challenging, there has recently been demonstrations of significant squeezing at frequencies below $100$ Hz \cite{Stefszky:2012}. We consider the application of higher-frequency squeezed light sources, which are easier to generate, in later sections of this paper. 

To begin, suppose that the squeezed light is generated via an optical parametric oscillator (OPO) \cite{Walls:2008}. Without loss of generality, we assume that $\textbf{k}_1$ points in the $z$-direction,  i.e. $\hat{E}(\textbf{r},t) \approx u_\text{trans}(x,y)\hat{E}(z,t)$. Then, in the momentum basis, the photon fields that form the inputs and outputs of the OPO, at times $t_{i}$ and $t_{0}$ respectively, are related by:
\begin{equation}
	\phihat(q, t_0) = \phihat(q,t_i) \cosh r - i e^{i\theta_\text{sq}} \phihatd(-q, t_i) \sinh r \, , \label{b(k)}
\end{equation}
where $q= k-k_1$, and $r$ and $\theta_\text{sq}$ are the squeezing parameter and angle, respectively. Strictly, $\phihat(q, t)$ and $\phihat^\dag(q, t)$ are defined in terms of the position-space photon field $\hat{E}(z,t)$:
\begin{subequations}
\label{b_and_bdag_Fourier}
\begin{eqnarray}
	\phihat(q, t) 	&=& \frac{1}{\sqrt{2\pi}}\int dz \, e^{-i (q + k_1) z} \hat{E}(z, t) \label{b_fourier}\\
	\phihatd(q, t) 	&=& \frac{1}{\sqrt{2\pi}} \int dz \, e^{i (q + k_1) z} \hat{E}^\dag(z, t).
\end{eqnarray}
\end{subequations}
Only the portion of the photon field under the pulse envelope $u_p(\textbf{r}, t)$ interacts with the atoms at a given time $t$. Furthermore, the physics of the atom-light interaction is determined by the temporal \emph{area} of the probe pulse. Consequently, the relevant mode of the photon field is
\begin{align}
	\bhat(t)	&\equiv \int \dr \, u^*_p(\boldr, t) e^{-i\boldk_1\cdot \boldr} \hat{E}(\boldr, t) \notag \\
			&= \int dq \, e^{i q c t}U_\text{prop}^*(q) \hat{\phi}(q, t) \notag \\
			&\approx \int dq \, U_\text{prop}^*(q) \hat{\phi}(q, t), \label{bt}
\end{align}
where 
\begin{equation}
	U_\mathrm{prop}(q) = \frac{1}{\sqrt{2\pi}}\int dz \, e^{-i q z} u_\mathrm{prop}(z).
\end{equation}
The second line of Eq.~(\ref{bt}) follows from Eq.~(\ref{b_fourier}) and $\int dx \, dy \, | u_\text{trans}(x,y) |^2 = 1$, and the dominant contribution to the integral occurs close to the carrier frequency $k_1$ (i.e. $q = 0$), thereby justifying the approximate expression in the third line \footnote{Actually, this phase factor $\exp(i q c t)$ is an artefact of writing the pulse shape $u_p(\textbf{r}, t) \exp(i \textbf{k}_1 \cdot \textbf{r})$ as a single frequency pulse with a slowly-varying envelope. Strictly, this pulse contains a range of frequencies about $k_1$ which exactly cancel $\exp(i q c t)$ in the above integral.}. It then follows from Eq.~(\ref{b(k)}) that
\begin{align}
	\hat{b}(t_0)	&= \left[ \int dq \, U_\text{prop}^*(q) \phihat(q, t_i) \right] \cosh r \notag \\
				-&\, i e^{i\theta_\text{sq}} \left[ \int dq \, U_\text{prop}^*(q)\phihatd(-q, t_i) \right] \sinh r. \label{bt_0_freq}
\end{align}
The first term in square brackets is clearly $\hat{b}(t_i)$. The second term in square brackets equals $[\hat{b}(t_i)]^\dag = \hat{b}^\dag(t_i)$ provided $U_\text{prop}(k)$ is real and symmetric, which in practice is easy to satisfy. Consequently, the photon mode output from the OPO is simply
\begin{align}
	\hat{b}(t_0)	&= \bhat(t_i)\cosh r - i e^{i\theta_\text{sq}} \bhatd(t_i) \sinh r, \label{single_mode_sq_opt_vac}
\end{align}
which for initial vacuum input is a single-mode squeezed state. Eq.~(\ref{bt_0_freq}) and Eq.~(\ref{single_mode_sq_opt_vac}) illustrate the relationship between the squeezing spectra, which is typically measured in optical squeezing experiments, and the temporal modes relevant for quantum state transfer. 

\begin{figure}[!t]
\centering
\includegraphics[width=\columnwidth]{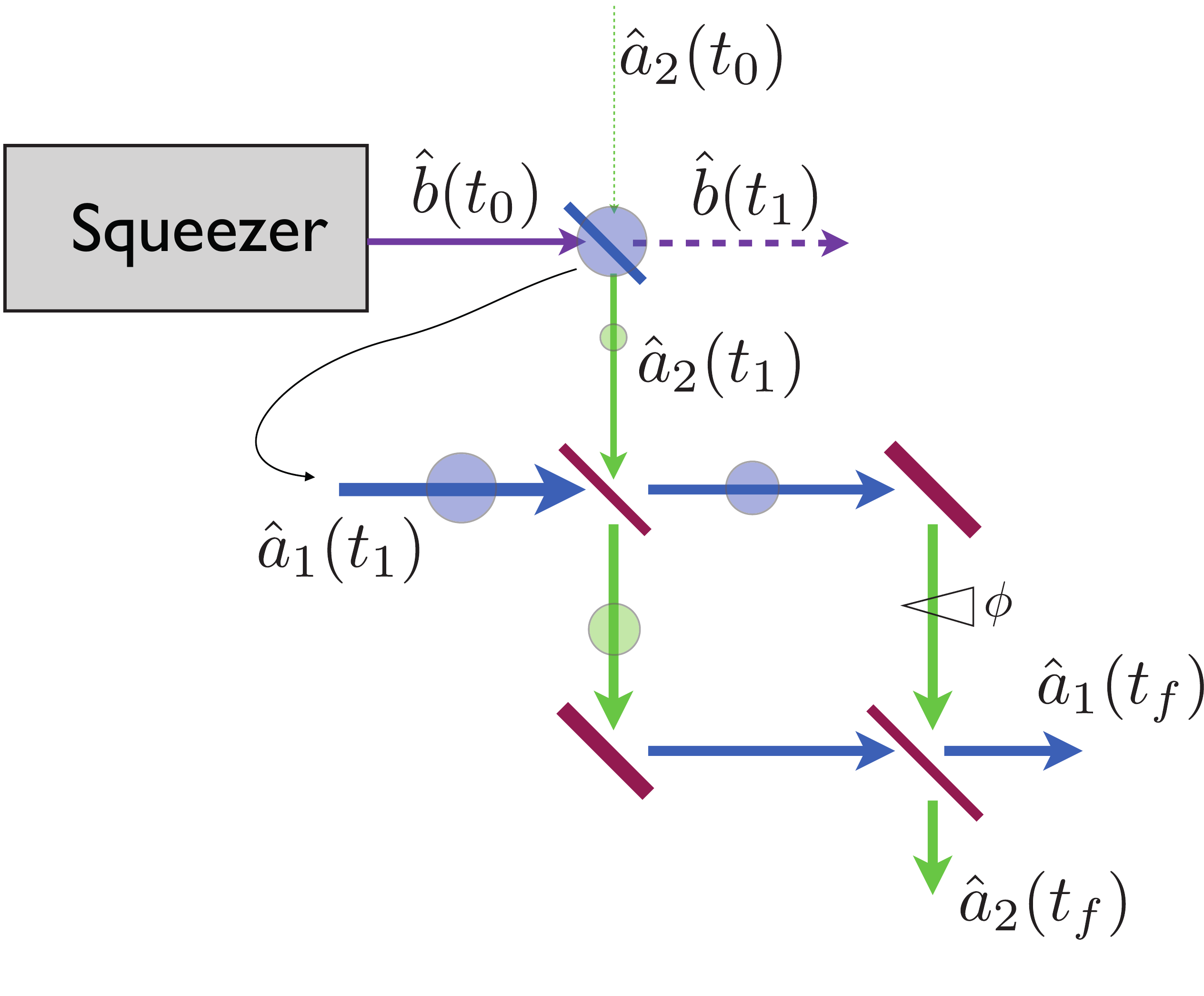}
\caption{A scheme for enhancing an atom interferometer with a single-mode squeezed optical vacuum. The squeezed light is used to outcouple a small number of atoms from a BEC. These outcoupled atoms and the remaining condensate atoms form the two inputs to a MZ atom interferometer.}
\label{fig:singlemode_scheme}
\end{figure}

Our scheme is summarized in Fig.~\ref{fig:singlemode_scheme}. An initial squeezed vacuum state $\hat{b}(t_0)$ is used to transfer a small number of atoms from mode $\ahat_1$ to $\ahat_2$ via the QST process (\ref{analytic_QST}), thereby transferring some or all of the quantum state from $\bhat(t_0)$ to $\ahat_2(t_1)$. The modes $\ahat_1(t_1)$ and $\ahat_2(t_1)$ then form the two input modes for a MZ atom interferometer (i.e. are coherently split, reflected and recombined), yielding the two outputs $\ahat_1(t_f)$ and $\ahat_2(t_f)$, used to construct the difference signal $\hat{S}_a$ [see Eq.~(\ref{sig1})]. Expectations are calculated with respect to the initial state $|\Psi(0)\rangle$, defined such that
\begin{subequations}
\begin{align}
	\hat{\phi}(q, t_i)|\Psi(0)\rangle 		&= \ahat_2(t_0)|\Psi(0)\rangle = 0 \\
	\ahat_1(t_1)|\Psi(0)\rangle 	&= \sqrt{N_{a_1}(t_1)}|\Psi(0)\rangle \, , \label{initialstate_a1}
\end{align}
\end{subequations}
where conservation of total atom number $N_t$ implies that $N_{a_1}(t_1) = N_t - \langle \hat{N}_{a_2}(t_1)\rangle$. Of course, in writing Eq.~(\ref{initialstate_a1}) we have assumed that the condensate is initially in a coherent state, and remains in a coherent state during the QST process. As shown below, this assumption is only valid when the number of outcoupled atoms $\langle \hat{N}_{a_2}(t_1)\rangle$ is much less than $N_t$.

\subsection{Complete quantum state transfer} \label{sec_single_mode_CQST}
\begin{figure}[!t]
\centering
\includegraphics[width=\columnwidth]{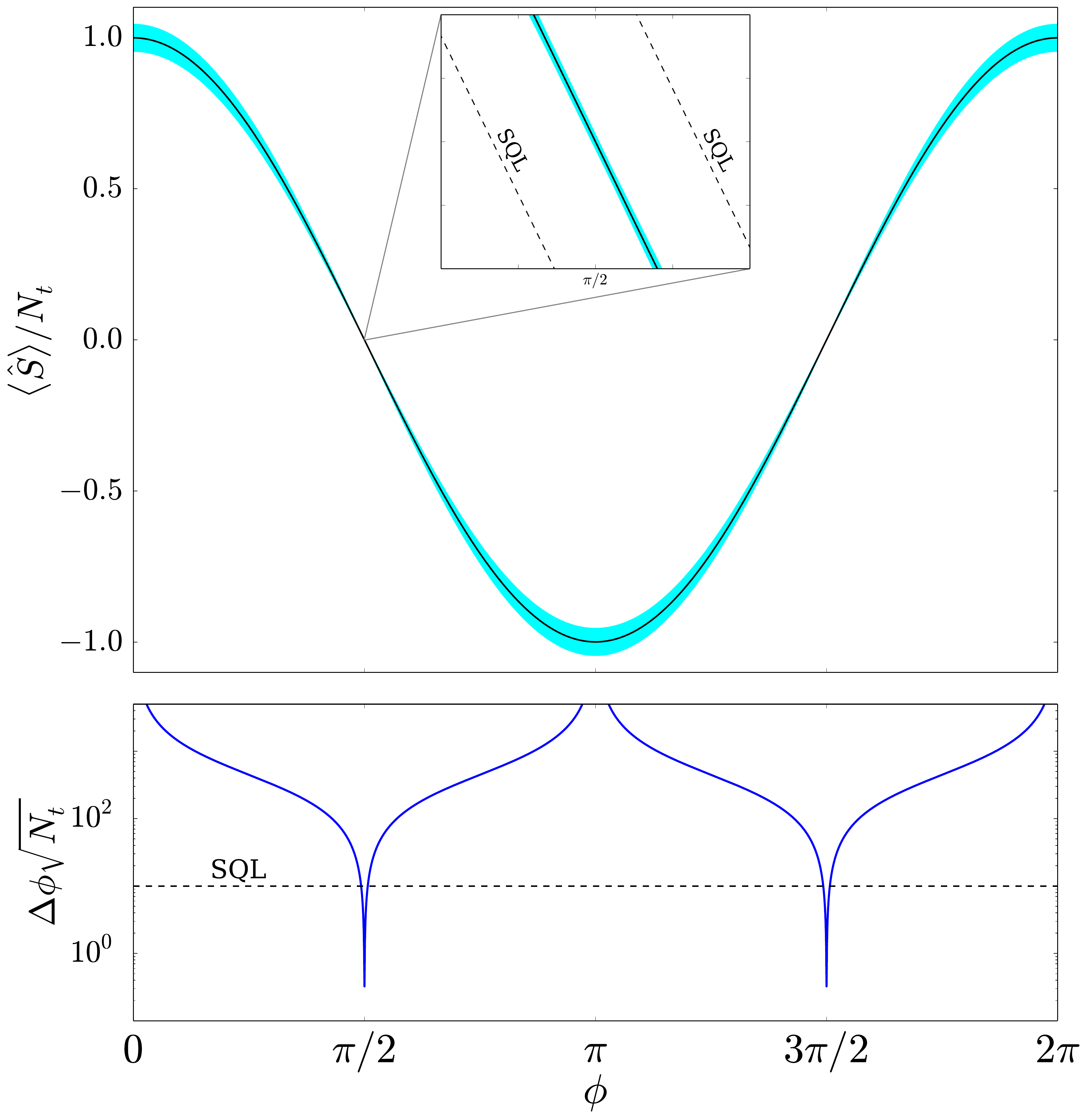}
\caption{ (Top).  The expectation value (black, solid line) and quantum uncertainty (light blue, shading) for the signal $\hat{S}$ normalized by total number of atoms ($N_t = 10^6$) for $\theta_\text{sq} = \pi/2$, and $r=r_\text{opt} \approx 3.8$. The inset shows that the variance is less than the SQL near $\phi = \pi/2$. (Bottom) The phase sensitivity for complete QST (blue, solid line), compared to the SQL.}
\label{fig:sigVsphi}
\end{figure}

We first consider the optimal regime of complete QST $\theta_\text{QST} = \pi$ (i.e. $\mathcal{Q} = 1$). In this case Eq.~(\ref{asol}) implies that $\hat{a}_{2}(t_1)=-i\hat{b}(t_0)$. The expectation of the difference signal (\ref{sig1}) simplifies to
\begin{equation}
	\langle \hat{S}_a \rangle = \left(N_t - 2\sinh^2 r\right) \cos \phi  \, . \label{avg_S_a}
\end{equation}
In order to achieve minimum phase sensitivity, the variance in the signal must attain a minimum when the slope of the output signal is a maximum [see Eq.~(\ref{deltaphidef})]. This occurs at phase $\phi = \pi/2$ and squeezing angle $\theta_\text{sq} = \pi/2$. Figure~\ref{fig:sigVsphi} shows the signal and phase sensitivity as a function of $\phi$ for $\theta_\text{sq} = \pi/2$. For $r=3.8$, we achieve an enhancement in sensitivity of approximately $30$ times better than the SQL. For comparison, we have shown the case where the two inputs to the MZ interferometer are a coherent state and a vacuum state, respectively. 

For these optimal values, the variance is simply
 \begin{equation}
 	V(S_a) = N_t e^{-2r} + 2 e^{-r}\sinh^3 r.
 \end{equation}
The minimum phase sensitivity, as a function of $r$ and $N_t$, is therefore 
\begin{align}
	\Delta \phi_\text{min} &= \frac{\sqrt{N_t e^{-2r} + 2 e^{-r}\sinh^3 r}}{ N_t - 2\sinh^2 r}  \label{sens_single_mode}\\
					&\approx \frac{e^{-r}}{\sqrt{N_t}}, \notag
\end{align}
where the approximate expression in the second line is only true in the limit $\langle \hat{N}_{b}(t_0)\rangle = \sinh^2 r \ll N_t$, where $\hat{N}_{b} = \hat{b}^\dag \hat{b}$.

\begin{figure}[t]
\centering
\includegraphics[width=\columnwidth]{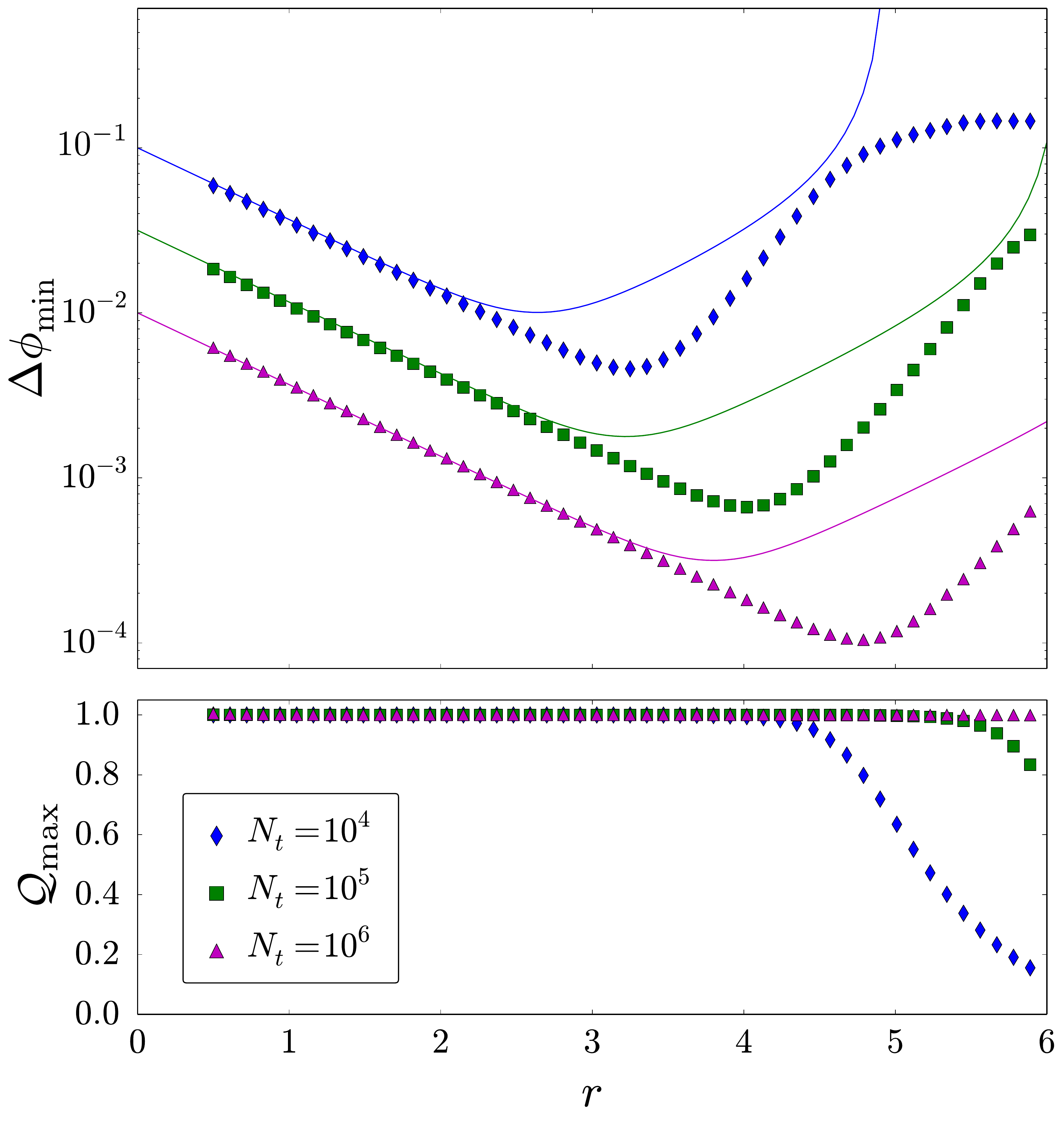}
\caption{(Top) Minimum phase sensitivity (i.e. at $\phi = \theta_\text{sq} = \pi/2$) as a function of squeezing parameter $r$ for initial atom numbers $N_t = 10^4$ (blue, top), $N_t = 10^5$ (green, middle) and $N_t = 10^6$ (magenta, bottom). The solid curves are the analytic solution [Eq.~(\ref{sens_single_mode})], while the points correspond to a TW numerical solution. The standard error in the TW solutions is no larger than the point width. There is good agreement between the analytics and numerics when $\langle \hat{N}_{b}(t_0) \rangle \ll N_t$, where the initial condensate is not significantly depleted. (Bottom) The maximum QST efficiency $\mathcal{Q}_\text{max} = \max_{t} \mathcal{Q}(t)$ as a function of $r$. Theoretically, complete QST ($\mathcal{Q} = 1$) is achievable provided less than $\sim 10$\% of the total condensate number is outcoupled. The analytics predict $\mathcal{Q} = 1$ always, so any deviation from this is due to depletion from the condensate mode $\hat{a}_1$. Note, however, that there exist regimes where mode $\hat{a}_1$ must be treated quantum mechanically even though $\mathcal{Q}_\text{max} = 1$.}
\label{fig:Dvsr}
\end{figure}

Figure~\ref{fig:Dvsr} shows the minimum interferometer sensitivity [Eq.~(\ref{sens_single_mode})] as a function of the squeezing parameter $r$, which determines the average number of input photons via $\langle \hat{N}_{b}(t_0) \rangle =\sinh^2 r$, for a range of initial BEC atom numbers. When $N_t \gg 1$, our analytic model predicts that an optimal squeezing parameter of $r_\text{opt} \approx \ln (4N_t)/4$ yields a minimum sensitivity of $\Delta \phi_\text{min}  \approx 1/N_t^{3/4}$. This is significantly less than the SQL, and furthermore is the best sensitivity possible in this undepleted regime provided $\sinh^2 r \ll N_t$ \cite{Pezze:2008, Lang:2013}. For $N_t=10^6$, this gives an enhancement in sensitivity of approximately $32$ compared with the SQL, which is equivalent to increasing the total number of atoms by a factor of $10^3$ at the SQL. For this value of $N_t$, the number of atoms outcoupled at $r=r_\text{opt}$ is $\sinh^2 r_\text{opt} \approx 500$, suggesting that the undepleted reservoir model is still reasonably valid in this regime. 

To include the effects of depletion from the condensate, we need to treat mode $\hat{a}_1$ quantum mechanically, and simulate the full quantum dynamics of the QST process, which are governed by Eqs~(\ref{3modeEOM}). This can be done via the truncated Wigner (TW) phase space method \cite{Steel:1998, Blakie:2008, Polkovnikov:2010, Opanchuk:2013}. Following standard methods \cite{Gardiner:2004b, Walls:2008}, the Heisenberg equations of motion are converted into a partial differential equation (PDE) for the Wigner quasi-probability distribution. Once third and higher-order derivatives are truncated (an uncontrolled approximation, but one that is typically valid provided the occupation per mode is not too small for appreciable time periods \cite{Sinatra:2002, Johnsson:2013}), this PDE takes the form of a Fokker-Planck equation, which can be efficiently simulated by a set of stochastic differential equations (SDEs) for complex numbers $\alpha_j(t)$. In our case, the set of SDEs corresponding to Eqs~(\ref{3modeEOM}) (with $\delta = 0$) is
\begin{subequations}
\label{wigner_SDEs}
\begin{align}
	i \dot{\alpha_1}	&= gf(t) \alpha_2 \beta^*, \\
	i \dot{\alpha_2}	&= gf(t) \alpha_1 \beta, \\
	i \dot{\beta}   	&= gf(t) \alpha_2 \alpha_1^*,
\end{align} 
\end{subequations}
where we have made the correspondences $\ahat_{i}(t) \rightarrow \alpha_{i}(t)$ and  $\bhat(t) \rightarrow \beta(t)$. The initial conditions for these SDEs are randomly sampled from the Wigner distribution corresponding to the initial quantum state \cite{Olsen:2009}. Specifically, just before the QST process, $\hat{a}_1$ is in a coherent state of mean number $N_t$, $\hat{a}_2$ is in a vacuum state, and $\hat{b}$ is in a single-mode squeezed vacuum state. The initial conditions corresponding to these initial states are
\begin{subequations}
\label{wigner_SDEs_initial}
\begin{align}
	\alpha_1(t_0) 		&= \sqrt{N_t} + \eta_{\alpha_1}, \\
	\alpha_{2}(t_0)		&= \eta_{\alpha_{2}}, \\
	\beta	(t_0)			&= \eta_{\beta} \cosh r  - i e^{i \theta_\text{sq}} \eta_{\beta}^* \sinh r.
\end{align}
\end{subequations}
The $\eta_i$ are complex, independent Gaussian noises satisfying $\overline{\eta_i} = 0$ and $\overline{\eta^*_i\eta_j} = \delta_{ij}$. The expectation value of some arbitrary operator function $h$ is then computed by averaging over solutions to Eqs~(\ref{wigner_SDEs}) with initial conditions~(\ref{wigner_SDEs_initial}):
\begin{equation}
	\langle \{ h(\ahatd_1, \ahatd_2, \bhatd, \ahat_1, \ahat_2, \bhat ) \}_\mathrm{sym} \rangle = \overline{h\left(\alpha^*_1, \alpha^*_2, \beta^*, \alpha_1, \alpha_2, \beta \right)} \, ,
\end{equation}
where ``sym'' denotes symmetric ordering \cite{Walls:2008}, and the overline denotes the average of simulated trajectories. 

Since beamsplitting and mirror operations are implemented by strong coherent optical fields, we can approximate them as linear, and the complex amplitudes $\alpha_1(t_f)$ and $\alpha_2(t_f)$ can be directly evolved from $\alpha_1(t_1)$ and $\alpha_2(t_1)$ by repeated application of Eqs~(\ref{coherent_BS_soln}). The mean $\langle \hat{S}_a \rangle$ and variance $V(S_a)$ can then be computed using the relations
\begin{subequations}
\label{N_Wigner_exp}
\begin{align}
	\langle \hat{N}_i \rangle 	&= \overline{|\alpha_i|^2} - 1/2, \\
	\langle \hat{N}_i^2 \rangle	&= \overline{|\alpha_i|^4} - \overline{|\alpha_i|^2}, \\
	\langle \hat{N}_1 \hat{N}_2 \rangle	&=  \overline{\left(|\alpha_1|^2 - 1/2\right)\left(|\alpha_2|^2 - 1/2\right)}.
\end{align}
\end{subequations}

\begin{figure}[t]
\centering
\includegraphics[width=\columnwidth]{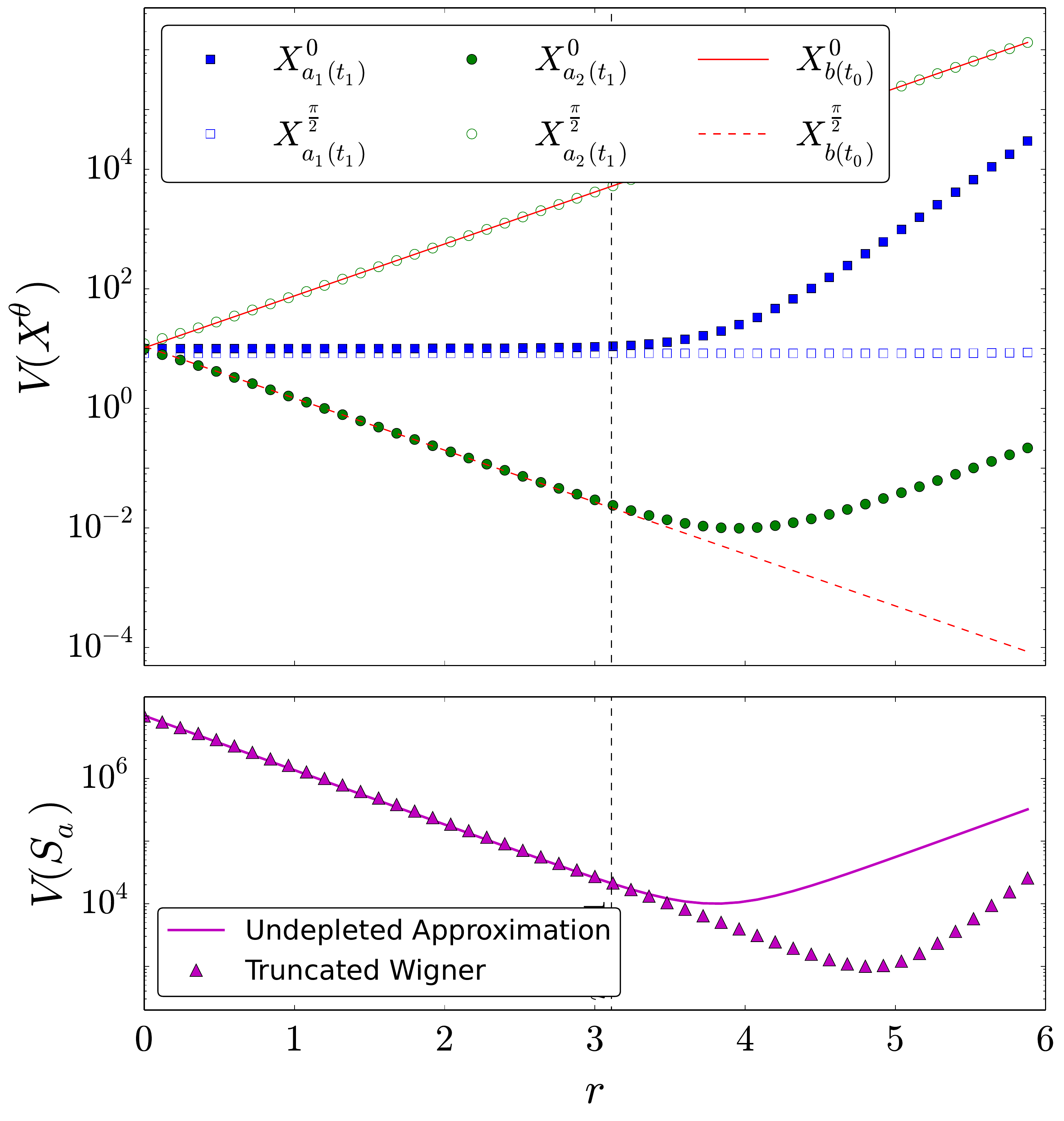}
\caption{(Top) TW simulations for the variance of atomic quadratures $\hat{X}_c^\theta = \exp(i \theta) \hat{c} + \exp(-i \theta) \hat{c}^\dag$, where $\hat{c}$ is an arbitrary mode, for an initial condensate number of $N_t = 10^6$. The quadratures for the initial squeezed optical input $\hat{b}_0$ are analytic, and can easily be computed from Eq.~(\ref{single_mode_sq_opt_vac}). According to the undepleted reservoir analytics, which assumes $\hat{a}_1$ is always coherent, the variance of any quadrature $\hat{X}_{a_1(t_1)}^\theta$ is unity, and $\hat{X}_{a_2(t_1)}^{0} = \hat{X}_{b(t_0)}^{\pi/2}$ and $\hat{X}_{a_2(t_1)}^{\pi/2} = \hat{X}_{b(t_0)}^{0}$ when QST is complete. However, even though $\mathcal{Q}=1$ for all values of $r$ shown here, TW simulations show that $\hat{X}_{a_1(t_1)}^{0}$ and $\hat{X}_{a_2(t_1)}^{0}$ diverge from the undepleted reservoir solution once $r \gtrsim 3.11$. (Bottom) Plot showing that the quantum noise (i.e. variance) on the signal $\hat{S}_a$ also diverges from the undepleted reservoir model for $r \gtrsim 3.11$.}
\label{fig:VarX}
\end{figure}

The effects of depletion on the minimum interferometer sensitivity, as numerically modelled by TW simulations, is shown in Fig.~\ref{fig:Dvsr}. Since the efficiency of the QST process does not uniquely depend upon the dynamics of the QST process (i.e. $\mathcal{Q}$ is not uniquely defined by the choice of $f(t)$), for simplicity simulations were performed with a uniform $f(t)$. For small $r$ we find good agreement between the undepleted reservoir approximation and the full quantum dynamics. However, as $r$ increases, the full quantum simulations actually predict a \emph{better} sensitivity than the undepleted reservoir model, reaching a minimum at an $r = r_\text{TW} > r_\text{opt}$. We can understand this feature by looking at the quadratures of the modes, for example $\hat{X}_{a_1}^\theta = \exp(i \theta) \hat{a}_1 + \exp(-i \theta) \hat{a}_1^\dag$. As shown in Fig.~\ref{fig:VarX}, the discrepancy is due to changes in $\hat{a}_1$ under depletion, evidenced by $V(\hat{X}_{a_1(t_1)}^{0})$ and $V(\hat{X}_{a_2(t_1)}^{0})$ deviating from the analytic solution. Although the state of $\hat{a}_1$ is coherent under the undepleted reservoir approximation, quantum depletion creates a state with decreased variance in $\hat{J}_x = [\ahat_1(t_1)\ahatd_2(t_1) + \ahat_2(t_1)\ahatd_1(t_1)]/2$ (and increased variance in $\hat{X}_{a_1(t_1)}^{0}$ and $\hat{X}_{a_2(t_1)}^{0}$). This gives a reduction in the sensitivity at $\phi = \pi/2$, since the noise in the signal is directly proportional to $\hat{J}_x$ by virtue of Eq.~(\ref{sig1}). 

For further increases in $r$, the effects of depletion become significant, and complete QST (i.e. $\mathcal{Q} = 1$) is no longer possible. The maximum possible QST efficiency, as a function of $r$, is shown in Fig.~\ref{fig:Dvsr}. Unsurprisingly, this is contrary to the undepleted reservoir model, where $\mathcal{Q} = 1$ always for $\theta_\text{QST} = \pi$. 

\subsection{Incomplete quantum state transfer and information recycling} \label{sec_single_mode_incomplete}
\begin{figure}[!t]
\centering
\includegraphics[width=\columnwidth]{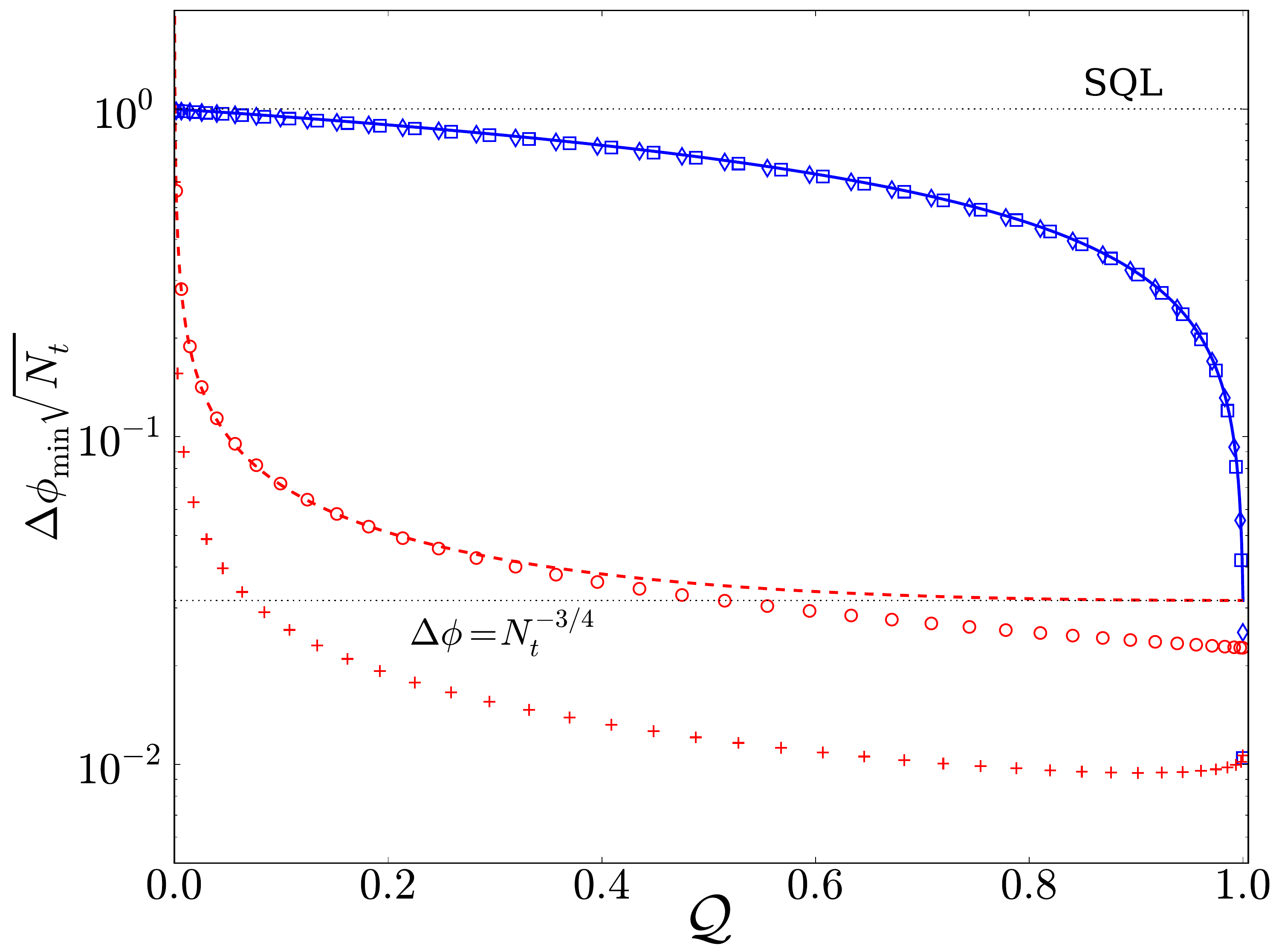}
\caption{Plots showing the QST efficiency dependence of the minimum phase sensitivity $\Delta \phi_\text{min}$ for an atom interferometer enhanced by single-mode squeezing, and initial atom number $N_t = 10^6$. For $r = r_\text{opt} \approx 3.8$, the sensitivity of the purely atomic signal $\hat{S}_a$ sharply degrades with decreasing $\mathcal{Q}$, and the TW simulations (blue diamonds) agree with the analytic undepleted reservoir prediction (solid blue line) except near $\mathcal{Q} = 1$. In contrast, the information-recycled signal $\hat{S} = \hat{S}_a -\mathcal{G}\hat{S}_b$ displays considerably better sensitivities and a slower degradation as $\mathcal{Q}$ decreases. For $r = r_\text{opt}$, the TW simulations (red circles) predict a slightly better sensitivity for $\mathcal{Q} \gtrsim 0.4$ than the analytic undepleted reservoir prediction (red dashed line). For comparison, the upper and lower horizontal black dotted lines show the SQL and the theoretical limit reached by perfect QST (i.e. $\Delta \phi = 1/N_t^{3/4}$), respectively. Interestingly, the TW simulations (red crosses) demonstrate that better sensitivities can be achieved for $r = r_\text{TW} \approx 4.8 > r_\text{opt}$; for $\hat{S}_a$ this is only true near $\mathcal{Q} = 1$ (blue squares).}
\label{fig:delta_phi_vs_theta_qst}
\end{figure}

In practice, it may be difficult to achieve the required coupling strength $g$ for complete QST. When $\theta_\text{QST} < \pi$, and optimal values $\phi = \pi/2$ and $\theta_\text{sq} = \pi/2$ are chosen, the undepleted reservoir model predicts a signal slope and variance of
\begin{subequations}
\begin{align}
	\frac{d\langle \hat{S}_a \rangle}{d\phi} 	&= - \left(N_t - 2\sin^2 \left(\theta_\text{QST}/2\right) \sinh^2 r\right), \\
	\intertext{and}
	V(S_a)							&=  N_t e^{-r} \left( \cosh r + \cos \theta_\text{QST} \sinh r \right) \notag \\
									&+ 2 e^{-r} \sin^4\left(\theta_\text{QST}/2\right) \sinh^3 r,
\end{align}
\end{subequations}
respectively. 
This gives a minimum phase sensitivity of
\begin{equation}
	\Delta \phi_\text{min} = \frac{\sqrt{N_t\left( 1 - 2 \mathcal{Q} e^{-r} \sinh r \right) + 2 \mathcal{Q}^2 e^{-r} \sinh^3 r}}{N_t - 2 \mathcal{Q} \sinh^2 r}.
\end{equation}
The blue curve in Fig.~\ref{fig:delta_phi_vs_theta_qst} shows $\Delta \phi_\text{min}$ vs. $\mathcal{Q}$ at optimum squeezing parameter ($r=r_\text{opt}$ when the undepleted reservoir approximation holds) for an initial condensate of $N = 10^6$ atoms. Although sub-SQL sensitivities are still possible for incomplete QST, the enhancement sharply reduces from the optimal $1/N_t^{3/4}$ scaling as $\mathcal{Q}$ decreases from unity. For instance, at $\mathcal{Q} = 0.5$, corresponding to $\theta_\text{QST} = \pi/2$ in the undepleted reservoir regime, the enhancement to the sensitivity beyond the SQL has dropped by a factor of 20 to $\sim\sqrt{2}$. 
The blue curve in Fig.~\ref{fig:delta_phi_vs_theta_qst} shows $\Delta \phi_\text{min}$ vs. $\mathcal{Q}$ at optimum squeezing parameter ($r=r_\text{opt}$ when the undepleted reservoir approximation holds) for an initial condensate of $N = 10^6$ atoms. Although sub-SQL sensitivities are still possible for incomplete QST, the enhancement sharply reduces from the optimal $1/N_t^{3/4}$ scaling as $\mathcal{Q}$ decreases from unity. For instance, at $\mathcal{Q} = 0.5$, corresponding to $\theta_\text{QST} = \pi/2$ in the undepleted reservoir regime, the enhancement to the sensitivity beyond the SQL has dropped by a factor of 20 to $\sim\sqrt{2}$. 

\begin{figure}[t]
\centering
\includegraphics[width=1.0\columnwidth]{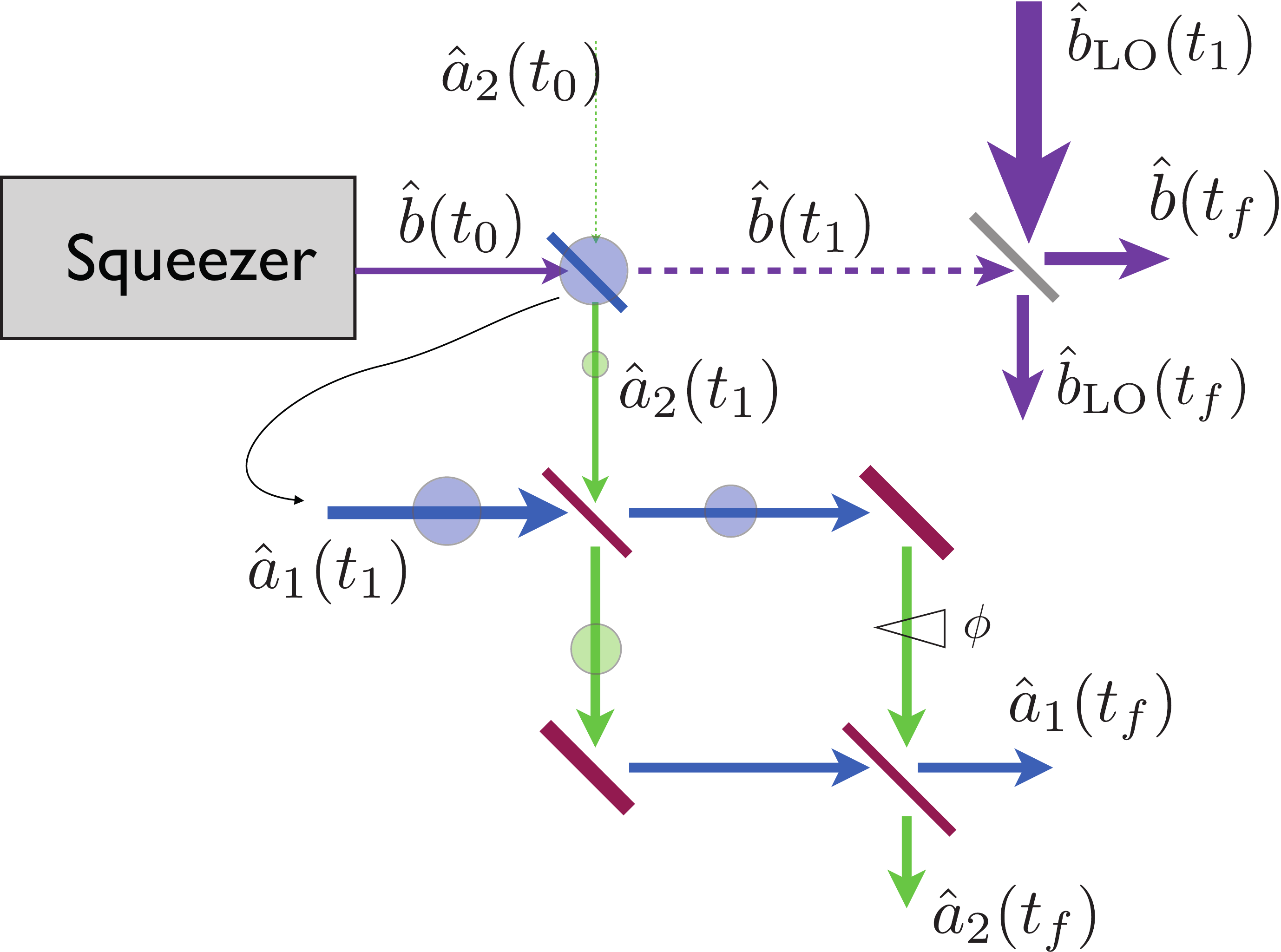}
\caption{An atom interferometer enhanced by both single-mode squeezed light and information recycling. The transmitted component of $\bhat(t_1)$ is interfered with a bright local oscillator, thereby allowing a homodyne measurement. The information from this measurement is then combined with the atomic signal. Information recycling gives an improvement to the sensitivity when the QST efficiency $\mathcal{Q}$ is less than unity.}
\label{singlemode_info_recyc_scheme}
\end{figure}

Fortunately, this degradation to the sensitivity due to incomplete QST can be ameliorated with the technique of \emph{information recycling} \cite{Haine:2013}. Specifically, after the atom-light beamsplitter, a quadrature of the transmitted field $\bhat(t_1)$ can be measured via homodyne detection by mixing the field $\bhat(t_1)$ with a bright local oscillator $\bhat_\text{LO}(t_1)$, which is assumed to be a large amplitude coherent state (see Fig.~\ref{singlemode_info_recyc_scheme}). In order to pick out the mode corresponding to $\bhat$, the local oscillator would need to be temporally shaped such that it is mode-matched to $u_p(\boldr,t)$.  The noise on this homodyne signal is correlated with the noise on the atomic signal $\hat{S}_a$, and hence can be combined with the atomic signal to reduce the overall noise of the phase measurement. That is, from the apparatus depicted in Fig.~\ref{singlemode_info_recyc_scheme} we construct the signal 
\begin{equation}
	\hat{S} = \hat{S}_a -\mathcal{G}\hat{S}_b, \label{info_recyc_signal}
\end{equation}
where 
\begin{subequations}
\begin{align}
	\hat{S}_b			&= \hat{N}_{b_\text{LO}}(t_f) - \hat{N}_{b}(t_f), \\
	\bhat(t_f) 			&= \frac{1}{\sqrt{2}}\big(\bhat(t_1)- i\bhat_\text{LO}(t_1)\big), \\
	\bhat_\text{LO}(t_f) 	&= \frac{1}{\sqrt{2}}\big(\bhat_\text{LO}(t_1)- i\bhat(t_1) \big), 
\end{align}
\end{subequations}
and $\mathcal{G}$ is an adjustable \emph{gain parameter}. 

In order to see more clearly how information recycling improves the phase sensitivity, it is instructive to first consider the optical analogy of the incomplete QST process - a beamsplitter. We know from the quantum optics literature that a single-mode squeezed state $\hat{b}_\text{sq}$ incident on a $50/50$ beamsplitter leads to entanglement between the two output fields. Specifically, the two beamsplitter outputs are $\ahat = (\hat{\vartheta} - i \hat{b}_\text{sq})/\sqrt{2}$ and $\bhat = (\hat{b}_\text{sq} - i \hat{\vartheta})/\sqrt{2}$, where $\hat{\vartheta}$ is a vacuum input. The entanglement leads to a variance $V\big( (\hat{X}_{a}^0-\hat{X}_{b}^{\pi/2})/\sqrt{2} \big) = \exp(-2r)$, which is less than for two uncorrelated coherent inputs, where we have defined the generalized quadrature of each output as $\hat{X}_{a(b)}^\theta = \ahat(\bhat) \exp(i\theta) +\ahatd(\bhatd) \exp(-i\theta)$. More generally, for an asymmetric beamsplitter with beamsplitting ratio (i.e. \emph{reflection coefficient}) $\sin^2(\theta_\text{QST}/2)$, the quantity
\begin{equation}
	V\left(\sin(\theta_\text{QST}/2) \hat{X}_{a}^0 - \cos (\theta_\text{QST}/2) \hat{X}_{b}^{\pi/2} \right) = e^{-2r} \label{V_BS_general}
\end{equation}
is less than for two uncorrelated states.  Essentially the same argument can be used to show that the information-recycled signal~(\ref{info_recyc_signal}) has a smaller variance, and therefore gives a smaller phase sensitivity, than the signal $\hat{S}_a$. Since $\bhat_\text{LO}(t_1)$ is a bright coherent state of complex amplitude $\sqrt{N_\text{LO}} \exp(i\theta_\text{LO})$, $N_\text{LO} \gg \langle \hat{N}_{b}(t_1) \rangle$ and therefore
\begin{equation}
\hat{S}_b \approx \sqrt{N_\text{LO}} \hat{X}_{b(t_1)}^{\theta_\text{LO}} \, .
\end{equation}
Similarly, $N_{a_1}(t_1) \gg \langle \hat{N}_{a_2}(t_1) \rangle$, and so at the most sensitive point of the atom interferometer, $\phi=\pi/2$,
\begin{equation}
	\hat{S}_a \approx \sqrt{N_{a_1}(t_1)}\hat{X}_{a_2(t_1)}^0 \, . \label{S_a_approx}
\end{equation} 
Thus, choosing $\theta_\text{LO} =\pi/2$ (i.e. we measure the phase quadrature of the transmitted photons), and 
\begin{equation}
\mathcal{G} = \sqrt{\frac{N_{a_1}(t_1)}{N_\text{LO}} \left(\frac{1 - \mathcal{Q}}{\mathcal{Q}}\right)} = \sqrt{\frac{N_{a_1}(t_1)}{N_\text{LO}}}\cot (\frac{\theta_\text{QST}}{2}),
\end{equation}
yields the combined (information-recycled) signal
\begin{equation}
	\hat{S} \approx \frac{\sqrt{N_{a_1}(t_1)}}{ \sin (\tfrac{\theta_\text{QST}}{2})}\left(\sin (\frac{\theta_\text{QST}}{2}) \hat{X}_{a_2(t_1)}^0 - \cos (\frac{\theta_\text{QST}}{2}) \hat{X}_{b(t_1)}^{\pi/2}  \right), \label{single_mode_S_approx}
\end{equation}
which can have smaller variance than $\hat{S}_a$. 

Figure~\ref{fig:delta_phi_vs_theta_qst} plots the minimum phase sensitivity $\Delta \phi_\text{min}$ as a function of QST efficiency $\mathcal{Q}$ corresponding to both the purely atomic signal $\hat{S}_a$ (solid, blue curve) and the information-recycled signal $\hat{S} = \hat{S}_a-\mathcal{G}\hat{S}_b$ (dashed, red curve). Although both instances suffer a degradation of sensitivity with poorer QST, this degradation is significantly arrested for the information-recycled signal. As a specific comparison, when $\mathcal{Q} = 0.5$, the analytic model predicts that information recycling gives $\Delta \phi_\text{min} \approx  0.035 / \sqrt{N_t}$, compared with $\Delta \phi_\text{min} \approx  0.71 / \sqrt{N_t}$ in the absence of information recycling. Even at $10\%$ QST efficiency, information recycling gives a sensitivity more than a factor of ten better than the SQL, whereas there is a negligible enhancement in the absence of information recycling. For very low levels of QST ($<0.05 \%$ for $r= r_\text{opt}$), the information recycling scheme gives sensitivities above the SQL. This is because $V(S)$ is very sensitive to slight imperfections in the estimates of the quadratures when QST is very low. Such imperfections arise due to the finite size of the condensate initially populating mode $\hat{a}_1$. For although the approximation (\ref{S_a_approx}) is exact in the limit of an infinitely large condensate, we are typically only working with $10^4 - 10^6$ atoms. Hence the deviation of $\hat{S_a}$ from a perfect quadrature measurement is considerable, which is the cause of the discrepancy at low values of $\mathcal{Q}$.

Figure~\ref{fig:delta_phi_vs_theta_qst} also compares the sensitivities predicted by the analytic undepleted reservoir solutions to TW simulations [see Eqs~(\ref{wigner_SDEs})]. Without information recycling, the agreement is excellent except near $\mathcal{Q} = 1$. In contrast, TW simulations predict \emph{better} sensitivities from the information-recycled signal $\hat{S}$ (by a factor between four and five), occurring at a squeezing parameter $r_\text{TW}$ larger than the analytic optimum $r_\text{opt}$. As discussed in Sec.~\ref{sec_single_mode_CQST}, this improvement is due to $\hat{a}_1$ deviating from a coherent state, leading to a reduced variance in $\hat{J}_x$ at the output.


\subsection{Effect of losses on phase sensitivity} \label{sec_losses}
Quantifying the effects of losses on the sensitivity is an important experimental consideration. The simplest method of accounting for losses is by introducing virtual beamsplitters with a transmission coefficient of $\eta$ that input vacuum noise $\hat{\vartheta}$ at various points within the interferometry scheme. Specifically, this maps some mode $\hat{c}$ to $\sqrt{\eta} \hat{c} + \sqrt{1-\eta}\hat{\vartheta}$.

Using this approach, we considered four types of losses:
\begin{enumerate}
	\item Losses in the generation and transmission of the squeezed optical state before the QST process - i.e. $\hat{c} = \hat{b}(t_0)$.
	\item Losses in the mode $\hat{a}_2$ after the QST process, due to imperfections in the QST process such as spontaneous emission - i.e. $\hat{c} = \hat{a}_2(t_1)$.
	\item Losses in the transmitted optical state, including detection inefficiency - i.e. $\hat{c} = \hat{b}(t_1)$.
	\item Symmetric losses within the atom interferometer, which also accounts for inefficient atom detection - i.e. $\hat{c} = [\hat{a}_1(t_1) - i \hat{a}_2(t_1)]/\sqrt{2}$ and $\hat{c} = [\hat{a}_2(t_1) - i \hat{a}_1(t_1)]/\sqrt{2}$.
\end{enumerate}

\begin{figure}[t]
\includegraphics[width=\columnwidth]{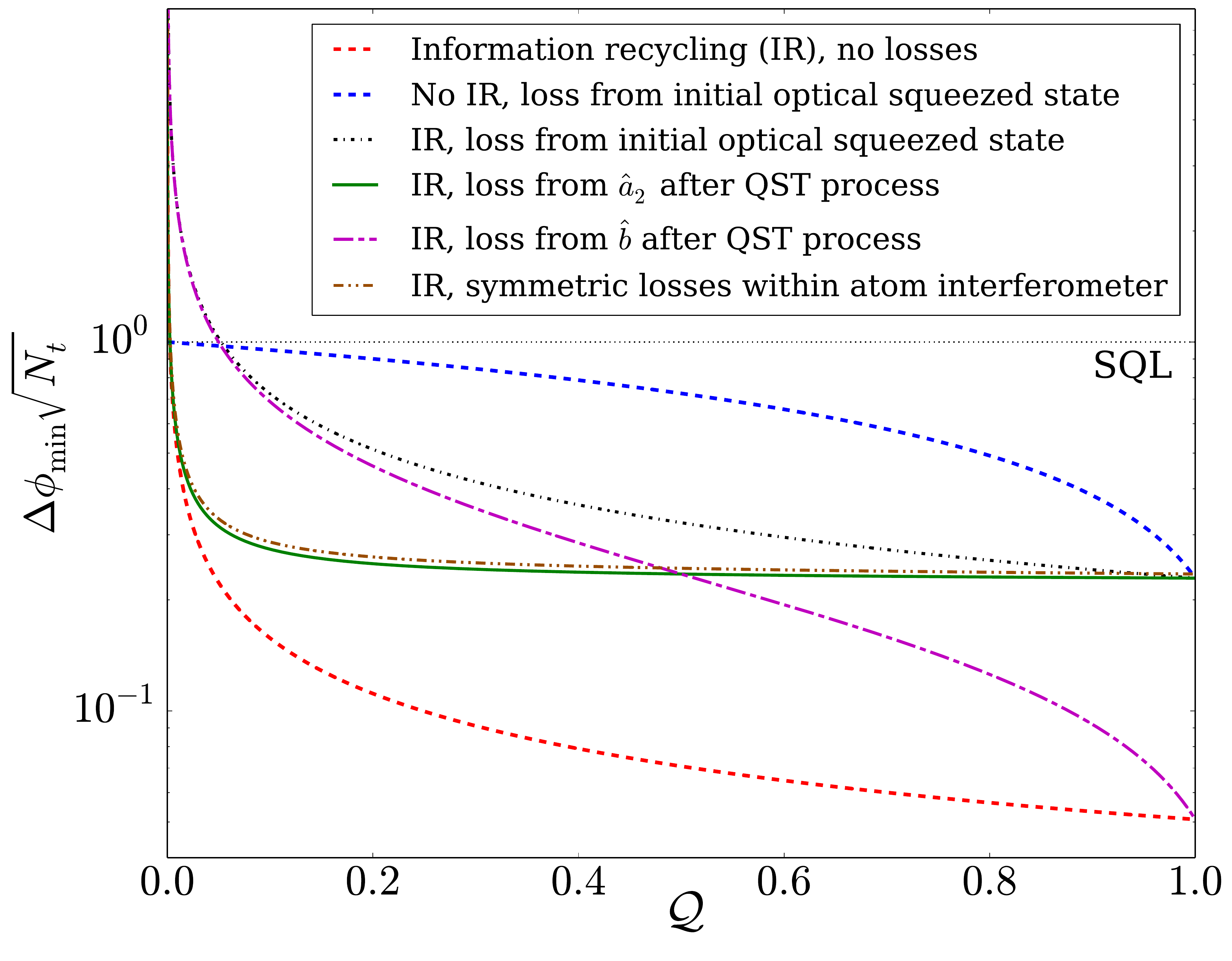} 
\caption{An illustration of the effects of losses ($\eta = 0.95$) on the phase sensitivity within the undepleted reservoir model for $N_t = 10^6$ and $r = 3$. Although losses clearly degrade the sensitivity, in most cases information recycling ameliorates this degradation. }
\label{loss_plot}
\end{figure}

Figure~\ref{loss_plot} illustrates the various effects of these losses on the sensitivity for an inefficiency of $\eta = 0.95$. Unsurprisingly, losses degrade the phase sensitivity, both with and without the inclusion of information recycling. Nevertheless, it is interesting that any type of loss is never worse than losses affecting the initial squeezed optical state. Furthermore, information recycling still delivers sensitivities below the SQL, and for values of $\mathcal{Q} > 5 - 10\%$ these are much better than what is possible without information recycling.

Since losses affect other squeezed-light enhanced atom interferometry schemes in a qualitatively similar fashion, we will not discuss losses further. We simply note that losses degrade the effects of squeezing, as they do in any optical squeezing experiment, and that if losses are not too great then information recycling can somewhat ameliorate the effects of this degradation.

\section{Enhanced atom interferometry with two-mode squeezed light} \label{sec_high_freq_sq}
\begin{figure*}[!htb]
\includegraphics[width=\textwidth]{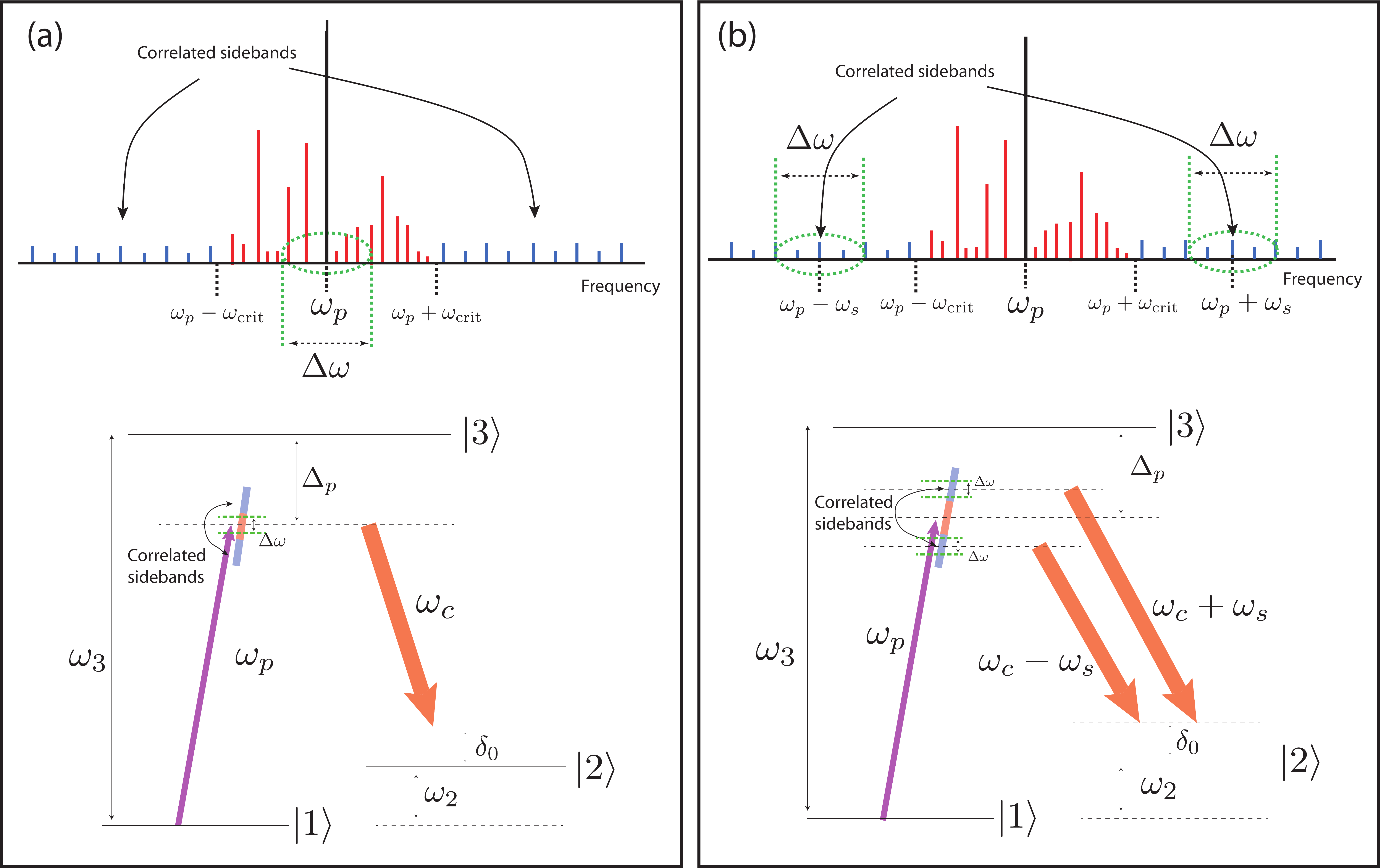} 
\caption{(a): Single-mode outcoupling scheme. If $2 \omega_\text{crit}>\Delta \omega$, then only the uncorrelated part of the spectrum of $\hat{E}_1$ takes part in the outcoupling process. No enhancement to the sensitivity is possible in this case. (b): An outcoupling method for utilizing high frequency squeezing. Two control beams (frequencies $\omega_c-\omega_s$ and $\omega_c+\omega_s$) are used to outcouple from the region of the spectrum of $\hat{E}_1$ centered around $\omega_p-\omega_s$ and $\omega_p+\omega_s$, respectively. Since $\Delta_p \gg \omega_s$, the coupling strengths of the two control beams are nearly identical. Here $\delta_0 = \omega_p - \omega_c - \omega_2$, which equals the change in the atomic kinetic energy $\hbar |\textbf{k}_1 - \textbf{k}_2|^2 / (2m)$ on two-photon resonance.}
\label{sidebands}
\end{figure*}

We now consider the case of high frequency squeezed light, which is experimentally less challenging to generate than low frequency squeezed light. We model this as a state of light that is dominated by uncorrelated classical noise at sideband frequencies below some frequency $\omega_\mathrm{crit}$, while containing quantum correlations above this frequency. Specifically, the mode with frequency $\omega_p + \omega$ is correlated with the mode with frequency $\omega_p - \omega$ only when $\omega>\omega_\mathrm{crit}$. Implementing the scheme from  Sec.~\ref{sec_single_mode_atom_int} would result in atoms being outcoupled without quantum correlations [see Fig.~\ref{sidebands}(a)]. We can, however, exploit this source of high-frequency squeezing in order to generate an atomic state that displays \emph{two-mode} squeezing. By adjusting the frequency of the classical control field from $\omega_c$ to $\omega_c+\omega_s$, where $\omega_s > \omega_\mathrm{crit}$, the mode that is on two-photon resonance $\delta = 0$ becomes far from the probe carrier frequency $\omega_p$, and outside the noisy low frequency region. Consequently, QST occurs within a spectral region of width $\Delta \omega$ centered around $\omega_p+\omega_s$. However, this does not yield an enhancement to the sensitivity of our atom interferometer, as a mode in this region displays no special quantum correlations in isolation; it only displays squeezing when considered in conjunction with the modes centered around $\omega_p - \omega_s$. We therefore include a second classical control beam of frequency $\omega_c-\omega_s$, which is on two-photon resonance and thereby causes QST at those frequencies between $\omega_p - (\omega_s -\Delta \omega/2)$ and $\omega_p -(\omega_s+\Delta \omega/2)$ [see Fig.~\ref{sidebands}(b)]. By making these two control fields counter-propagating (\ie wavevectors $\boldk_2$ and $-\boldk_2$), the atom-light interaction results in two correlated modes of outcoupled atoms with different momenta, which can be easily distinguished. 

\begin{figure}[t]
\includegraphics[width=0.8\columnwidth]{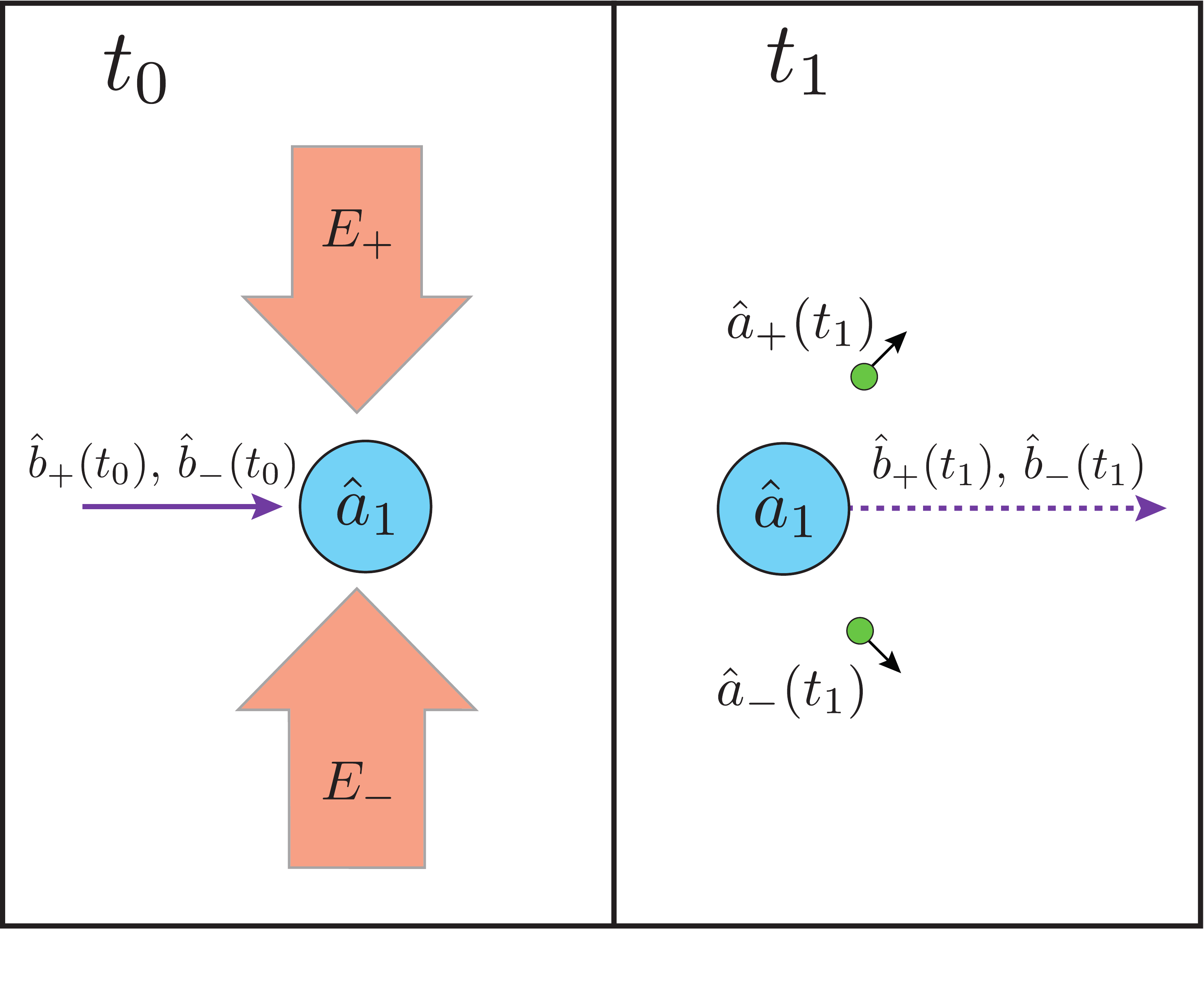} 
\caption{QST scheme for a two-mode squeezed optical input. Two counter-propagating classical control beams $E_{+}$ (frequency $\omega_c +\omega_s$) and $E_{-}$ (frequency $\omega_c -\omega_s$) are used to implement a Raman transition with $\hat{E}_1$. Due to energy-momentum resonance, $E_{\pm}$ is resonant with the region of $\hat{E}_1$ centered at frequency $\omega_p \pm \omega_s$. This outcouples atoms from mode $\hat{a}_1$, and results in QST between the optical modes $\bhat_{\pm}$ and atomic modes $\ahat_{\pm}$. Note that $\ahat_{\pm }$ recoils with momentum $\hbar\mathbf{k}_1\pm \hbar\mathbf{k}_2$, where $\pm\mathbf{k}_2$ is the wavevector of $E_{\pm}$. } 
\label{twomode_scheme}
\end{figure}

More precisely, modifying Hamiltonian~(\ref{full_Ham}) such that there are two counter-propagating classical control fields at frequencies $\omega_c \pm \omega_s$ and wavevectors $\pm\textbf{k}_2$, respectively, both shaped by the pulse envelope $u_c(\textbf{r},t) = u_\text{trans}(x,y)u_\text{prop}(z-ct)$, gives
\begin{align}
	\hat{\mathcal{H}} 	&= \sum_{j=1,2} \int  d\textbf{r} \, \psihatd_j(\boldr) H_j(\textbf{r}) \psihat_j(\boldr) +\hat{\mathcal{H}}_\text{light} \notag\\
					+&\, \hbar g \int d\textbf{r}  \left[ \psihat_1(\boldr) \psihatd_2(\boldr) \hat{E}_1(\boldr) u^*_c(\textbf{r},t)\left(e^{- i [\boldk_2\cdot \boldr - (\omega_c + \omega_s) t]} \right. \right. \notag \\
					+&\, \left. \left. e^{- i [-\boldk_2\cdot \boldr - (\omega_c - \omega_s) t]} \right) + h.c. \right]. \label{full_Ham_two_controls}
\end{align} 
As in Sec.~\ref{sec_theoretical_model}, we assume that the entire atomic population is initially in hyperfine state $|1\rangle$ and in spatial mode $u_0(\boldr)$. Conservation of energy and momentum implies that the modes resonant with the atom-light coupling are:
\begin{subequations}
\label{two_mode_expansions}
\begin{eqnarray}
\ahat_1 &=& \int \dr \, u_0^*(\boldr) \psihat_1(\boldr), \\
\ahat_{\pm} &=& \int \dr \, u_0^*(\boldr)e^{i(\boldk_{\pm} \mp \boldk_2)\cdot\boldr} \psihat_2(\boldr), \label{a_pm_exp}\\
\bhat_\pm(t) &=& \int \dr \, u_c^*(\textbf{r},t) e^{-i\boldk_{\pm}\cdot \boldr} \hat{E}_1(\boldr), \label{bpmdef}
\end{eqnarray}
\end{subequations}
where $\boldk_{\pm} = (\omega_p\pm \omega_s)\hat{\bold{z}}/c$. Note that the mode expansion in Eq.~(\ref{a_pm_exp}) is approximate, and only valid in the regime where the two mode functions $u_0(\textbf{r})\exp[i (\textbf{k}_{\pm} \mp \textbf{k}_2) \cdot \textbf{r}]$ are approximately orthonormal. This is true provided the wavelength of the control fields is much smaller than the spatial extent of the condensate, or equivalently that the momentum kick imparted to the atoms is larger than the momentum width of the atomic cloud. Such a condition is easily satisfied in typical atom interferometers with Bose-condensed sources. Substituting Eqs~(\ref{two_mode_expansions}) into \eq{full_Ham_two_controls} gives $\hat{\mathcal{H}} \approx  \hat{\mathcal{H}}_0 +  \hat{\mathcal{H}}_\text{int}'$, where
\begin{align}
	\hat{\mathcal{H}}_0	&= \Big(\hbar\omega_2 + \frac{\hbar^2}{2m}|\boldk_1-\boldk_2|^2 \Big)\left(\ahatd_{+}\ahat_{+} + \ahatd_{-}\ahat_{-}\right) \notag \\
					&+ \hbar (\omega_p+\omega_s)\bhatd_+ \bhat_+  + \hbar (\omega_p-\omega_s)\bhatd_- \bhat_-, \\		
	 \hat{\mathcal{H}}_\text{int}'	&= \hbar g f(t) \ahat_1 (\ahatd_{+}e^{i(\omega_c+\omega_s) t} + \ahatd_{-} e^{i(\omega_c-\omega_s) t} ) (\bhat_+ + \bhat_- ) \notag \\
	 				& + h.c. \, ,
\end{align}
where we have chosen the orientation of our control and probe fields such that $|\boldk_{+} -\boldk_2|^2 = |\boldk_{-} +\boldk_2|^2\equiv |\boldk_{1} +\boldk_2|^2$, where $\boldk_1 = (\omega_p/c) \bold{\hat{z}}$.  Making the transformation $\hat{\mathcal{H}} \to \hat{\mathcal{H}}_\text{int} = \hat{U}^\dag \hat{\mathcal{H}}' \hat{U}$ with $\hat{U} = \exp(-i \hat{\mathcal{H}}_0 t/\hbar)$ yields
\begin{eqnarray}
	\hat{\mathcal{H}}_\text{int} 	&=& \hbar gf(t) \left[ \ahatd_{+} \ahat_1 \left(\bhat_+ e^{-i\delta t} + \bhat_- e^{i(2 \omega_s - \delta)t}\right) \right. \nonumber \\
	&+&   \left. \ahatd_{-} \ahat_1  \left(\bhat_+ e^{-i(2 \omega_s + \delta)t} +  \bhat_- e^{-i\delta t}\right) + h.c. \right],
\end{eqnarray}
where $\delta = \omega_p - \omega_c - \omega_2 - \hbar |\boldk_1-\boldk_2|^2/(2m)$. If we assume $\omega_s \gg \Delta \omega$, and adjust $\omega_c$ such that $\delta=0$, the contribution from the fast rotating terms can be neglected, and we obtain
\begin{equation}
	\hat{\mathcal{H}}_\text{int} \approx \hbar g f(t) \left[ \hat{a}_1 \left( \hat{a}_+^\dag \hat{b}_+ + \hat{a}_-^\dag \hat{b}_-\right) + h.c. \right].
\end{equation}
The Heisenberg equations of motion for this Hamiltonian are [\emph{c.f.} Eqs~(\ref{3modeEOM})]:
\begin{subequations}
\label{EOM_two_mode_squeezy}
\begin{align}
	i\dot{\ahat}_1 		&= g f(t) \left(\ahat_{+} \bhatd_+  + \ahat_{-} \bhatd_-  \right),  \\
	i\dot{\ahat}_{\pm} 	&= g f(t) \ahat_1\bhat_\pm,  \\
	i\dot{\bhat}_\pm 	&= g f(t) \ahat_1^\dag \ahat_{\pm}.
\end{align}
\end{subequations}
In the regime of low depletion from the condensate, we make the undepleted reservoir approximation $\hat{a}_1 \to \sqrt{N_{a_1}}$ as in Sec.~\ref{sec_QST}, and obtain
\begin{subequations}
\label{two_mode_QST_BS}
\begin{eqnarray}
	\ahat_{\pm}(t_1) &=& \ahat_{\pm}(t_0) \cos (\frac{ \theta_\text{QST}}{2}) - i\bhat_\pm(t_0)\sin (\frac{ \theta_\text{QST}}{2}),  \\ 
	\bhat_\pm(t_1) &=& \bhat_{\pm}(t_0) \cos (\frac{ \theta_\text{QST}}{2}) - i\ahat_{\pm}(t_0)\sin (\frac{ \theta_\text{QST}}{2}).
\end{eqnarray}
\end{subequations}
Note that the `+' and `-' modes decouple, and so the QST process depicted in Fig.~\ref{twomode_scheme} can be conceptualized as two independent atom-light beamsplitters with the same `reflectivity'. 

It is not difficult to define input states $\bhat_{\pm}(t_0)$ that are correlated, and more specifically correspond to a two-mode squeezed vacuum state. The argument follows that laid out in Sec.~\ref{sec_single_mode_atom_int}. First, we assume $\hat{E}_1(\textbf{r},t) \approx u_\text{trans}(x,y)\hat{E}_1(z,t)$, and rewrite \eq{bpmdef} in terms of the momentum modes of $\hat{E}_1(z,t)$ [\emph{c.f.} Eq.~(\ref{bt})]:
\begin{align}
	\bhat_\pm(t) = \int dq \, U_\text{prop}^*(q \mp q_s) \phihat(q,t),
\end{align}
where $q_s = |\textbf{k}_\pm - \mathbf{k}_1| = \omega_s/c$. Therefore, since \eq{b(k)} is the optical state output from the OPO:
\begin{align}
	\hat{b}_\pm(t_0)	&= \left[ \int dq \, U_\text{prop}^*(q \mp q_s) \hat{\phi}(q, t_i) \right] \cosh r \notag \\
					-&\, i e^{i\theta_\text{sq}} \left[ \int dq \, U_\text{prop}^*(q \mp q_s)\hat{\phi}^\dag(-q, t_i) \right] \sinh r.
\end{align}
The term in the first set of square brackets is $\hat{b}_\pm(t_i)$. Since $U_\text{prop}(q)$ is assumed real and symmetric,
\begin{align}
	 \int dq \, U_\text{prop}^*(q \mp q_s)\hat{\phi}^\dag(-q, t_i)	&= \int dq \, U_\text{prop}(q \pm q_s) \phihatd(q,t) \notag \\
	 										&= \bhat_\mp^\dag(t_i).
\end{align}
Hence, we arrive at
\begin{equation}
	\hat{b}_\pm(t_0) = \bhat_\pm(t_i) \cosh r - i e^{i\theta_\text{sq}} \bhatd_\mp(t_i) \sinh r. \label{twomodesqz}	
\end{equation}
Since the initial quantum state is chosen such that
\begin{subequations}
\label{initialstate_two_mode}
\begin{align}
	\hat{\phi}(q, t_i)|\Psi(0)\rangle 	&= \ahat_\pm(t_0)|\Psi(0)\rangle = 0, \\
	\ahat_1(t_0)|\Psi(0)\rangle 	&= \sqrt{N_{a_1}(t_0)}|\Psi(0)\rangle, 
\end{align}
\end{subequations}
the modes $\hat{b}_\pm(t_0)$ comprise a two-mode squeezed vacuum state.

\begin{figure}[t]
\includegraphics[width=1.0\columnwidth]{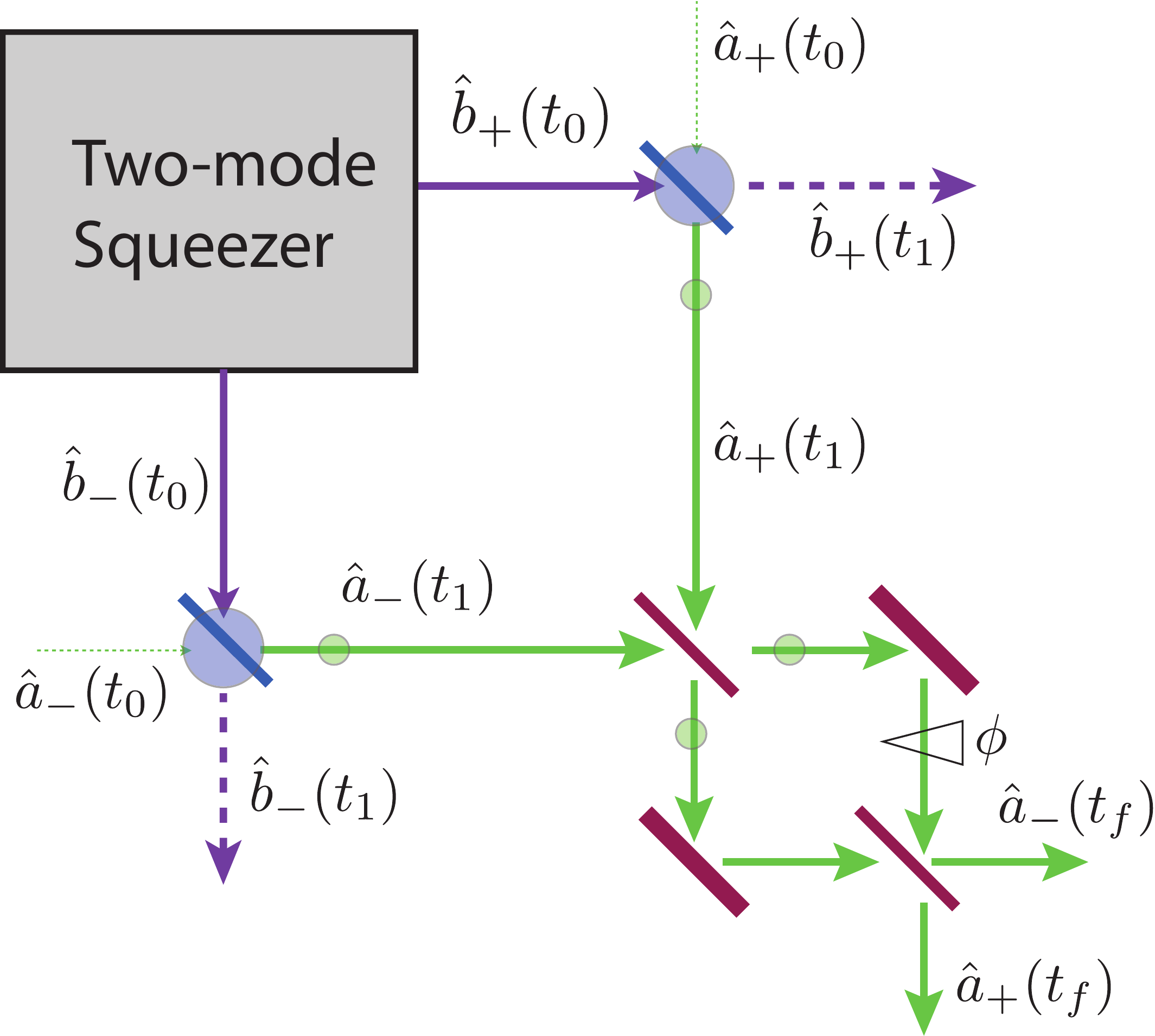} 
\caption{Analogous optical circuit for an atom interferometer enhanced with two-mode squeezed light. The atom-light coupling (partially) maps the quantum state of the optical modes $\bhat_+(t_0)$ and $\bhat_-(t_0)$ onto the atomic modes $\hat{a}_+(t_1)$ and $\hat{a}_-(t_1)$, respectively. These two atomic modes form the inputs to a Mach-Zehnder atom interferometer. The modes $\bhat_+$ and $\bhat_-$ are co-propagating, but shown here to be spatially separated for the purposes of visual clarity. Likewise, the blue atom-light beamsplitter represents the condensate mode $\ahat_1$, which is common to both QST processes (see Fig.~\ref{twomode_scheme}).} 
\label{twomode_BS_scheme}
\end{figure}

Our scheme that utilizes this two-mode squeezed optical vacuum is shown in Fig.~\ref{twomode_BS_scheme}. The modes $\bhat_+(t_0)$ and $\bhat_-(t_0)$ form the inputs for the QST process, transferring all or part of their quantum state to the modes $\ahat_{+}(t_1)$ and $\ahat_{-}(t_1)$ respectively [see Eqs~(\ref{two_mode_QST_BS})]. These two atomic modes form the input for the atom interferometer, where after the usual Mach-Zehnder interferometry sequence (beam split, reflect, beam split), the atom number difference at the outputs is measured:
\begin{align}
\Delta \hat{N}_a(t_f)	&\equiv \hat{N}_{a_+}(t_f) -\hat{N}_{a_-}(t_f) \notag \\
				&= 2 \hat{J}_z \cos \phi  + 2 \hat{J}_x \sin \phi . \label{sig_twomode}
\end{align}
Here we have used the pseudo-spin operators $\jhat_k \equiv \frac12\v{a}^\dag \sigma_k \v{a}$, where $\v{a} = (\ahat_+(t_1), \ahat_-(t_1))^T$ and $\sigma_k$ are the set of Pauli spin matrices; this notation is convenient for some of the expressions below.
However, $\Delta \hat{N}_a$ is a poor choice for our signal, since the average number difference of a two-mode squeezed vacuum state is zero, and hence $\langle \Delta \hat{N}_a \rangle = 0$ even for the case of incomplete QST. Consequently, we choose a signal based on the \emph{fluctuations} of the output number difference:
\begin{equation}
	\hat{S}_{a} = \left(\hat{N}_{a_+}(t_f) - \hat{N}_{a_-}(t_f)\right)^2. \label{2_mode_bad_sig}
\end{equation}

\subsection{Complete quantum state transfer} \label{sec_two_mode_CQST}
When QST is perfect (i.e. $\theta_\text{QST} = \pi$) and $\hat{a}_\pm(t_1) = -i \hat{b}_\pm(t_0)$, the average and variance of the signal (\ref{2_mode_bad_sig}) are
\begin{subequations}
\label{CQST_two_mode_quantities}
\begin{align}
	\langle \hat{S}_a \rangle 				&= \sinh^2(2r) \sin^2 \phi, \\
	V(\hat{S}_a)						&= \frac{1}{2}\sinh^2(2r)\left[ 1 + \cosh(4r) \left( 3 - 4 \cos(2\phi) \right) \right] \notag \\
									&+ \sinh^4(2r) \cos(4\phi),
\end{align}
\end{subequations}
respectively. Note that these expressions are independent of the squeezing angle $\theta_\text{sq}$. Equations~(\ref{CQST_two_mode_quantities}) yield a phase sensitivity of
\begin{equation}
	\Delta \phi = \frac{\sqrt{1 + \cosh(4 r) \tan^2 \phi}}{\sinh(2r)}. \label{2_mode_sensitive_anal}
\end{equation}
A minimum sensitivity of $\Delta \phi_\text{min} = 1/\sinh(2r)$ occurs at $\phi = \pi$. However, since the total number of atoms detected at the interferometer output is $N_t = \langle \hat{N}_{a_+}(t_f) + \hat{N}_{a_-}(t_f) \rangle = 2 \sinh^2 r$, we can rewrite 
\begin{equation} \label{two_mode_sen_CQST}
	\Delta \phi_\text{min} = \frac{1}{\sqrt{N_t(N_t  +2)}},
\end{equation}
which is approximately the Heisenberg limit $1/N_t$ in the limit of large $N_t$.

Na\"ively, this interferometer scheme compares favourably to the single-mode squeezed-light enhanced scheme considered in Sec.~\ref{sec_single_mode_atom_int}. For instance, in order to achieve a sensitivity of $\Delta \phi \sim 10^{-5}$, which is the sensitivity obtained by an ideal atom interferometer of $10^6$ atoms enhanced by single-mode squeezed light with $r = r_\text{TW} \approx 4.8$  (see Fig.~\ref{fig:Dvsr}), we need to outcouple $N_t \sim 10^5$ atoms, which requires a squeezing parameter of $r \sim 6.1$. However, the story is not quite as simple once the effects of incomplete QST and depletion are considered.

\subsection{Incomplete quantum state transfer and information recycling} \label{sec_two_mode_incomplete}
\begin{figure}[!t]
\includegraphics[width=1.0\columnwidth]{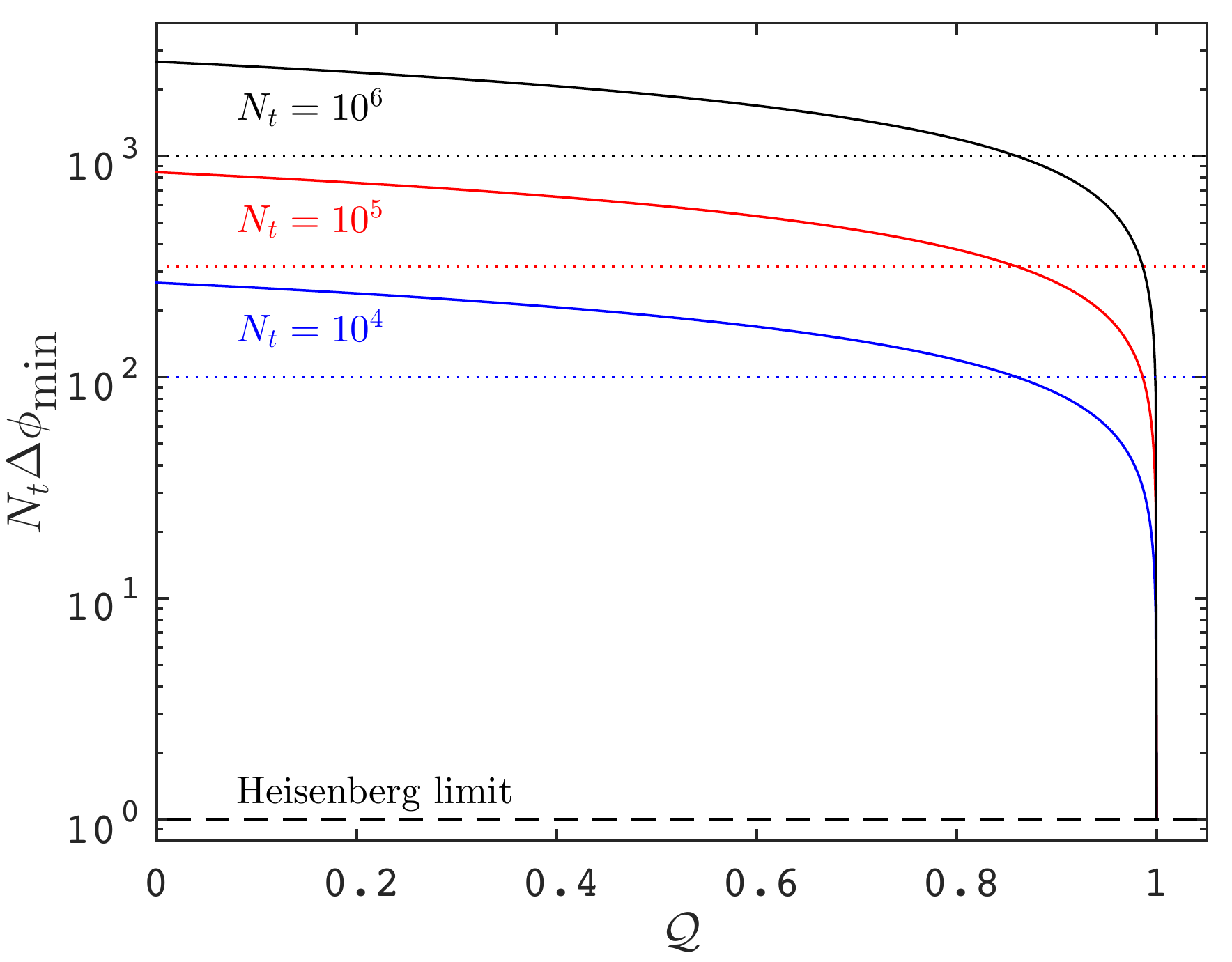} 
\caption{Plots demonstrating the QST efficiency dependence of the minimum phase sensitivity $\Delta \phi_\text{min}$ for an atom interferometer enhanced by two-mode optical squeezing [Eq.~(\ref{two_mode_sens_incomplete})]. The horizontal dashed lines that intercept each curve at $\mathcal{Q} = (2 + \sqrt{10})/6$ indicate the SQL for $N_t = 10^6$ (black, top), $10^5$ (red, middle) and $10^4$ (blue, bottom). The Heisenberg scaling is quickly lost for small perturbations of $\mathcal{Q}$ from unity. In contract, when information recycling is applied $\Delta \phi_\text{min}$ remains at the Heisenberg limit for all $\mathcal{Q}$ [see Eq.~(\ref{two_mode_phi_min_recyc})].} 
\label{twomode_Deltaphi}
\end{figure}

For incomplete QST, the minimum phase sensitivity (which occurs at $\phi = \tan^{-1}[ (V(\jhat_z^2) / V(\jhat_x^2))^{1/4}]$) is \cite{Haine:2014b}
\begin{widetext}
\begin{align}
	(\Delta \phi_\text{min})^2 	&= \frac{2\sqrt{V(\jhat_z^2) V(\jhat_x^2)} + C(\jhat_x^2, \jhat_z^2) + \langle   (\jhat_x\jhat_z + \jhat_z\jhat_x)^2 \rangle}{4( \langle  \jhat_z^2 \rangle - \langle  \jhat_x^2 \rangle)^2}, \label{app_sens} \\
				&= \frac{\sqrt{(1-\mathcal{Q})(1 + 5(1-\mathcal{Q})N_t)\left( 4 + 2(N_t - 2 \mathcal{Q}) + \frac{N_t \left( 9 - \mathcal{Q} \left( 42 - 37 \mathcal{Q} \right) \right) - 64 \mathcal{Q}^2 (1 - \mathcal{Q}) + 7 \mathcal{Q} + 1}{4(N_t + 2\mathcal{Q})^2} \right)}}{N_t (N_t + 2\mathcal{Q})} \notag \\
				&+ \frac{ \left( 1 + \frac{1 - \mathcal{Q}}{2(N_t + 2 \mathcal{Q})}\right) + N_t (1 - \mathcal{Q})\left( 3 + \frac{2 N_t + 5 - \mathcal{Q}}{2 (N_t + 2 \mathcal{Q})}\right)}{N_t (N_t + 2\mathcal{Q})}, \label{two_mode_sens_incomplete}	
\end{align}
\end{widetext}
where $V(\hat X) = \langle   \hat{X}^2 \rangle - \langle   \hat{X} \rangle^2$ is the variance of $\hat X$, $C(\hat X,\hat Y) = \langle   \hat{X} \hat{Y} + \hat{Y} \hat{X} \rangle - 2\langle   \hat{X} \rangle \langle   \hat{Y} \rangle$ is the symmetrized covariance of $\hat X$ and $\hat Y$, $\mathcal{Q} = \sin^2 (\theta_\text{QST}/2)$ and $N_t =  2 \mathcal{Q} \sinh^2 r$. As shown in Fig.~\ref{twomode_Deltaphi}, even small decreases from complete QST result in a rapid degradation of sensitivity. Indeed, provided $N_t \gg 1$ and $(1-\mathcal{Q}) \gg 1/N_t$, both of which are easily satisfied in practice, then
\begin{equation}
	\Delta \phi_\text{min} \approx \sqrt{\frac{(1-\mathcal{Q})(4 + \sqrt{10})}{N_t}}.
\end{equation}
Although sensitivities below the SQL are obtainable for $\mathcal{Q} > (2 + \sqrt{10})/6 \approx 0.86$, Heisenberg scaling is lost for extremely small perturbations from complete QST.   


As for the interferometer enhanced by single-mode squeezed light, we can arrest the loss of sensitivity at incomplete QST via information recycling. Specifically, we directly measure the number of photons output after the QST process [i.e. $\hat{N}_{b_\pm} = \hat{b}_\pm^\dag(t_1) \hat{b}_\pm(t_1)$] via photodetection, and subtract these counts from the number of atoms outcoupled during the QST process (i.e. $\hat{N}_{a_\pm}$). Explicitly, we construct the signal
\begin{equation}
	\hat{S} = \left( \Delta \hat{N}_a(t_f) - \Delta \hat{N}_b(t_1) \right)^2.
\end{equation}
where $\Delta \hat{N}_b \equiv \hat{N}_{b_+} - \hat{N}_{b_-}$
Thus, at the optimum phase $\phi = \pi$, the minimum sensitivity with information recycling takes the remarkably simple form:
\begin{align} \label{two_mode_phi_min_recyc}
	\Delta \phi_\text{min} 	&= \frac{\left(\sinh r \sin (\theta_\text{QST}/2)\right)^{-1}}{\sqrt{2\left[ 1 + \cosh(2r) \sin^2 (\theta_\text{QST}/2)\right]} } \notag \\
				&= \frac{1}{\sqrt{N_t \left(N_t + 1 + \mathcal{Q}\right)}},
\end{align}
which for large $N_t$ is approximately the Heisenberg limit, but most importantly is almost independent of $\mathcal{Q}$ and approximately equal to the sensitivity when $\mathcal{Q} = 1$ [see Eq.~(\ref{two_mode_sen_CQST})].

Unlike the atom interferometry scheme enhanced with single-mode squeezed light, information recycling almost completely removes any degradation due to incomplete QST. However, this does \emph{not} imply that the QST efficiency does not matter, since $N_t \propto \mathcal{Q}$. Nevertheless, Eq.~(\ref{two_mode_phi_min_recyc}) represents the minimum sensitivity attainable for a fixed $N_t$, and certainly gives better sensitivities than for the purely atomic signal $\hat{S}_a$. Perhaps surprisingly, this remains true even under the effects of depletion, as shown in the next subsection.

\subsection{Effects of depletion} \label{sec_loss_and_depletion}
The analytics presented in this section thus far have assumed the undepleted reservoir approximation $\hat{a}_1 \to \sqrt{N_{a_1}}$. However, this approximation breaks down for even moderate values of $r$. Thus, following Sec.~\ref{sec_single_mode_atom_int} we treated mode $\hat{a}_1$ as a quantum dynamical degree of freedom by performing numerical TW simulations of the QST process. Under TW, the operators are mapped to complex stochastic amplitudes via the correspondences $\hat{a}_i \to \alpha_i$ and $\hat{b}_i \to \beta_i$. These stochastic variables evolve according to the SDEs [\emph{c.f.} Eqs~(\ref{EOM_two_mode_squeezy})]:
\begin{subequations}
\label{eqs_trunc_Wigner}
\begin{align}
	i\dot{\alpha}_1 		&= g f(t) \left(\alpha_{+} \beta_+^*  + \alpha_{-} \beta_-^*  \right),  \\
	i\dot{\alpha}_{\pm} 	&= g f(t) \alpha_1\beta_\pm,  \\
	i\dot{\beta}_\pm 	&= g f(t) \alpha_1^* \alpha_{\pm},
\end{align}
\end{subequations}
with initial conditions
\begin{subequations}
\begin{align}
	\alpha_1(0) 		&= \sqrt{N_{a_1}(t_0)} + \eta_{\alpha_1}, \\
	\alpha_{\pm}(0)	&= \eta_{\alpha_{\pm}}, \\
	\beta_\pm	(0)		&= \eta_{\beta_\pm} \cosh r  - i e^{i \theta_\text{sq}} \eta_{\beta_\mp}^* \sinh r.
\end{align}
\end{subequations}
The $\eta_i$ are complex, independent Gaussian noises with zero mean and variance $1/2$. Without loss of generality, we assumed a uniform $f(t)$ for the TW simulations presented below.

\begin{figure}[t!]
\includegraphics[width=1.0\columnwidth]{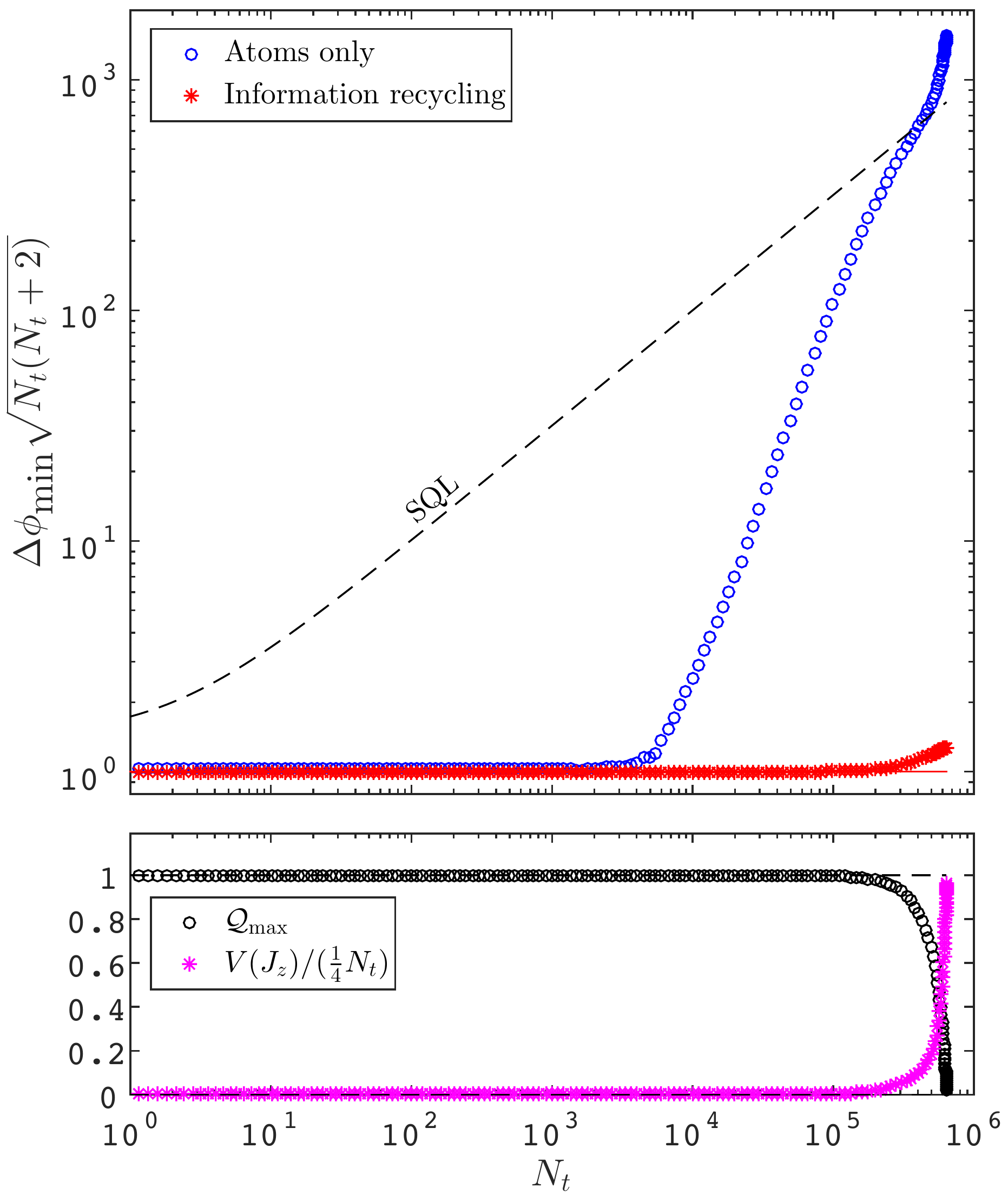} 
\caption{(Top) TW calculations of the minimum phase sensitivity, normalized to the sensitivity for complete QST [see Eq.~(\ref{two_mode_sen_CQST})], with (red asterisks) and without (blue circles) information recycling, assuming an initial BEC of $N_{a_1}(t_0) = 10^6$ atoms. $N_t$ was varied by adjusting the number of input photons $N_b(t_0) = 2 \sinh^2 r $ from zero to $\sim 10^8$. The standard error of each point is on the order of the point width, and the solid red line is the analytic solution (\ref{two_mode_sen_CQST}), which is approximately the Heisenberg limit. The TW simulations reveal that the undepleted reservoir approximation ceases to hold for relatively small values of $N_t$; however, the negative effect of depletion is (almost) completely removed when information recycling is used. (Bottom) TW calculations of the variance in $\hat{J}_z$ and the maximum possible QST efficiency, $\mathcal{Q}_\text{max} = \max_t \mathcal{Q}(t)$, as a function of the total number of atoms detected at the outputs. $V(J_z)$ has been normalized to $N_t/4$, which is the variance in $\hat{J}_z$ for two uncorrelated coherent states each with a mean number of $N_t/2$. In the undepleted reservoir approximation, the analytic solution predicts $V(J_z) = 0$ always. However, as $N_t$ increases beyond $\sim 10^5$, a full quantum treatment of mode $\hat{a}_1$ shows that there are increasingly large fluctuations in the number difference of the two MZ input ports, which suppress correlations between $\hat{a}_+(t_1)$ and $\hat{a}_-(t_1)$.
}
\label{twomodes_plot}
\end{figure}

The phase sensitivity without information recycling was calculated using Eq.~(\ref{app_sens}) and the following expressions:
\begin{subequations}
\begin{align}
	\langle \hat{J}_x^2 \rangle	&= \overline{\mathcal{J}_x^2} - \frac{1}{8}, \\
	\langle \hat{J}_z^2 \rangle	&= \overline{\mathcal{J}_z^2} - \frac{1}{8}, \\
	\langle \hat{J}_x^4 \rangle	&= \overline{\mathcal{J}_x^4} - \frac{5}{4} \overline{\mathcal{J}_x^2} + \frac{1}{16},\\
	\langle \hat{J}_z^4 \rangle	&= \overline{\mathcal{J}_z^4} - \frac{5}{4} \overline{\mathcal{J}_z^2} + \frac{1}{16}, \\
	\langle \hat{J}_x^2 \hat{J}_z^2 + \hat{J}_z^2 \hat{J}_x^2 \rangle &= 2 \overline{\mathcal{J}_x^2 \mathcal{J}_z^2} + \frac{5}{4}\overline{\mathcal{J}_x^2} + \frac{1}{4}\overline{\mathcal{J}_z^2} - \overline{|\alpha_+|^2 |\alpha_-|^2}, \\
	\langle ( \hat{J}_x \hat{J}_z + \hat{J}_z \hat{J}_x )^2 \rangle &= 4 \overline{\mathcal{J}_x^2 \mathcal{J}_z^2} - \frac{5}{2}\overline{\mathcal{J}_x^2} - \frac{3}{2}\overline{\mathcal{J}_z^2} + \overline{|\alpha_+|^2 |\alpha_-|^2}, 
\end{align}
\end{subequations}
where $\mathcal{J}_z = |\alpha_+(t_1)|^2 - |\alpha_-(t_1)|^2$ and $\mathcal{J}_x = \alpha_+(t_1) \alpha_-^*(t_1) + \alpha_+^*(t_1) \alpha_-(t_1)$. The phase sensitivity with information recycling was computed using $\Delta \phi_\text{min} = 1/\sqrt{4 \langle \hat{J}_x^2\rangle}$ \cite{Haine:2014b}.

Figure~\ref{twomodes_plot} shows that the effects of depletion cause a serious degradation to the sensitivity, even when QST is complete. The cause of this degradation is shown in the bottom panel of Fig.~\ref{twomodes_plot}, which plots the variance in $\hat{J}_z$ (which is proportional to the number difference of the two input ports of the atom interferometer). A large, non-zero variance implies that there exist low (but not negligible) probability trajectories where the quantum state of a photon in $\hat{b}_+$ is successfully mapped to an atom in mode $\hat{a}_+$, but $\hat{b}_-$ is not mapped to $\hat{a}_-$ (and similarly for an exchange of `+' and `-'). The inclusion of unequal atom numbers in the two paths of the atom interferometer inevitably degrades the phase sensitivity. Fortunately, the effect is entirely reversed with information recycling, as measurement of the transmitted photons provides information about the full quantum correlations within the system.

The bottom panel of Fig.~\ref{twomodes_plot} shows that the maximum possible QST efficiency rapidly drops below 100\% once $N_t$ is $\gtrsim 10 \%$ of the initial number of atoms populating mode $\hat{a}_1$. This is what we intuitively expect; it is impossible to outcouple more than $N_t = N_{a_1}(t_0)$ atoms from the condensate, and so increasing $r$ only results in a lower $\mathcal{Q}$. Furthermore, in this regime depletion causes a slight increase to the minimum phase sensitivity of the information-recycled signal. Here, the finite size of the BEC necessarily truncates the atom number probability distribution for the MZ input state at $N = N_{a_1}(t_0)$. Consequently, at best the input atomic state will have an atom number probability distribution corresponding to a two-mode squeezed optical state with a truncated `tail'. When $N_t \approx N_{a_1}(t_0)$, this distribution will be somewhere between a twin-Fock state and a two-mode squeezed vacuum state, and as discussed in \cite{Haine:2014b} this yields a minimum phase sensitivity of $\Delta \phi \lesssim \sqrt{2}/ N_t$. This is consistent with the TW simulations shown in the top panel of Fig.~\ref{twomodes_plot}.


\section{A single input enhancement with two-mode squeezed-light and information recycling} \label{sec_high_to_low_sq}
Although the previously presented two-mode squeezed light enhanced atom interferometry scheme has many advantages (e.g. it utilizes high frequency squeezing and can attain Heisenberg scaling), one disadvantage is that the number of atoms detected at the output is proportional to the number of atoms outcoupled during the QST process. Consequently, a high atom number interferometer requires both a large squeezing parameter and good QST efficiency. In contrast, the number of detected atoms in the scheme enhanced by single-mode squeezed light is \emph{independent} of $r$ and $\mathcal{Q}$; it only depends on the atom number of the initial condensate.   

\begin{figure}[!t]
\includegraphics[width=1.0\columnwidth]{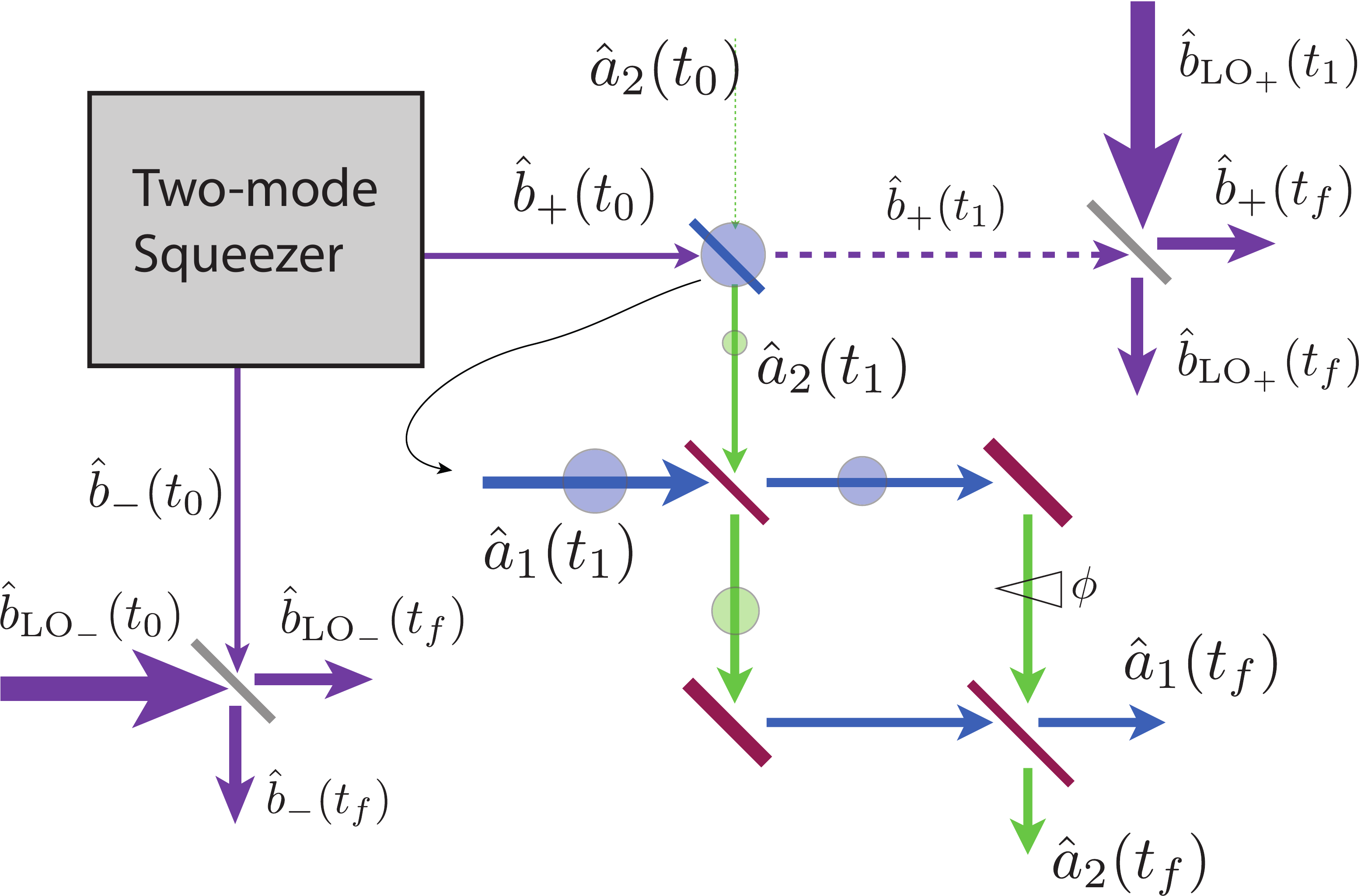} 
\caption{Analgous optical circuit for modified atom interferometry utilizing two-mode optical squeezing. Only one of the modes (which corresponds to photons in a particular frequency band) is coupled to the atoms, while the other mode is measured directly. $\bhat_+$ and $\bhat_-$ are co-propagaiting, but shown here to be spatially separated to highlight their distiniguishability.} 
\label{fig:hybrid}
\end{figure}

In this section we present an alternative scheme based on high frequency squeezing [i.e. a two-mode squeezed state; see Eq.~(\ref{twomodesqz})] that also utilizes all the atoms in the condensate, independent of the squeezing parameter and QST efficiency. This scheme is summarized in Fig.~\ref{fig:hybrid}. Instead of using two control beams, this scheme uses just one, detuned from two-photon resonance by an amount $\omega_s > \omega_\text{crit}$. This ensures that the QST process only occurs for a region of probe frequencies centered about $\omega_p +\omega_s$. This corresponds to the optical mode $\hat{b}_+(t_0)$ interacting with the initial condensate mode $\hat{a}_1$, leading to outcoupled atoms in mode $\hat{a}_2$. Since there is only one control beam, the portion of the spectrum correlated with this region (i.e. the region centered around $\omega_p-\omega_s$) does not couple to the atoms, and is therefore transmitted. This transmitted light is described by the mode $\hat{b}_-$. By monitoring this transmitted light with homodyne detection, the quantum correlations of the two-mode squeezed state can still be used to enhance the sensitivity via information recycling. Furthermore, any incomplete QST can similarly be ameliorated by monitoring any transmitted light in the region centered around $\omega_p+\omega_s$. The outcoupled atomic mode and the remaining condensate are then used as the two inputs to an atom interferometer. The relevant (information-recycled) signal is then a linear combination of the atomic number difference at the output of the atom interferometer and the photon number differences of the transmitted light centered around $\omega_p+\omega_s$ and $\omega_p-\omega_s$. We call this scheme a \emph{single input} enhancement with two-mode squeezed light, since only one mode of the two-mode squeezed optical vacuum directly interacts with the atoms. This is in contrast to the scheme considered in Sec.~\ref{sec_high_freq_sq}, which we could describe as a \emph{double input} enhancement with two-mode squeezed light, since both modes of the two-mode squeezed optical vacuum couple to the BEC.

We compute the phase sensitivity of this interferometry scheme from the information-recycled signal 
\begin{equation}
	\hat{S} = \sqrt{\mathcal{Q}} \hat{S}_a -\mathcal{G}_+ \sqrt{1 - \mathcal{Q}}\hat{S}_{b_+} + \mathcal{G}_- \hat{S}_{b_-} \, ,
\end{equation}
where $\hat{S}_a$ is given by Eq.~(\ref{sig1}), 
\begin{equation}
	\hat{S}_{b_\pm} = \bhatd_{\text{LO}_\pm}(t_f)\bhat_{\text{LO}_\pm}(t_f) - \bhatd_\pm(t_f)\bhat_\pm(t_f), 
\end{equation}
and
\begin{equation}
\mathcal{G}_\pm = \sqrt{\frac{\langle\ahatd_1(t_1)\ahat_1(t_1) \rangle}{\langle \bhatd_{\text{LO}_\pm}(t_1)\bhat_{\text{LO}_\pm}(t_1)\rangle}} \, .
\end{equation}
 Note that the photon modes at time $t_f$ are related to those at $t_1$ by the simple beamsplitting relations:
 \begin{subequations}
 \begin{align}
 	\bhat_\pm(t_f) 	&= \frac{1}{\sqrt{2}}\left(\bhat_\pm(t_1) - i\bhat_{\text{LO}_\pm}(t_1)\right) \\ 
  	\bhat_{\text{LO}_\pm}(t_f) &= \frac{1}{\sqrt{2}}\left(\bhat_{\text{LO}_\pm}(t_1) - i\bhat_{\pm}(t_1)\right),
 \end{align}
 \end{subequations}
 and of course the trivial relation $\bhat_-(t_1) = \bhat_-(t_0)$ holds. Our justification for this choice of signal follows an argument similar to that used to justify the choice of $\hat{S}$ in Sec.~\ref{sec_single_mode_incomplete}. In brief, in the regime where depletion from the condensate is minimal during the QST process, the undepleted reservoir approximation $\hat{a}_1 \to \sqrt{N_{a_1}}$ gives the familiar atom-light beamsplitter relations:
\begin{subequations}
\begin{eqnarray}
	 \ahat_2(t_1) &=& \ahat_2(t_0) \cos (\frac{\theta_\text{QST}}{2}) - i\bhat_+(t_0)\sin (\frac{\theta_\text{QST}}{2}),  \\ 
	\bhat_+(t_1) &=& \bhat_+(t_0) \cos (\frac{\theta_\text{QST}}{2}) - i\ahat_2(t_0)\sin (\frac{\theta_\text{QST}}{2}),
 \end{eqnarray} 
 \end{subequations}
and $\mathcal{Q} = \sin^2 (\theta_\text{QST}/2)$. Since $\hat{S}_a \approx \sqrt{N_{a_1}(t_1)} \hat{X}_{a_2(t_1)}^0$ and $\hat{S}_{b_\pm} \approx \sqrt{N_{\text{LO}_\pm}} \hat{X}_{b_\pm(t_1)}^{\theta_{\text{LO}_\pm}}$, then for $\theta_{\text{LO}_\pm} = \pi/2$:
\begin{align}
	\hat{S}	&\approx \sqrt{N_{a_1}(t_1)}\left( \hat{X}_{b_-(t_0)}^{\pi/2} - \hat{X}_{b_+(t_0)}^{\pi/2} \right),
\end{align}
which for optimal squeezing angle $\theta_\text{sq} = \pi/2$ has a variance $V(S) = 2 \sqrt{N_{a_1}(t_1)} \exp(-2r)$, which can be smaller than $V(S_a)$.

\begin{figure}[!t]
\includegraphics[width=1.0\columnwidth]{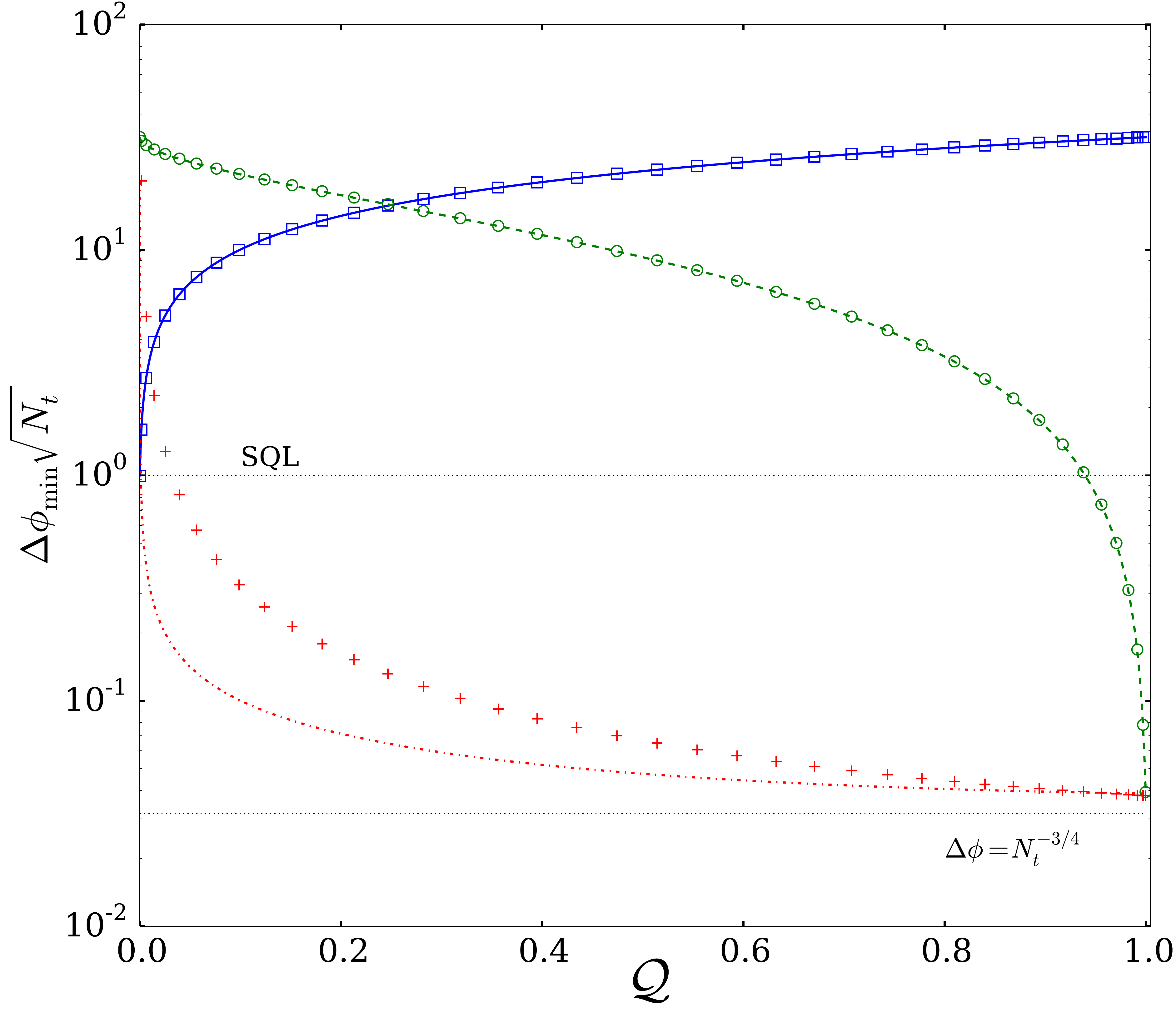} 
\caption{The minimum phase sensitivity $\Delta \phi_\text{min}$ (i.e. for $\phi = \theta_\text{sq} = \pi/2$) for a single input two-mode optical squeezed vacuum enhanced atom interferometer, corresponding to a purely atomic signal $\hat{S}_a$ (undepleted reservoir approximation: solid blue curve; TW: blue squares), partial information-recycled signal $\hat{S}_a - \mathcal{G}_-\hat{S}_{b_-}$ (undepleted reservoir approximation: green dashed curve; TW: green circles), and complete information-recycled signal $\hat{S} = \sqrt{\mathcal{Q}} \hat{S}_a -\mathcal{G}_+ \sqrt{1 - \mathcal{Q}}\hat{S}_{b_+} + \mathcal{G}_- \hat{S}_{b_-}$ (undepleted reservoir approximation: red dot-dashed curve; TW: red crosses). All these plots assumed $N_t = 10^6$ and $r = r_\text{opt} \approx 3.8$. For convenience, these curves have been normalized relative to the SQL $1/\sqrt{N_t}$.   The upper and lower horizontal dotted lines show the SQL, and the theoretical limit reached by a single-mode squeezed optical vacuum enhancement with perfect QST (i.e. $1/N_t^{3/4}$), respectively.} 
\label{fig:twomode_hybrid_deltaphi}
\end{figure}

 
The dependence of the minimum sensitivity on the QST efficiency is shown in Fig.~\ref{fig:twomode_hybrid_deltaphi}. There are clear similarities to the plots shown in Fig.~\ref{fig:delta_phi_vs_theta_qst} corresponding to the scheme enhanced by single-mode squeezed optical vacuum. Nevertheless, we highlight some important differences. Firstly, the pure atomic signal $\hat{S}_a$ always gives a sensitivity worse than the SQL. Secondly, when the QST efficiency is close to $100\%$, it is sufficient to use a partial information-recycled signal $\hat{S}_a - \mathcal{G}_-\hat{S}_{b_-}$. Thirdly, even for the signal $\hat{S}$ the sensitivity remains above the analytic limit $\Delta \phi = 1/N_t^{3/4}$ derived for the single-mode squeezed optical vacuum enhanced atom interferometer, since for finite levels of squeezing measurements of $\hat{S}_{b_-}$ are not perfectly correlated with $\sqrt{\mathcal{Q}} \hat{S}_a -\mathcal{G}_+\sqrt{1-\mathcal{Q}}\hat{S}_{b_+}$. Finally, the TW simulations predict that the effects of depletion lead to a poorer sensitivity than that given by the analytic sensitivity for the information-recycled signal $\hat{S}$.  

\section{Summary and discussion.} \label{sec_summary}

In this paper we have shown how squeezed light can be used to enhance the sensitivity of atom interferometers. We have specifically considered three schemes: (Sec.~\ref{sec_single_mode_atom_int}) enhancement with a single-mode squeezed optical vacuum (i.e. low frequency squeezing), (Sec.~\ref{sec_high_freq_sq}) double input enhancement with two-mode squeezed optical vacuum (i.e. high frequency squeezing), and (Sec.~\ref{sec_high_to_low_sq}) single input enhancement with two-mode squeezed optical vacuum. We have shown that all three schemes give sensitivities below the SQL - even when the effects of depletion from the initial condensate and incomplete QST are included. Furthermore, we have demonstrated that information recycling provides a further enhancement to the sensitivity when QST between the atoms and light is incomplete. 

Table~\ref{summary_table} provides a quantitative comparison of the sensitivities for the three schemes for complete QST, and for incomplete QST ($\mathcal{Q} = 0.2$) with and without information recycling. This is a concise demonstration of the different sensitivities obtainable (including the effects of depletion), the degrading effects of incomplete QST, and how information recycling ameliorates this degradation. The choice of $\mathcal{Q} = 0.2$ is not unrealistic given current technology. For example, the 780~nm D2 transition of $^{87}$Rb can be addressed with a Rabi frequency of $\Omega \sim 20$ MHz and detunings $\Delta_p \approx \Delta_c \sim 70$ GHz. Since the electric dipole moment of this transition is $d_{12} \sim 2 \times 10^{-29}$~C/m, Eq.~(\ref{effective_coupling}) implies that an effective coupling of $g \sim 65 \times 10^{-3}$~m$^{3/2}$/s is obtainable. If we assume a condensate of Gaussian spatial profile $u_0(\textbf{r})$, with transverse width $R_\perp \sim 15 \, \mu$m, and further assume the pulse envelope $u_p(\text{r},t)$ is also Gaussian with the same transverse width $R_\perp$ and pulse duration $T \sim 1$ ms, then
\begin{equation}
	\theta_\text{QST} = \frac{4}{3}(2\pi)^{1/4} g \sqrt{\frac{N T}{c A_\perp}} \sim \frac{3 \pi}{10},
\end{equation} 
where $A_\perp = \pi R_\perp^2$ and we have used $N = 10^6$. This gives a QST efficiency of $\mathcal{Q} \sim 20$\%.

\begin{table}[!t]
\begin{ruledtabular}
\setlength{\extrarowheight}{4pt}
\footnotetext{Note that $\hat{S}$ gives $\Delta \phi = 3.8 \times 10^{-5}$ for $\mathcal{Q} = 100$\%.}
\begin{tabular}{m{3.4cm} |  m{1.6cm}  m{1.6cm}  m{1.6cm}}
		\multirow{3}{3.4cm}{\centering Interferometer scheme}	&	\multicolumn{3}{c}{$\Delta \phi_\text{min}$} \\
	\cline{2-4}
	 	&   \centering $\hat{S}_a$ and $\mathcal{Q} = 100\%$	&  \centering $\hat{S}_a$ and $\mathcal{Q} = 20\%$		&  \multicolumn{1}{m{1.6cm}}{\centering $\hat{S}$ and $\mathcal{Q} = 20\%$} \\
	 \hline 
	\centering Enhancement with single-mode squeezed state, $r = 4.8$, and $N_t = 10^6$ (Sec.~\ref{sec_single_mode_atom_int})		&	\centering $1 \times 10^{-5}$	& \centering $ 9 \times 10^{-4}$	& \multicolumn{1}{m{1.6cm}}{\centering  $1.9 \times 10^{-5}$} \\
	\centering Double input enhancement with two-mode squeezed state and $N_t = 6.2 \times 10^5$ (Sec.~\ref{sec_high_freq_sq})		&	\centering $1.6 \times 10^{-6}$	& \centering $2.1 \times 10^{-3}$	&  \multicolumn{1}{m{1.6cm}}{ \centering $2 \times 10^{-6}$} \\
	\centering Single input enhancement with two-mode squeezed state, $r = 3.8$ and $N_t = 10^6$ (Sec.~\ref{sec_high_to_low_sq})$^\text{a}$ 	&	\centering $3.2 \times 10^{-2}$	& \centering $ 1.4 \times 10^{-2}$	& \multicolumn{1}{m{1.6cm}}{\centering $ 1.6 \times 10^{-4}$}
\end{tabular}
\end{ruledtabular}
\caption{Numerical values of the minimum phase sensitivities for the three different squeezed-light enhanced atom interferometry schemes, in three distinct scenarios. These values were obtained via TW simulations.
All three schemes assumed an initial condensate of $N_{a_1}(t_0) = 10^6$ atoms; a large but achievable \cite{van_der_Stam:2007} number of atoms. All these atoms can be detected at the outputs for the two schemes utilizing a single port enhancement (i.e. the first and third rows of the table). In contrast, $N_t$ depends on the value of $r$ for the double input enhancement (i.e. the middle row of the table); the choice $N_t \sim 6.2 \times 10^5$ corresponds to $\mathcal{Q} \approx 0.2$ and $r \approx 7.8$. 
For comparison, an atom interferometer operating at the SQL with $N_{a_1}(t_0) = 10^6$ atoms has a sensitivity of $\Delta \phi = 10^{-3}$.}
\label{summary_table}
\end{table}

Furthermore, losses due to spontaneous emission during the QST process are small in this regime. An estimate for the (time-dependent) rate at which atoms in the excited state $| 3 \rangle $ are lost due to spontaneous emission is $\Gamma(t) = \Omega_\text{eff}(t) \gamma / \Delta_p$, where $\gamma = 2\pi \times 6.07$ MHz is the natural linewidth for $^{87}$Rb and $\Omega_\text{eff}$ is the effective Rabi frequency between modes $\hat{a}_1$ and $\hat{a}_2$ during the QST process. Assuming a beamsplitter-like coupling [such as described by Eqs~(\ref{coherent_BS_soln})], then $N_{a_1}(t_1) = N_t \cos^2( \int dt' \, \Omega_\text{eff}(t')/2) = N_t - \langle \hat{N}_b(t_0) \rangle \sin^2(\theta_\text{QST}/2)$, where the final equality follows from conservation of atom number. Therefore, the total loss is 
\begin{equation}
	l 	= \int dt' \, \Gamma(t') = \frac{\gamma}{\Delta_p} \cos^{-1} \sqrt{1 - \frac{\langle \hat{N}_b(t_0) \rangle}{N_t} \sin^2(\frac{\theta_\text{QST}}{2})}.  
\end{equation}
For a squeezing parameter of $r = 3.5$ (i.e. $\hat{N}_b(t_0) \sim 270$) and the parameters specified above, $l \sim 6.5 \times 10^{-7}$. However, $l$ is an estimate of the fraction of the total atoms lost; what is more important is the number of atoms lost compared to the total number of atoms in $\hat{a}_2$, which is $l N_t / N_{a_2} \sim 1 \%$. 

As mentioned throughout the paper, there are some important differences between the three schemes. The single-mode squeezed optical vacuum enhancement has the advantages of conceptual simplicity and full utilization of all the atoms in the condensate. However, the generation of a single-mode squeezed optical vacuum occurs at low frequencies which, while possible, is technically challenging. The double input enhancement with two-mode squeezed optical vacuum uses technically less challenging high-frequency squeezing, and its approximate Heisenberg scaling - independent of the QST efficiency with information recycling - is clearly the best of the three schemes. However, achieving a large $N_t$, and therefore a small \emph{absolute} sensitivity, requires a relatively large squeezing parameter $r$ and good QST efficiency $\mathcal{Q}$. The single input enhancement with two-mode squeezed optical vacuum can be thought of as a compromise between the former two schemes. Although it attains a minimum sensitivity similar in magnitude to the single-mode squeezed optical vacuum enhancement, it utilizes both high-frequency squeezing and all the atoms in the initially prepared condensate. 

Ultimately, however, all three schemes attain sensitivities that are substantially below the SQL, that are robust to imperfect QST once information recycling is incorporated, and are strongly compatible with the practical requirements of current state-of-the-art atom interferometry. This provides a compelling case for the further development of atom interferometers that are enhanced by squeezed optical states and/or information recycling. 

\section*{Acknowledgements}
We would like to acknowledge useful discussions with Warwick Bowen, John Close, Joel Corney, Saleh Rahimi-Keshari and Nicholas Robins. Numerical simulations were performed using XMDS2 \cite{Dennis:2012} on the University of Queensland (UQ) School of Mathematics and Physics computer `Obelix', with thanks to Elliott Hilaire and Ian Mortimer for computing support. SSS acknowledges the support of Ian P.~McCulloch and the Australian Research Council (ARC) Centre of Excellence for Engineered Quantum Systems (project no. CE110001013). SAH acknowledges the support of ARC Project DE130100575.

\bibliography{squeezy_bib}

\begin{thebibliography}{83}%
\makeatletter
\providecommand \@ifxundefined [1]{%
 \@ifx{#1\undefined}
}%
\providecommand \@ifnum [1]{%
 \ifnum #1\expandafter \@firstoftwo
 \else \expandafter \@secondoftwo
 \fi
}%
\providecommand \@ifx [1]{%
 \ifx #1\expandafter \@firstoftwo
 \else \expandafter \@secondoftwo
 \fi
}%
\providecommand \natexlab [1]{#1}%
\providecommand \enquote  [1]{``#1''}%
\providecommand \bibnamefont  [1]{#1}%
\providecommand \bibfnamefont [1]{#1}%
\providecommand \citenamefont [1]{#1}%
\providecommand \href@noop [0]{\@secondoftwo}%
\providecommand \href [0]{\begingroup \@sanitize@url \@href}%
\providecommand \@href[1]{\@@startlink{#1}\@@href}%
\providecommand \@@href[1]{\endgroup#1\@@endlink}%
\providecommand \@sanitize@url [0]{\catcode `\\12\catcode `\$12\catcode
  `\&12\catcode `\#12\catcode `\^12\catcode `\_12\catcode `\%12\relax}%
\providecommand \@@startlink[1]{}%
\providecommand \@@endlink[0]{}%
\providecommand \url  [0]{\begingroup\@sanitize@url \@url }%
\providecommand \@url [1]{\endgroup\@href {#1}{\urlprefix }}%
\providecommand \urlprefix  [0]{URL }%
\providecommand \Eprint [0]{\href }%
\providecommand \doibase [0]{http://dx.doi.org/}%
\providecommand \selectlanguage [0]{\@gobble}%
\providecommand \bibinfo  [0]{\@secondoftwo}%
\providecommand \bibfield  [0]{\@secondoftwo}%
\providecommand \translation [1]{[#1]}%
\providecommand \BibitemOpen [0]{}%
\providecommand \bibitemStop [0]{}%
\providecommand \bibitemNoStop [0]{.\EOS\space}%
\providecommand \EOS [0]{\spacefactor3000\relax}%
\providecommand \BibitemShut  [1]{\csname bibitem#1\endcsname}%
\let\auto@bib@innerbib\@empty
\bibitem [{\citenamefont {Peters}\ \emph {et~al.}(1999)\citenamefont {Peters},
  \citenamefont {Chung},\ and\ \citenamefont {Chu}}]{Peters:1999}%
  \BibitemOpen
  \bibfield  {author} {\bibinfo {author} {\bibfnamefont {A.}~\bibnamefont
  {Peters}}, \bibinfo {author} {\bibfnamefont {K.~Y.}\ \bibnamefont {Chung}}, \
  and\ \bibinfo {author} {\bibfnamefont {S.}~\bibnamefont {Chu}},\ }\href
  {http://dx.doi.org/10.1038/23655} {\bibfield  {journal} {\bibinfo  {journal}
  {Nature}\ }\textbf {\bibinfo {volume} {400}},\ \bibinfo {pages} {849}
  (\bibinfo {year} {1999})}\BibitemShut {NoStop}%
\bibitem [{\citenamefont {M\"uller}\ \emph
  {et~al.}(2008{\natexlab{a}})\citenamefont {M\"uller}, \citenamefont {Chiow},
  \citenamefont {Herrmann}, \citenamefont {Chu},\ and\ \citenamefont
  {Chung}}]{Muller:2008c}%
  \BibitemOpen
  \bibfield  {author} {\bibinfo {author} {\bibfnamefont {H.}~\bibnamefont
  {M\"uller}}, \bibinfo {author} {\bibfnamefont {S.-w.}\ \bibnamefont {Chiow}},
  \bibinfo {author} {\bibfnamefont {S.}~\bibnamefont {Herrmann}}, \bibinfo
  {author} {\bibfnamefont {S.}~\bibnamefont {Chu}}, \ and\ \bibinfo {author}
  {\bibfnamefont {K.-Y.}\ \bibnamefont {Chung}},\ }\href {\doibase
  10.1103/PhysRevLett.100.031101} {\bibfield  {journal} {\bibinfo  {journal}
  {Phys. Rev. Lett.}\ }\textbf {\bibinfo {volume} {100}},\ \bibinfo {pages}
  {031101} (\bibinfo {year} {2008}{\natexlab{a}})}\BibitemShut {NoStop}%
\bibitem [{\citenamefont {Altin}\ \emph {et~al.}(2013)\citenamefont {Altin},
  \citenamefont {Johnsson}, \citenamefont {Negnevitsky}, \citenamefont
  {Dennis}, \citenamefont {Anderson}, \citenamefont {Debs}, \citenamefont
  {Szigeti}, \citenamefont {Hardman}, \citenamefont {Bennetts}, \citenamefont
  {McDonald}, \citenamefont {Turner}, \citenamefont {Close},\ and\
  \citenamefont {Robins}}]{Altin:2013}%
  \BibitemOpen
  \bibfield  {author} {\bibinfo {author} {\bibfnamefont {P.~A.}\ \bibnamefont
  {Altin}}, \bibinfo {author} {\bibfnamefont {M.~T.}\ \bibnamefont {Johnsson}},
  \bibinfo {author} {\bibfnamefont {V.}~\bibnamefont {Negnevitsky}}, \bibinfo
  {author} {\bibfnamefont {G.~R.}\ \bibnamefont {Dennis}}, \bibinfo {author}
  {\bibfnamefont {R.~P.}\ \bibnamefont {Anderson}}, \bibinfo {author}
  {\bibfnamefont {J.~E.}\ \bibnamefont {Debs}}, \bibinfo {author}
  {\bibfnamefont {S.~S.}\ \bibnamefont {Szigeti}}, \bibinfo {author}
  {\bibfnamefont {K.~S.}\ \bibnamefont {Hardman}}, \bibinfo {author}
  {\bibfnamefont {S.}~\bibnamefont {Bennetts}}, \bibinfo {author}
  {\bibfnamefont {G.~D.}\ \bibnamefont {McDonald}}, \bibinfo {author}
  {\bibfnamefont {L.~D.}\ \bibnamefont {Turner}}, \bibinfo {author}
  {\bibfnamefont {J.~D.}\ \bibnamefont {Close}}, \ and\ \bibinfo {author}
  {\bibfnamefont {N.~P.}\ \bibnamefont {Robins}},\ }\href
  {http://stacks.iop.org/1367-2630/15/i=2/a=023009} {\bibfield  {journal}
  {\bibinfo  {journal} {New J. Phys.}\ }\textbf {\bibinfo {volume} {15}},\
  \bibinfo {pages} {023009} (\bibinfo {year} {2013})}\BibitemShut {NoStop}%
\bibitem [{\citenamefont {Canuel}\ \emph {et~al.}(2006)\citenamefont {Canuel},
  \citenamefont {Leduc}, \citenamefont {Holleville}, \citenamefont {Gauguet},
  \citenamefont {Fils}, \citenamefont {Virdis}, \citenamefont {Clairon},
  \citenamefont {Dimarcq}, \citenamefont {Bord\'e}, \citenamefont {Landragin},\
  and\ \citenamefont {Bouyer}}]{Canuel:2006}%
  \BibitemOpen
  \bibfield  {author} {\bibinfo {author} {\bibfnamefont {B.}~\bibnamefont
  {Canuel}}, \bibinfo {author} {\bibfnamefont {F.}~\bibnamefont {Leduc}},
  \bibinfo {author} {\bibfnamefont {D.}~\bibnamefont {Holleville}}, \bibinfo
  {author} {\bibfnamefont {A.}~\bibnamefont {Gauguet}}, \bibinfo {author}
  {\bibfnamefont {J.}~\bibnamefont {Fils}}, \bibinfo {author} {\bibfnamefont
  {A.}~\bibnamefont {Virdis}}, \bibinfo {author} {\bibfnamefont
  {A.}~\bibnamefont {Clairon}}, \bibinfo {author} {\bibfnamefont
  {N.}~\bibnamefont {Dimarcq}}, \bibinfo {author} {\bibfnamefont {C.~J.}\
  \bibnamefont {Bord\'e}}, \bibinfo {author} {\bibfnamefont {A.}~\bibnamefont
  {Landragin}}, \ and\ \bibinfo {author} {\bibfnamefont {P.}~\bibnamefont
  {Bouyer}},\ }\href {\doibase 10.1103/PhysRevLett.97.010402} {\bibfield
  {journal} {\bibinfo  {journal} {Phys. Rev. Lett.}\ }\textbf {\bibinfo
  {volume} {97}},\ \bibinfo {pages} {010402} (\bibinfo {year}
  {2006})}\BibitemShut {NoStop}%
\bibitem [{\citenamefont {Lenef}\ \emph {et~al.}(1997)\citenamefont {Lenef},
  \citenamefont {Hammond}, \citenamefont {Smith}, \citenamefont {Chapman},
  \citenamefont {Rubenstein},\ and\ \citenamefont {Pritchard}}]{Lenef:1997}%
  \BibitemOpen
  \bibfield  {author} {\bibinfo {author} {\bibfnamefont {A.}~\bibnamefont
  {Lenef}}, \bibinfo {author} {\bibfnamefont {T.~D.}\ \bibnamefont {Hammond}},
  \bibinfo {author} {\bibfnamefont {E.~T.}\ \bibnamefont {Smith}}, \bibinfo
  {author} {\bibfnamefont {M.~S.}\ \bibnamefont {Chapman}}, \bibinfo {author}
  {\bibfnamefont {R.~A.}\ \bibnamefont {Rubenstein}}, \ and\ \bibinfo {author}
  {\bibfnamefont {D.~E.}\ \bibnamefont {Pritchard}},\ }\href {\doibase
  10.1103/PhysRevLett.78.760} {\bibfield  {journal} {\bibinfo  {journal} {Phys.
  Rev. Lett.}\ }\textbf {\bibinfo {volume} {78}},\ \bibinfo {pages} {760}
  (\bibinfo {year} {1997})}\BibitemShut {NoStop}%
\bibitem [{\citenamefont {Gustavson}\ \emph {et~al.}(1997)\citenamefont
  {Gustavson}, \citenamefont {Bouyer},\ and\ \citenamefont
  {Kasevich}}]{Gustavson:1997}%
  \BibitemOpen
  \bibfield  {author} {\bibinfo {author} {\bibfnamefont {T.~L.}\ \bibnamefont
  {Gustavson}}, \bibinfo {author} {\bibfnamefont {P.}~\bibnamefont {Bouyer}}, \
  and\ \bibinfo {author} {\bibfnamefont {M.~A.}\ \bibnamefont {Kasevich}},\
  }\href {\doibase 10.1103/PhysRevLett.78.2046} {\bibfield  {journal} {\bibinfo
   {journal} {Phys. Rev. Lett.}\ }\textbf {\bibinfo {volume} {78}},\ \bibinfo
  {pages} {2046} (\bibinfo {year} {1997})}\BibitemShut {NoStop}%
\bibitem [{\citenamefont {Snadden}\ \emph {et~al.}(1998)\citenamefont
  {Snadden}, \citenamefont {McGuirk}, \citenamefont {Bouyer}, \citenamefont
  {Haritos},\ and\ \citenamefont {Kasevich}}]{Snadden:1998}%
  \BibitemOpen
  \bibfield  {author} {\bibinfo {author} {\bibfnamefont {M.~J.}\ \bibnamefont
  {Snadden}}, \bibinfo {author} {\bibfnamefont {J.~M.}\ \bibnamefont
  {McGuirk}}, \bibinfo {author} {\bibfnamefont {P.}~\bibnamefont {Bouyer}},
  \bibinfo {author} {\bibfnamefont {K.~G.}\ \bibnamefont {Haritos}}, \ and\
  \bibinfo {author} {\bibfnamefont {M.~A.}\ \bibnamefont {Kasevich}},\ }\href
  {\doibase 10.1103/PhysRevLett.81.971} {\bibfield  {journal} {\bibinfo
  {journal} {Phys. Rev. Lett.}\ }\textbf {\bibinfo {volume} {81}},\ \bibinfo
  {pages} {971} (\bibinfo {year} {1998})}\BibitemShut {NoStop}%
\bibitem [{\citenamefont {McGuirk}\ \emph {et~al.}(2002)\citenamefont
  {McGuirk}, \citenamefont {Foster}, \citenamefont {Fixler}, \citenamefont
  {Snadden},\ and\ \citenamefont {Kasevich}}]{McGuirk:2002}%
  \BibitemOpen
  \bibfield  {author} {\bibinfo {author} {\bibfnamefont {J.~M.}\ \bibnamefont
  {McGuirk}}, \bibinfo {author} {\bibfnamefont {G.~T.}\ \bibnamefont {Foster}},
  \bibinfo {author} {\bibfnamefont {J.~B.}\ \bibnamefont {Fixler}}, \bibinfo
  {author} {\bibfnamefont {M.~J.}\ \bibnamefont {Snadden}}, \ and\ \bibinfo
  {author} {\bibfnamefont {M.~A.}\ \bibnamefont {Kasevich}},\ }\href {\doibase
  10.1103/PhysRevA.65.033608} {\bibfield  {journal} {\bibinfo  {journal} {Phys.
  Rev. A}\ }\textbf {\bibinfo {volume} {65}},\ \bibinfo {pages} {033608}
  (\bibinfo {year} {2002})}\BibitemShut {NoStop}%
\bibitem [{\citenamefont {Vengalattore}\ \emph {et~al.}(2007)\citenamefont
  {Vengalattore}, \citenamefont {Higbie}, \citenamefont {Leslie}, \citenamefont
  {Guzman}, \citenamefont {Sadler},\ and\ \citenamefont
  {Stamper-Kurn}}]{Vengalattore:2007}%
  \BibitemOpen
  \bibfield  {author} {\bibinfo {author} {\bibfnamefont {M.}~\bibnamefont
  {Vengalattore}}, \bibinfo {author} {\bibfnamefont {J.~M.}\ \bibnamefont
  {Higbie}}, \bibinfo {author} {\bibfnamefont {S.~R.}\ \bibnamefont {Leslie}},
  \bibinfo {author} {\bibfnamefont {J.}~\bibnamefont {Guzman}}, \bibinfo
  {author} {\bibfnamefont {L.~E.}\ \bibnamefont {Sadler}}, \ and\ \bibinfo
  {author} {\bibfnamefont {D.~M.}\ \bibnamefont {Stamper-Kurn}},\ }\href
  {\doibase 10.1103/PhysRevLett.98.200801} {\bibfield  {journal} {\bibinfo
  {journal} {Phys. Rev. Lett.}\ }\textbf {\bibinfo {volume} {98}},\ \bibinfo
  {pages} {200801} (\bibinfo {year} {2007})}\BibitemShut {NoStop}%
\bibitem [{\citenamefont {Gupta}\ \emph {et~al.}(2002)\citenamefont {Gupta},
  \citenamefont {Dieckmann}, \citenamefont {Hadzibabic},\ and\ \citenamefont
  {Pritchard}}]{Gupta:2002}%
  \BibitemOpen
  \bibfield  {author} {\bibinfo {author} {\bibfnamefont {S.}~\bibnamefont
  {Gupta}}, \bibinfo {author} {\bibfnamefont {K.}~\bibnamefont {Dieckmann}},
  \bibinfo {author} {\bibfnamefont {Z.}~\bibnamefont {Hadzibabic}}, \ and\
  \bibinfo {author} {\bibfnamefont {D.~E.}\ \bibnamefont {Pritchard}},\ }\href
  {\doibase 10.1103/PhysRevLett.89.140401} {\bibfield  {journal} {\bibinfo
  {journal} {Phys. Rev. Lett.}\ }\textbf {\bibinfo {volume} {89}},\ \bibinfo
  {pages} {140401} (\bibinfo {year} {2002})}\BibitemShut {NoStop}%
\bibitem [{\citenamefont {Bouchendira}\ \emph {et~al.}(2011)\citenamefont
  {Bouchendira}, \citenamefont {Clad\'e}, \citenamefont {Guellati-Kh\'elifa},
  \citenamefont {Nez},\ and\ \citenamefont {Biraben}}]{Bouchendira:2011}%
  \BibitemOpen
  \bibfield  {author} {\bibinfo {author} {\bibfnamefont {R.}~\bibnamefont
  {Bouchendira}}, \bibinfo {author} {\bibfnamefont {P.}~\bibnamefont
  {Clad\'e}}, \bibinfo {author} {\bibfnamefont {S.}~\bibnamefont
  {Guellati-Kh\'elifa}}, \bibinfo {author} {\bibfnamefont {F.}~\bibnamefont
  {Nez}}, \ and\ \bibinfo {author} {\bibfnamefont {F.}~\bibnamefont
  {Biraben}},\ }\href {\doibase 10.1103/PhysRevLett.106.080801} {\bibfield
  {journal} {\bibinfo  {journal} {Phys. Rev. Lett.}\ }\textbf {\bibinfo
  {volume} {106}},\ \bibinfo {pages} {080801} (\bibinfo {year}
  {2011})}\BibitemShut {NoStop}%
\bibitem [{\citenamefont {Fixler}\ \emph {et~al.}(2007)\citenamefont {Fixler},
  \citenamefont {Foster}, \citenamefont {McGuirk},\ and\ \citenamefont
  {Kasevich}}]{Fixler:2007}%
  \BibitemOpen
  \bibfield  {author} {\bibinfo {author} {\bibfnamefont {J.~B.}\ \bibnamefont
  {Fixler}}, \bibinfo {author} {\bibfnamefont {G.~T.}\ \bibnamefont {Foster}},
  \bibinfo {author} {\bibfnamefont {J.~M.}\ \bibnamefont {McGuirk}}, \ and\
  \bibinfo {author} {\bibfnamefont {M.~A.}\ \bibnamefont {Kasevich}},\ }\href
  {\doibase 10.1126/science.1135459} {\bibfield  {journal} {\bibinfo  {journal}
  {Science}\ }\textbf {\bibinfo {volume} {315}},\ \bibinfo {pages} {74}
  (\bibinfo {year} {2007})}\BibitemShut {NoStop}%
\bibitem [{\citenamefont {Lamporesi}\ \emph {et~al.}(2008)\citenamefont
  {Lamporesi}, \citenamefont {Bertoldi}, \citenamefont {Cacciapuoti},
  \citenamefont {Prevedelli},\ and\ \citenamefont {Tino}}]{Lamporesi:2008}%
  \BibitemOpen
  \bibfield  {author} {\bibinfo {author} {\bibfnamefont {G.}~\bibnamefont
  {Lamporesi}}, \bibinfo {author} {\bibfnamefont {A.}~\bibnamefont {Bertoldi}},
  \bibinfo {author} {\bibfnamefont {L.}~\bibnamefont {Cacciapuoti}}, \bibinfo
  {author} {\bibfnamefont {M.}~\bibnamefont {Prevedelli}}, \ and\ \bibinfo
  {author} {\bibfnamefont {G.~M.}\ \bibnamefont {Tino}},\ }\href {\doibase
  10.1103/PhysRevLett.100.050801} {\bibfield  {journal} {\bibinfo  {journal}
  {Phys. Rev. Lett.}\ }\textbf {\bibinfo {volume} {100}},\ \bibinfo {pages}
  {050801} (\bibinfo {year} {2008})}\BibitemShut {NoStop}%
\bibitem [{\citenamefont {Andia}\ \emph {et~al.}(2013)\citenamefont {Andia},
  \citenamefont {Jannin}, \citenamefont {Nez}, \citenamefont {Biraben},
  \citenamefont {Guellati-Kh\'elifa},\ and\ \citenamefont
  {Clad\'e}}]{Andia:2013}%
  \BibitemOpen
  \bibfield  {author} {\bibinfo {author} {\bibfnamefont {M.}~\bibnamefont
  {Andia}}, \bibinfo {author} {\bibfnamefont {R.}~\bibnamefont {Jannin}},
  \bibinfo {author} {\bibfnamefont {F.}~\bibnamefont {Nez}}, \bibinfo {author}
  {\bibfnamefont {F.}~\bibnamefont {Biraben}}, \bibinfo {author} {\bibfnamefont
  {S.}~\bibnamefont {Guellati-Kh\'elifa}}, \ and\ \bibinfo {author}
  {\bibfnamefont {P.}~\bibnamefont {Clad\'e}},\ }\href {\doibase
  10.1103/PhysRevA.88.031605} {\bibfield  {journal} {\bibinfo  {journal} {Phys.
  Rev. A}\ }\textbf {\bibinfo {volume} {88}},\ \bibinfo {pages} {031605}
  (\bibinfo {year} {2013})}\BibitemShut {NoStop}%
\bibitem [{\citenamefont {Rosi}\ \emph {et~al.}(2014)\citenamefont {Rosi},
  \citenamefont {Sorrentino}, \citenamefont {Cacciapuoti}, \citenamefont
  {Prevedelli},\ and\ \citenamefont {Tino}}]{Rosi:2014}%
  \BibitemOpen
  \bibfield  {author} {\bibinfo {author} {\bibfnamefont {G.}~\bibnamefont
  {Rosi}}, \bibinfo {author} {\bibfnamefont {F.}~\bibnamefont {Sorrentino}},
  \bibinfo {author} {\bibfnamefont {L.}~\bibnamefont {Cacciapuoti}}, \bibinfo
  {author} {\bibfnamefont {M.}~\bibnamefont {Prevedelli}}, \ and\ \bibinfo
  {author} {\bibfnamefont {G.~M.}\ \bibnamefont {Tino}},\ }\href
  {http://dx.doi.org/10.1038/nature13433} {\bibfield  {journal} {\bibinfo
  {journal} {Nature}\ }\textbf {\bibinfo {volume} {510}},\ \bibinfo {pages}
  {518} (\bibinfo {year} {2014})}\BibitemShut {NoStop}%
\bibitem [{\citenamefont {Fray}\ \emph {et~al.}(2004)\citenamefont {Fray},
  \citenamefont {Diez}, \citenamefont {H\"ansch},\ and\ \citenamefont
  {Weitz}}]{Fray:2004}%
  \BibitemOpen
  \bibfield  {author} {\bibinfo {author} {\bibfnamefont {S.}~\bibnamefont
  {Fray}}, \bibinfo {author} {\bibfnamefont {C.~A.}\ \bibnamefont {Diez}},
  \bibinfo {author} {\bibfnamefont {T.~W.}\ \bibnamefont {H\"ansch}}, \ and\
  \bibinfo {author} {\bibfnamefont {M.}~\bibnamefont {Weitz}},\ }\href
  {\doibase 10.1103/PhysRevLett.93.240404} {\bibfield  {journal} {\bibinfo
  {journal} {Phys. Rev. Lett.}\ }\textbf {\bibinfo {volume} {93}},\ \bibinfo
  {pages} {240404} (\bibinfo {year} {2004})}\BibitemShut {NoStop}%
\bibitem [{\citenamefont {Dimopoulos}\ \emph {et~al.}(2007)\citenamefont
  {Dimopoulos}, \citenamefont {Graham}, \citenamefont {Hogan},\ and\
  \citenamefont {Kasevich}}]{Dimopoulos:2007}%
  \BibitemOpen
  \bibfield  {author} {\bibinfo {author} {\bibfnamefont {S.}~\bibnamefont
  {Dimopoulos}}, \bibinfo {author} {\bibfnamefont {P.~W.}\ \bibnamefont
  {Graham}}, \bibinfo {author} {\bibfnamefont {J.~M.}\ \bibnamefont {Hogan}}, \
  and\ \bibinfo {author} {\bibfnamefont {M.~A.}\ \bibnamefont {Kasevich}},\
  }\href {\doibase 10.1103/PhysRevLett.98.111102} {\bibfield  {journal}
  {\bibinfo  {journal} {Phys. Rev. Lett.}\ }\textbf {\bibinfo {volume} {98}},\
  \bibinfo {pages} {111102} (\bibinfo {year} {2007})}\BibitemShut {NoStop}%
\bibitem [{\citenamefont {Schlippert}\ \emph {et~al.}(2014)\citenamefont
  {Schlippert}, \citenamefont {Hartwig}, \citenamefont {Albers}, \citenamefont
  {Richardson}, \citenamefont {Schubert}, \citenamefont {Roura}, \citenamefont
  {Schleich}, \citenamefont {Ertmer},\ and\ \citenamefont
  {Rasel}}]{Schlippert:2014}%
  \BibitemOpen
  \bibfield  {author} {\bibinfo {author} {\bibfnamefont {D.}~\bibnamefont
  {Schlippert}}, \bibinfo {author} {\bibfnamefont {J.}~\bibnamefont {Hartwig}},
  \bibinfo {author} {\bibfnamefont {H.}~\bibnamefont {Albers}}, \bibinfo
  {author} {\bibfnamefont {L.~L.}\ \bibnamefont {Richardson}}, \bibinfo
  {author} {\bibfnamefont {C.}~\bibnamefont {Schubert}}, \bibinfo {author}
  {\bibfnamefont {A.}~\bibnamefont {Roura}}, \bibinfo {author} {\bibfnamefont
  {W.~P.}\ \bibnamefont {Schleich}}, \bibinfo {author} {\bibfnamefont
  {W.}~\bibnamefont {Ertmer}}, \ and\ \bibinfo {author} {\bibfnamefont {E.~M.}\
  \bibnamefont {Rasel}},\ }\href {\doibase 10.1103/PhysRevLett.112.203002}
  {\bibfield  {journal} {\bibinfo  {journal} {Phys. Rev. Lett.}\ }\textbf
  {\bibinfo {volume} {112}},\ \bibinfo {pages} {203002} (\bibinfo {year}
  {2014})}\BibitemShut {NoStop}%
\bibitem [{\citenamefont {Amelino-Camelia}\ \emph {et~al.}(2009)\citenamefont
  {Amelino-Camelia}, \citenamefont {L\"ammerzahl}, \citenamefont {Mercati},\
  and\ \citenamefont {Tino}}]{Amelino-Camelia:2009}%
  \BibitemOpen
  \bibfield  {author} {\bibinfo {author} {\bibfnamefont {G.}~\bibnamefont
  {Amelino-Camelia}}, \bibinfo {author} {\bibfnamefont {C.}~\bibnamefont
  {L\"ammerzahl}}, \bibinfo {author} {\bibfnamefont {F.}~\bibnamefont
  {Mercati}}, \ and\ \bibinfo {author} {\bibfnamefont {G.~M.}\ \bibnamefont
  {Tino}},\ }\href {\doibase 10.1103/PhysRevLett.103.171302} {\bibfield
  {journal} {\bibinfo  {journal} {Phys. Rev. Lett.}\ }\textbf {\bibinfo
  {volume} {103}},\ \bibinfo {pages} {171302} (\bibinfo {year}
  {2009})}\BibitemShut {NoStop}%
\bibitem [{\citenamefont {Tino}\ and\ \citenamefont
  {Vetrano}(2007)}]{Tino:2007}%
  \BibitemOpen
  \bibfield  {author} {\bibinfo {author} {\bibfnamefont {G.~M.}\ \bibnamefont
  {Tino}}\ and\ \bibinfo {author} {\bibfnamefont {F.}~\bibnamefont {Vetrano}},\
  }\href {http://stacks.iop.org/0264-9381/24/i=9/a=001} {\bibfield  {journal}
  {\bibinfo  {journal} {Class. Quantum Grav.}\ }\textbf {\bibinfo {volume}
  {24}},\ \bibinfo {pages} {2167} (\bibinfo {year} {2007})}\BibitemShut
  {NoStop}%
\bibitem [{\citenamefont {Dimopoulos}\ \emph {et~al.}(2008)\citenamefont
  {Dimopoulos}, \citenamefont {Graham}, \citenamefont {Hogan}, \citenamefont
  {Kasevich},\ and\ \citenamefont {Rajendran}}]{Dimopoulos:2008}%
  \BibitemOpen
  \bibfield  {author} {\bibinfo {author} {\bibfnamefont {S.}~\bibnamefont
  {Dimopoulos}}, \bibinfo {author} {\bibfnamefont {P.~W.}\ \bibnamefont
  {Graham}}, \bibinfo {author} {\bibfnamefont {J.~M.}\ \bibnamefont {Hogan}},
  \bibinfo {author} {\bibfnamefont {M.~A.}\ \bibnamefont {Kasevich}}, \ and\
  \bibinfo {author} {\bibfnamefont {S.}~\bibnamefont {Rajendran}},\ }\href
  {\doibase 10.1103/PhysRevD.78.122002} {\bibfield  {journal} {\bibinfo
  {journal} {Phys. Rev. D}\ }\textbf {\bibinfo {volume} {78}},\ \bibinfo
  {pages} {122002} (\bibinfo {year} {2008})}\BibitemShut {NoStop}%
\bibitem [{\citenamefont {Robins}\ \emph {et~al.}(2013)\citenamefont {Robins},
  \citenamefont {Altin}, \citenamefont {Debs},\ and\ \citenamefont
  {Close}}]{Robins:2013}%
  \BibitemOpen
  \bibfield  {author} {\bibinfo {author} {\bibfnamefont {N.~P.}\ \bibnamefont
  {Robins}}, \bibinfo {author} {\bibfnamefont {P.~A.}\ \bibnamefont {Altin}},
  \bibinfo {author} {\bibfnamefont {J.~E.}\ \bibnamefont {Debs}}, \ and\
  \bibinfo {author} {\bibfnamefont {J.~D.}\ \bibnamefont {Close}},\ }\href
  {\doibase http://dx.doi.org/10.1016/j.physrep.2013.03.006} {\bibfield
  {journal} {\bibinfo  {journal} {Physics Reports}\ }\textbf {\bibinfo {volume}
  {529}},\ \bibinfo {pages} {265} (\bibinfo {year} {2013})}\BibitemShut
  {NoStop}%
\bibitem [{\citenamefont {Kheruntsyan}\ \emph {et~al.}(2005)\citenamefont
  {Kheruntsyan}, \citenamefont {Olsen},\ and\ \citenamefont
  {Drummond}}]{Kheruntsyan:2005b}%
  \BibitemOpen
  \bibfield  {author} {\bibinfo {author} {\bibfnamefont {K.~V.}\ \bibnamefont
  {Kheruntsyan}}, \bibinfo {author} {\bibfnamefont {M.~K.}\ \bibnamefont
  {Olsen}}, \ and\ \bibinfo {author} {\bibfnamefont {P.~D.}\ \bibnamefont
  {Drummond}},\ }\href {\doibase 10.1103/PhysRevLett.95.150405} {\bibfield
  {journal} {\bibinfo  {journal} {Phys. Rev. Lett.}\ }\textbf {\bibinfo
  {volume} {95}},\ \bibinfo {pages} {150405} (\bibinfo {year}
  {2005})}\BibitemShut {NoStop}%
\bibitem [{\citenamefont {Pu}\ and\ \citenamefont {Meystre}(2000)}]{Pu:2000}%
  \BibitemOpen
  \bibfield  {author} {\bibinfo {author} {\bibfnamefont {H.}~\bibnamefont
  {Pu}}\ and\ \bibinfo {author} {\bibfnamefont {P.}~\bibnamefont {Meystre}},\
  }\href {\doibase 10.1103/PhysRevLett.85.3987} {\bibfield  {journal} {\bibinfo
   {journal} {Phys. Rev. Lett.}\ }\textbf {\bibinfo {volume} {85}},\ \bibinfo
  {pages} {3987} (\bibinfo {year} {2000})}\BibitemShut {NoStop}%
\bibitem [{\citenamefont {L{\"u}cke}\ \emph {et~al.}(2011)\citenamefont
  {L{\"u}cke}, \citenamefont {Scherer}, \citenamefont {Kruse}, \citenamefont
  {Pezz{\'e}}, \citenamefont {Deuretzbacher}, \citenamefont {Hyllus},
  \citenamefont {Topic}, \citenamefont {Peise}, \citenamefont {Ertmer},
  \citenamefont {Arlt}, \citenamefont {Santos}, \citenamefont {Smerzi},\ and\
  \citenamefont {Klempt}}]{Lucke:2011}%
  \BibitemOpen
  \bibfield  {author} {\bibinfo {author} {\bibfnamefont {B.}~\bibnamefont
  {L{\"u}cke}}, \bibinfo {author} {\bibfnamefont {M.}~\bibnamefont {Scherer}},
  \bibinfo {author} {\bibfnamefont {J.}~\bibnamefont {Kruse}}, \bibinfo
  {author} {\bibfnamefont {L.}~\bibnamefont {Pezz{\'e}}}, \bibinfo {author}
  {\bibfnamefont {F.}~\bibnamefont {Deuretzbacher}}, \bibinfo {author}
  {\bibfnamefont {P.}~\bibnamefont {Hyllus}}, \bibinfo {author} {\bibfnamefont
  {O.}~\bibnamefont {Topic}}, \bibinfo {author} {\bibfnamefont
  {J.}~\bibnamefont {Peise}}, \bibinfo {author} {\bibfnamefont
  {W.}~\bibnamefont {Ertmer}}, \bibinfo {author} {\bibfnamefont
  {J.}~\bibnamefont {Arlt}}, \bibinfo {author} {\bibfnamefont {L.}~\bibnamefont
  {Santos}}, \bibinfo {author} {\bibfnamefont {A.}~\bibnamefont {Smerzi}}, \
  and\ \bibinfo {author} {\bibfnamefont {C.}~\bibnamefont {Klempt}},\
  }\href@noop {} {\bibfield  {journal} {\bibinfo  {journal} {Science}\ }\textbf
  {\bibinfo {volume} {334}},\ \bibinfo {pages} {773} (\bibinfo {year}
  {2011})}\BibitemShut {NoStop}%
\bibitem [{\citenamefont {Gross}\ \emph {et~al.}(2011)\citenamefont {Gross},
  \citenamefont {Strobel}, \citenamefont {Nicklas}, \citenamefont {Zibold},
  \citenamefont {Bar-Gill}, \citenamefont {Kurizki},\ and\ \citenamefont
  {Oberthaler}}]{Gross:2011}%
  \BibitemOpen
  \bibfield  {author} {\bibinfo {author} {\bibfnamefont {C.}~\bibnamefont
  {Gross}}, \bibinfo {author} {\bibfnamefont {H.}~\bibnamefont {Strobel}},
  \bibinfo {author} {\bibfnamefont {E.}~\bibnamefont {Nicklas}}, \bibinfo
  {author} {\bibfnamefont {T.}~\bibnamefont {Zibold}}, \bibinfo {author}
  {\bibfnamefont {N.}~\bibnamefont {Bar-Gill}}, \bibinfo {author}
  {\bibfnamefont {G.}~\bibnamefont {Kurizki}}, \ and\ \bibinfo {author}
  {\bibfnamefont {M.~K.}\ \bibnamefont {Oberthaler}},\ }\href
  {http://dx.doi.org/10.1038/nature10654} {\bibfield  {journal} {\bibinfo
  {journal} {Nature}\ }\textbf {\bibinfo {volume} {480}},\ \bibinfo {pages}
  {219} (\bibinfo {year} {2011})}\BibitemShut {NoStop}%
\bibitem [{\citenamefont {Jaskula}\ \emph {et~al.}(2010)\citenamefont
  {Jaskula}, \citenamefont {Bonneau}, \citenamefont {Partridge}, \citenamefont
  {Krachmalnicoff}, \citenamefont {Deuar}, \citenamefont {Kheruntsyan},
  \citenamefont {Aspect}, \citenamefont {Boiron},\ and\ \citenamefont
  {Westbrook}}]{Jaskula:2010}%
  \BibitemOpen
  \bibfield  {author} {\bibinfo {author} {\bibfnamefont {J.-C.}\ \bibnamefont
  {Jaskula}}, \bibinfo {author} {\bibfnamefont {M.}~\bibnamefont {Bonneau}},
  \bibinfo {author} {\bibfnamefont {G.~B.}\ \bibnamefont {Partridge}}, \bibinfo
  {author} {\bibfnamefont {V.}~\bibnamefont {Krachmalnicoff}}, \bibinfo
  {author} {\bibfnamefont {P.}~\bibnamefont {Deuar}}, \bibinfo {author}
  {\bibfnamefont {K.~V.}\ \bibnamefont {Kheruntsyan}}, \bibinfo {author}
  {\bibfnamefont {A.}~\bibnamefont {Aspect}}, \bibinfo {author} {\bibfnamefont
  {D.}~\bibnamefont {Boiron}}, \ and\ \bibinfo {author} {\bibfnamefont {C.~I.}\
  \bibnamefont {Westbrook}},\ }\href {\doibase 10.1103/PhysRevLett.105.190402}
  {\bibfield  {journal} {\bibinfo  {journal} {Phys. Rev. Lett.}\ }\textbf
  {\bibinfo {volume} {105}},\ \bibinfo {pages} {190402} (\bibinfo {year}
  {2010})}\BibitemShut {NoStop}%
\bibitem [{\citenamefont {Bucker}\ \emph {et~al.}(2011)\citenamefont {Bucker},
  \citenamefont {Grond}, \citenamefont {Manz}, \citenamefont {Berrada},
  \citenamefont {Betz}, \citenamefont {Koller}, \citenamefont {Hohenester},
  \citenamefont {Schumm}, \citenamefont {Perrin},\ and\ \citenamefont
  {Schmiedmayer}}]{Bucker:2011}%
  \BibitemOpen
  \bibfield  {author} {\bibinfo {author} {\bibfnamefont {R.}~\bibnamefont
  {Bucker}}, \bibinfo {author} {\bibfnamefont {J.}~\bibnamefont {Grond}},
  \bibinfo {author} {\bibfnamefont {S.}~\bibnamefont {Manz}}, \bibinfo {author}
  {\bibfnamefont {T.}~\bibnamefont {Berrada}}, \bibinfo {author} {\bibfnamefont
  {T.}~\bibnamefont {Betz}}, \bibinfo {author} {\bibfnamefont {C.}~\bibnamefont
  {Koller}}, \bibinfo {author} {\bibfnamefont {U.}~\bibnamefont {Hohenester}},
  \bibinfo {author} {\bibfnamefont {T.}~\bibnamefont {Schumm}}, \bibinfo
  {author} {\bibfnamefont {A.}~\bibnamefont {Perrin}}, \ and\ \bibinfo {author}
  {\bibfnamefont {J.}~\bibnamefont {Schmiedmayer}},\ }\href
  {http://dx.doi.org/10.1038/nphys1992} {\bibfield  {journal} {\bibinfo
  {journal} {Nat Phys}\ }\textbf {\bibinfo {volume} {7}},\ \bibinfo {pages}
  {608} (\bibinfo {year} {2011})}\BibitemShut {NoStop}%
\bibitem [{\citenamefont {Haine}\ and\ \citenamefont
  {Ferris}(2011)}]{Haine:2011}%
  \BibitemOpen
  \bibfield  {author} {\bibinfo {author} {\bibfnamefont {S.~A.}\ \bibnamefont
  {Haine}}\ and\ \bibinfo {author} {\bibfnamefont {A.~J.}\ \bibnamefont
  {Ferris}},\ }\href {\doibase 10.1103/PhysRevA.84.043624} {\bibfield
  {journal} {\bibinfo  {journal} {Phys. Rev. A}\ }\textbf {\bibinfo {volume}
  {84}},\ \bibinfo {pages} {043624} (\bibinfo {year} {2011})}\BibitemShut
  {NoStop}%
\bibitem [{\citenamefont {Lewis-Swan}\ and\ \citenamefont
  {Kheruntsyan}(2014)}]{Lewis-Swan:2014}%
  \BibitemOpen
  \bibfield  {author} {\bibinfo {author} {\bibfnamefont {R.~J.}\ \bibnamefont
  {Lewis-Swan}}\ and\ \bibinfo {author} {\bibfnamefont {K.~V.}\ \bibnamefont
  {Kheruntsyan}},\ }\href {http://dx.doi.org/10.1038/ncomms4752} {\bibfield
  {journal} {\bibinfo  {journal} {Nat Commun}\ }\textbf {\bibinfo {volume} {5}}
  (\bibinfo {year} {2014})}\BibitemShut {NoStop}%
\bibitem [{\citenamefont {Kitagawa}\ and\ \citenamefont
  {Ueda}(1993)}]{Kitagawa:1993}%
  \BibitemOpen
  \bibfield  {author} {\bibinfo {author} {\bibfnamefont {M.}~\bibnamefont
  {Kitagawa}}\ and\ \bibinfo {author} {\bibfnamefont {M.}~\bibnamefont
  {Ueda}},\ }\href {\doibase 10.1103/PhysRevA.47.5138} {\bibfield  {journal}
  {\bibinfo  {journal} {Phys. Rev. A}\ }\textbf {\bibinfo {volume} {47}},\
  \bibinfo {pages} {5138} (\bibinfo {year} {1993})}\BibitemShut {NoStop}%
\bibitem [{\citenamefont {S\o{}ndberg~S\o{}rensen}(2002)}]{Sorensen:2002}%
  \BibitemOpen
  \bibfield  {author} {\bibinfo {author} {\bibfnamefont {A.}~\bibnamefont
  {S\o{}ndberg~S\o{}rensen}},\ }\href {\doibase 10.1103/PhysRevA.65.043610}
  {\bibfield  {journal} {\bibinfo  {journal} {Phys. Rev. A}\ }\textbf {\bibinfo
  {volume} {65}},\ \bibinfo {pages} {043610} (\bibinfo {year}
  {2002})}\BibitemShut {NoStop}%
\bibitem [{\citenamefont {Gross}\ \emph {et~al.}(2010)\citenamefont {Gross},
  \citenamefont {Zibold}, \citenamefont {Nicklas}, \citenamefont {Est{\`e}ve},\
  and\ \citenamefont {Oberthaler}}]{Gross:2010}%
  \BibitemOpen
  \bibfield  {author} {\bibinfo {author} {\bibfnamefont {C.}~\bibnamefont
  {Gross}}, \bibinfo {author} {\bibfnamefont {T.}~\bibnamefont {Zibold}},
  \bibinfo {author} {\bibfnamefont {E.}~\bibnamefont {Nicklas}}, \bibinfo
  {author} {\bibfnamefont {J.}~\bibnamefont {Est{\`e}ve}}, \ and\ \bibinfo
  {author} {\bibfnamefont {M.~K.}\ \bibnamefont {Oberthaler}},\ }\href
  {http://dx.doi.org/10.1038/nature08919} {\bibfield  {journal} {\bibinfo
  {journal} {Nature}\ }\textbf {\bibinfo {volume} {464}},\ \bibinfo {pages}
  {1165} (\bibinfo {year} {2010})}\BibitemShut {NoStop}%
\bibitem [{\citenamefont {Riedel}\ \emph {et~al.}(2010)\citenamefont {Riedel},
  \citenamefont {B{\"o}hi}, \citenamefont {Li}, \citenamefont {H{\"a}nsch},
  \citenamefont {Sinatra},\ and\ \citenamefont {Treutlein}}]{Riedel:2010}%
  \BibitemOpen
  \bibfield  {author} {\bibinfo {author} {\bibfnamefont {M.~F.}\ \bibnamefont
  {Riedel}}, \bibinfo {author} {\bibfnamefont {P.}~\bibnamefont {B{\"o}hi}},
  \bibinfo {author} {\bibfnamefont {Y.}~\bibnamefont {Li}}, \bibinfo {author}
  {\bibfnamefont {T.~W.}\ \bibnamefont {H{\"a}nsch}}, \bibinfo {author}
  {\bibfnamefont {A.}~\bibnamefont {Sinatra}}, \ and\ \bibinfo {author}
  {\bibfnamefont {P.}~\bibnamefont {Treutlein}},\ }\href
  {http://dx.doi.org/10.1038/nature08988} {\bibfield  {journal} {\bibinfo
  {journal} {Nature}\ }\textbf {\bibinfo {volume} {464}},\ \bibinfo {pages}
  {1170} (\bibinfo {year} {2010})}\BibitemShut {NoStop}%
\bibitem [{\citenamefont {Johnsson}\ and\ \citenamefont
  {Haine}(2007)}]{Johnsson:2007a}%
  \BibitemOpen
  \bibfield  {author} {\bibinfo {author} {\bibfnamefont {M.~T.}\ \bibnamefont
  {Johnsson}}\ and\ \bibinfo {author} {\bibfnamefont {S.~A.}\ \bibnamefont
  {Haine}},\ }\href {\doibase 10.1103/PhysRevLett.99.010401} {\bibfield
  {journal} {\bibinfo  {journal} {Phys. Rev. Lett.}\ }\textbf {\bibinfo
  {volume} {99}},\ \bibinfo {pages} {010401} (\bibinfo {year}
  {2007})}\BibitemShut {NoStop}%
\bibitem [{\citenamefont {Haine}\ and\ \citenamefont
  {Johnsson}(2009)}]{Haine:2009}%
  \BibitemOpen
  \bibfield  {author} {\bibinfo {author} {\bibfnamefont {S.~A.}\ \bibnamefont
  {Haine}}\ and\ \bibinfo {author} {\bibfnamefont {M.~T.}\ \bibnamefont
  {Johnsson}},\ }\href {\doibase 10.1103/PhysRevA.80.023611} {\bibfield
  {journal} {\bibinfo  {journal} {Phys. Rev. A}\ }\textbf {\bibinfo {volume}
  {80}},\ \bibinfo {pages} {023611} (\bibinfo {year} {2009})}\BibitemShut
  {NoStop}%
\bibitem [{\citenamefont {Haine}\ \emph
  {et~al.}(2014{\natexlab{a}})\citenamefont {Haine}, \citenamefont {Lau},
  \citenamefont {Anderson},\ and\ \citenamefont {Johnsson}}]{Haine:2014}%
  \BibitemOpen
  \bibfield  {author} {\bibinfo {author} {\bibfnamefont {S.~A.}\ \bibnamefont
  {Haine}}, \bibinfo {author} {\bibfnamefont {J.}~\bibnamefont {Lau}}, \bibinfo
  {author} {\bibfnamefont {R.~P.}\ \bibnamefont {Anderson}}, \ and\ \bibinfo
  {author} {\bibfnamefont {M.~T.}\ \bibnamefont {Johnsson}},\ }\href {\doibase
  10.1103/PhysRevA.90.023613} {\bibfield  {journal} {\bibinfo  {journal} {Phys.
  Rev. A}\ }\textbf {\bibinfo {volume} {90}},\ \bibinfo {pages} {023613}
  (\bibinfo {year} {2014}{\natexlab{a}})}\BibitemShut {NoStop}%
\bibitem [{\citenamefont {Jing}\ \emph {et~al.}(2000)\citenamefont {Jing},
  \citenamefont {Chen},\ and\ \citenamefont {Ge}}]{Jing:2000}%
  \BibitemOpen
  \bibfield  {author} {\bibinfo {author} {\bibfnamefont {H.}~\bibnamefont
  {Jing}}, \bibinfo {author} {\bibfnamefont {J.-L.}\ \bibnamefont {Chen}}, \
  and\ \bibinfo {author} {\bibfnamefont {M.-L.}\ \bibnamefont {Ge}},\ }\href
  {\doibase 10.1103/PhysRevA.63.015601} {\bibfield  {journal} {\bibinfo
  {journal} {Phys. Rev. A}\ }\textbf {\bibinfo {volume} {63}},\ \bibinfo
  {pages} {015601} (\bibinfo {year} {2000})}\BibitemShut {NoStop}%
\bibitem [{\citenamefont {Fleischhauer}\ and\ \citenamefont
  {Gong}(2002)}]{Fleischhauer:2002b}%
  \BibitemOpen
  \bibfield  {author} {\bibinfo {author} {\bibfnamefont {M.}~\bibnamefont
  {Fleischhauer}}\ and\ \bibinfo {author} {\bibfnamefont {S.}~\bibnamefont
  {Gong}},\ }\href {\doibase 10.1103/PhysRevLett.88.070404} {\bibfield
  {journal} {\bibinfo  {journal} {Phys. Rev. Lett.}\ }\textbf {\bibinfo
  {volume} {88}},\ \bibinfo {pages} {070404} (\bibinfo {year}
  {2002})}\BibitemShut {NoStop}%
\bibitem [{\citenamefont {Haine}\ and\ \citenamefont
  {Hope}(2005{\natexlab{a}})}]{Haine:2005}%
  \BibitemOpen
  \bibfield  {author} {\bibinfo {author} {\bibfnamefont {S.~A.}\ \bibnamefont
  {Haine}}\ and\ \bibinfo {author} {\bibfnamefont {J.~J.}\ \bibnamefont
  {Hope}},\ }\href {\doibase 10.1103/PhysRevA.72.033601} {\bibfield  {journal}
  {\bibinfo  {journal} {Phys. Rev. A}\ }\textbf {\bibinfo {volume} {72}},\
  \bibinfo {pages} {033601} (\bibinfo {year} {2005}{\natexlab{a}})}\BibitemShut
  {NoStop}%
\bibitem [{\citenamefont {Haine}\ and\ \citenamefont
  {Hope}(2005{\natexlab{b}})}]{Haine:2005b}%
  \BibitemOpen
  \bibfield  {author} {\bibinfo {author} {\bibfnamefont {S.~A.}\ \bibnamefont
  {Haine}}\ and\ \bibinfo {author} {\bibfnamefont {J.~J.}\ \bibnamefont
  {Hope}},\ }\href {\doibase 10.1002/lapl.200510052} {\bibfield  {journal}
  {\bibinfo  {journal} {Laser Phys. Lett.}\ }\textbf {\bibinfo {volume} {2}},\
  \bibinfo {pages} {597} (\bibinfo {year} {2005}{\natexlab{b}})}\BibitemShut
  {NoStop}%
\bibitem [{\citenamefont {Haine}\ \emph {et~al.}(2006)\citenamefont {Haine},
  \citenamefont {Olsen},\ and\ \citenamefont {Hope}}]{Haine:2006b}%
  \BibitemOpen
  \bibfield  {author} {\bibinfo {author} {\bibfnamefont {S.~A.}\ \bibnamefont
  {Haine}}, \bibinfo {author} {\bibfnamefont {M.~K.}\ \bibnamefont {Olsen}}, \
  and\ \bibinfo {author} {\bibfnamefont {J.~J.}\ \bibnamefont {Hope}},\ }\href
  {\doibase 10.1103/PhysRevLett.96.133601} {\bibfield  {journal} {\bibinfo
  {journal} {Phys. Rev. Lett.}\ }\textbf {\bibinfo {volume} {96}},\ \bibinfo
  {pages} {133601} (\bibinfo {year} {2006})}\BibitemShut {NoStop}%
\bibitem [{\citenamefont {Hammerer}\ \emph {et~al.}(2010)\citenamefont
  {Hammerer}, \citenamefont {S\o{}rensen},\ and\ \citenamefont
  {Polzik}}]{Hammerer:2010}%
  \BibitemOpen
  \bibfield  {author} {\bibinfo {author} {\bibfnamefont {K.}~\bibnamefont
  {Hammerer}}, \bibinfo {author} {\bibfnamefont {A.~S.}\ \bibnamefont
  {S\o{}rensen}}, \ and\ \bibinfo {author} {\bibfnamefont {E.~S.}\ \bibnamefont
  {Polzik}},\ }\href {\doibase 10.1103/RevModPhys.82.1041} {\bibfield
  {journal} {\bibinfo  {journal} {Rev. Mod. Phys.}\ }\textbf {\bibinfo {volume}
  {82}},\ \bibinfo {pages} {1041} (\bibinfo {year} {2010})}\BibitemShut
  {NoStop}%
\bibitem [{\citenamefont {Caves}(1981)}]{Caves:1981}%
  \BibitemOpen
  \bibfield  {author} {\bibinfo {author} {\bibfnamefont {C.~M.}\ \bibnamefont
  {Caves}},\ }\href {\doibase 10.1103/PhysRevD.23.1693} {\bibfield  {journal}
  {\bibinfo  {journal} {Phys. Rev. D}\ }\textbf {\bibinfo {volume} {23}},\
  \bibinfo {pages} {1693} (\bibinfo {year} {1981})}\BibitemShut {NoStop}%
\bibitem [{\citenamefont {Debs}\ \emph {et~al.}(2011)\citenamefont {Debs},
  \citenamefont {Altin}, \citenamefont {Barter}, \citenamefont {D\"oring},
  \citenamefont {Dennis}, \citenamefont {McDonald}, \citenamefont {Anderson},
  \citenamefont {Close},\ and\ \citenamefont {Robins}}]{Debs:2011}%
  \BibitemOpen
  \bibfield  {author} {\bibinfo {author} {\bibfnamefont {J.~E.}\ \bibnamefont
  {Debs}}, \bibinfo {author} {\bibfnamefont {P.~A.}\ \bibnamefont {Altin}},
  \bibinfo {author} {\bibfnamefont {T.~H.}\ \bibnamefont {Barter}}, \bibinfo
  {author} {\bibfnamefont {D.}~\bibnamefont {D\"oring}}, \bibinfo {author}
  {\bibfnamefont {G.~R.}\ \bibnamefont {Dennis}}, \bibinfo {author}
  {\bibfnamefont {G.}~\bibnamefont {McDonald}}, \bibinfo {author}
  {\bibfnamefont {R.~P.}\ \bibnamefont {Anderson}}, \bibinfo {author}
  {\bibfnamefont {J.~D.}\ \bibnamefont {Close}}, \ and\ \bibinfo {author}
  {\bibfnamefont {N.~P.}\ \bibnamefont {Robins}},\ }\href {\doibase
  10.1103/PhysRevA.84.033610} {\bibfield  {journal} {\bibinfo  {journal} {Phys.
  Rev. A}\ }\textbf {\bibinfo {volume} {84}},\ \bibinfo {pages} {033610}
  (\bibinfo {year} {2011})}\BibitemShut {NoStop}%
\bibitem [{\citenamefont {Szigeti}\ \emph {et~al.}(2012)\citenamefont
  {Szigeti}, \citenamefont {Debs}, \citenamefont {Hope}, \citenamefont
  {Robins},\ and\ \citenamefont {Close}}]{Szigeti:2012}%
  \BibitemOpen
  \bibfield  {author} {\bibinfo {author} {\bibfnamefont {S.~S.}\ \bibnamefont
  {Szigeti}}, \bibinfo {author} {\bibfnamefont {J.~E.}\ \bibnamefont {Debs}},
  \bibinfo {author} {\bibfnamefont {J.~J.}\ \bibnamefont {Hope}}, \bibinfo
  {author} {\bibfnamefont {N.~P.}\ \bibnamefont {Robins}}, \ and\ \bibinfo
  {author} {\bibfnamefont {J.~D.}\ \bibnamefont {Close}},\ }\href
  {http://stacks.iop.org/1367-2630/14/i=2/a=023009} {\bibfield  {journal}
  {\bibinfo  {journal} {New J. Phys.}\ }\textbf {\bibinfo {volume} {14}},\
  \bibinfo {pages} {023009} (\bibinfo {year} {2012})}\BibitemShut {NoStop}%
\bibitem [{\citenamefont {Haine}\ \emph {et~al.}(2004)\citenamefont {Haine},
  \citenamefont {Ferris}, \citenamefont {Close},\ and\ \citenamefont
  {Hope}}]{Haine:2004}%
  \BibitemOpen
  \bibfield  {author} {\bibinfo {author} {\bibfnamefont {S.~A.}\ \bibnamefont
  {Haine}}, \bibinfo {author} {\bibfnamefont {A.~J.}\ \bibnamefont {Ferris}},
  \bibinfo {author} {\bibfnamefont {J.~D.}\ \bibnamefont {Close}}, \ and\
  \bibinfo {author} {\bibfnamefont {J.~J.}\ \bibnamefont {Hope}},\ }\href
  {\doibase 10.1103/PhysRevA.69.013605} {\bibfield  {journal} {\bibinfo
  {journal} {Phys. Rev. A}\ }\textbf {\bibinfo {volume} {69}},\ \bibinfo
  {pages} {013605} (\bibinfo {year} {2004})}\BibitemShut {NoStop}%
\bibitem [{\citenamefont {Szigeti}\ \emph {et~al.}(2009)\citenamefont
  {Szigeti}, \citenamefont {Hush}, \citenamefont {Carvalho},\ and\
  \citenamefont {Hope}}]{Szigeti:2009}%
  \BibitemOpen
  \bibfield  {author} {\bibinfo {author} {\bibfnamefont {S.~S.}\ \bibnamefont
  {Szigeti}}, \bibinfo {author} {\bibfnamefont {M.~R.}\ \bibnamefont {Hush}},
  \bibinfo {author} {\bibfnamefont {A.~R.~R.}\ \bibnamefont {Carvalho}}, \ and\
  \bibinfo {author} {\bibfnamefont {J.~J.}\ \bibnamefont {Hope}},\ }\href
  {\doibase 10.1103/PhysRevA.80.013614} {\bibfield  {journal} {\bibinfo
  {journal} {Phys. Rev. A}\ }\textbf {\bibinfo {volume} {80}},\ \bibinfo {eid}
  {013614} (\bibinfo {year} {2009})}\BibitemShut {NoStop}%
\bibitem [{\citenamefont {Szigeti}\ \emph {et~al.}(2010)\citenamefont
  {Szigeti}, \citenamefont {Hush}, \citenamefont {Carvalho},\ and\
  \citenamefont {Hope}}]{Szigeti:2010}%
  \BibitemOpen
  \bibfield  {author} {\bibinfo {author} {\bibfnamefont {S.~S.}\ \bibnamefont
  {Szigeti}}, \bibinfo {author} {\bibfnamefont {M.~R.}\ \bibnamefont {Hush}},
  \bibinfo {author} {\bibfnamefont {A.~R.~R.}\ \bibnamefont {Carvalho}}, \ and\
  \bibinfo {author} {\bibfnamefont {J.~J.}\ \bibnamefont {Hope}},\ }\href
  {\doibase 10.1103/PhysRevA.82.043632} {\bibfield  {journal} {\bibinfo
  {journal} {Phys. Rev. A}\ }\textbf {\bibinfo {volume} {82}},\ \bibinfo
  {pages} {043632} (\bibinfo {year} {2010})}\BibitemShut {NoStop}%
\bibitem [{\citenamefont {Szigeti}\ \emph {et~al.}(2013)\citenamefont
  {Szigeti}, \citenamefont {Adlong}, \citenamefont {Hush}, \citenamefont
  {Carvalho},\ and\ \citenamefont {Hope}}]{Szigeti:2013}%
  \BibitemOpen
  \bibfield  {author} {\bibinfo {author} {\bibfnamefont {S.~S.}\ \bibnamefont
  {Szigeti}}, \bibinfo {author} {\bibfnamefont {S.~J.}\ \bibnamefont {Adlong}},
  \bibinfo {author} {\bibfnamefont {M.~R.}\ \bibnamefont {Hush}}, \bibinfo
  {author} {\bibfnamefont {A.~R.~R.}\ \bibnamefont {Carvalho}}, \ and\ \bibinfo
  {author} {\bibfnamefont {J.~J.}\ \bibnamefont {Hope}},\ }\href {\doibase
  10.1103/PhysRevA.87.013626} {\bibfield  {journal} {\bibinfo  {journal} {Phys.
  Rev. A}\ }\textbf {\bibinfo {volume} {87}},\ \bibinfo {pages} {013626}
  (\bibinfo {year} {2013})}\BibitemShut {NoStop}%
\bibitem [{\citenamefont {Hush}\ \emph {et~al.}(2013)\citenamefont {Hush},
  \citenamefont {Szigeti}, \citenamefont {Carvalho},\ and\ \citenamefont
  {Hope}}]{Hush:2013}%
  \BibitemOpen
  \bibfield  {author} {\bibinfo {author} {\bibfnamefont {M.~R.}\ \bibnamefont
  {Hush}}, \bibinfo {author} {\bibfnamefont {S.~S.}\ \bibnamefont {Szigeti}},
  \bibinfo {author} {\bibfnamefont {A.~R.~R.}\ \bibnamefont {Carvalho}}, \ and\
  \bibinfo {author} {\bibfnamefont {J.~J.}\ \bibnamefont {Hope}},\ }\href
  {http://stacks.iop.org/1367-2630/15/i=11/a=113060} {\bibfield  {journal}
  {\bibinfo  {journal} {New J. Phys.}\ }\textbf {\bibinfo {volume} {15}},\
  \bibinfo {pages} {113060} (\bibinfo {year} {2013})}\BibitemShut {NoStop}%
\bibitem [{\citenamefont {Haine}(2013)}]{Haine:2013}%
  \BibitemOpen
  \bibfield  {author} {\bibinfo {author} {\bibfnamefont {S.~A.}\ \bibnamefont
  {Haine}},\ }\href {\doibase 10.1103/PhysRevLett.110.053002} {\bibfield
  {journal} {\bibinfo  {journal} {Phys. Rev. Lett.}\ }\textbf {\bibinfo
  {volume} {110}},\ \bibinfo {pages} {053002} (\bibinfo {year}
  {2013})}\BibitemShut {NoStop}%
\bibitem [{\citenamefont {Sinatra}\ \emph {et~al.}(1995)\citenamefont
  {Sinatra}, \citenamefont {Castelli}, \citenamefont {Lugiato}, \citenamefont
  {Grangier},\ and\ \citenamefont {Poizat}}]{Sinatra:1995}%
  \BibitemOpen
  \bibfield  {author} {\bibinfo {author} {\bibfnamefont {A.}~\bibnamefont
  {Sinatra}}, \bibinfo {author} {\bibfnamefont {F.}~\bibnamefont {Castelli}},
  \bibinfo {author} {\bibfnamefont {L.~A.}\ \bibnamefont {Lugiato}}, \bibinfo
  {author} {\bibfnamefont {P.}~\bibnamefont {Grangier}}, \ and\ \bibinfo
  {author} {\bibfnamefont {J.~P.}\ \bibnamefont {Poizat}},\ }\href
  {http://stacks.iop.org/1355-5111/7/i=3/a=016} {\bibfield  {journal} {\bibinfo
   {journal} {Quantum Semiclass. Opt.}\ }\textbf {\bibinfo {volume} {7}},\
  \bibinfo {pages} {405} (\bibinfo {year} {1995})}\BibitemShut {NoStop}%
\bibitem [{\citenamefont {Brion}\ \emph {et~al.}(2007)\citenamefont {Brion},
  \citenamefont {Pedersen},\ and\ \citenamefont {M{\o}lmer}}]{Brion:2007}%
  \BibitemOpen
  \bibfield  {author} {\bibinfo {author} {\bibfnamefont {E.}~\bibnamefont
  {Brion}}, \bibinfo {author} {\bibfnamefont {L.~H.}\ \bibnamefont {Pedersen}},
  \ and\ \bibinfo {author} {\bibfnamefont {K.}~\bibnamefont {M{\o}lmer}},\
  }\href {http://stacks.iop.org/1751-8121/40/i=5/a=011} {\bibfield  {journal}
  {\bibinfo  {journal} {J. Phys. A: Math. Theor.}\ }\textbf {\bibinfo {volume}
  {40}},\ \bibinfo {pages} {1033} (\bibinfo {year} {2007})}\BibitemShut
  {NoStop}%
\bibitem [{\citenamefont {Paulisch}\ \emph {et~al.}(2014)\citenamefont
  {Paulisch}, \citenamefont {Rui}, \citenamefont {Ng},\ and\ \citenamefont
  {Englert}}]{Paulisch:2014}%
  \BibitemOpen
  \bibfield  {author} {\bibinfo {author} {\bibfnamefont {V.}~\bibnamefont
  {Paulisch}}, \bibinfo {author} {\bibfnamefont {H.}~\bibnamefont {Rui}},
  \bibinfo {author} {\bibfnamefont {H.}~\bibnamefont {Ng}}, \ and\ \bibinfo
  {author} {\bibfnamefont {B.-G.}\ \bibnamefont {Englert}},\ }\href {\doibase
  10.1140/epjp/i2014-14012-8} {\bibfield  {journal} {\bibinfo  {journal} {Eur.
  Phys. J. Plus}\ }\textbf {\bibinfo {volume} {129}},\ \bibinfo {eid} {12}
  (\bibinfo {year} {2014}),\ 10.1140/epjp/i2014-14012-8}\BibitemShut {NoStop}%
\bibitem [{\citenamefont {Jamison}\ \emph {et~al.}(2011)\citenamefont
  {Jamison}, \citenamefont {Kutz},\ and\ \citenamefont {Gupta}}]{Jamison:2011}%
  \BibitemOpen
  \bibfield  {author} {\bibinfo {author} {\bibfnamefont {A.~O.}\ \bibnamefont
  {Jamison}}, \bibinfo {author} {\bibfnamefont {J.~N.}\ \bibnamefont {Kutz}}, \
  and\ \bibinfo {author} {\bibfnamefont {S.}~\bibnamefont {Gupta}},\ }\href
  {\doibase 10.1103/PhysRevA.84.043643} {\bibfield  {journal} {\bibinfo
  {journal} {Phys. Rev. A}\ }\textbf {\bibinfo {volume} {84}},\ \bibinfo
  {pages} {043643} (\bibinfo {year} {2011})}\BibitemShut {NoStop}%
\bibitem [{\citenamefont {Altin}\ \emph
  {et~al.}(2011{\natexlab{a}})\citenamefont {Altin}, \citenamefont {McDonald},
  \citenamefont {D{\"o}ring}, \citenamefont {Debs}, \citenamefont {Barter},
  \citenamefont {Close}, \citenamefont {Robins}, \citenamefont {Haine},
  \citenamefont {Hanna},\ and\ \citenamefont {Anderson}}]{Altin:2011}%
  \BibitemOpen
  \bibfield  {author} {\bibinfo {author} {\bibfnamefont {P.~A.}\ \bibnamefont
  {Altin}}, \bibinfo {author} {\bibfnamefont {G.}~\bibnamefont {McDonald}},
  \bibinfo {author} {\bibfnamefont {D.}~\bibnamefont {D{\"o}ring}}, \bibinfo
  {author} {\bibfnamefont {J.~E.}\ \bibnamefont {Debs}}, \bibinfo {author}
  {\bibfnamefont {T.~H.}\ \bibnamefont {Barter}}, \bibinfo {author}
  {\bibfnamefont {J.~D.}\ \bibnamefont {Close}}, \bibinfo {author}
  {\bibfnamefont {N.~P.}\ \bibnamefont {Robins}}, \bibinfo {author}
  {\bibfnamefont {S.~A.}\ \bibnamefont {Haine}}, \bibinfo {author}
  {\bibfnamefont {T.~M.}\ \bibnamefont {Hanna}}, \ and\ \bibinfo {author}
  {\bibfnamefont {R.~P.}\ \bibnamefont {Anderson}},\ }\href
  {http://stacks.iop.org/1367-2630/13/i=6/a=065020} {\bibfield  {journal}
  {\bibinfo  {journal} {New J. Phys.}\ }\textbf {\bibinfo {volume} {13}},\
  \bibinfo {pages} {065020} (\bibinfo {year} {2011}{\natexlab{a}})}\BibitemShut
  {NoStop}%
\bibitem [{\citenamefont {Altin}\ \emph
  {et~al.}(2011{\natexlab{b}})\citenamefont {Altin}, \citenamefont {McDonald},
  \citenamefont {D{\"o}ring}, \citenamefont {Debs}, \citenamefont {Barter},
  \citenamefont {Close}, \citenamefont {Robins}, \citenamefont {Haine},
  \citenamefont {Hanna},\ and\ \citenamefont {Anderson}}]{Altin:2011b}%
  \BibitemOpen
  \bibfield  {author} {\bibinfo {author} {\bibfnamefont {P.~A.}\ \bibnamefont
  {Altin}}, \bibinfo {author} {\bibfnamefont {G.}~\bibnamefont {McDonald}},
  \bibinfo {author} {\bibfnamefont {D.}~\bibnamefont {D{\"o}ring}}, \bibinfo
  {author} {\bibfnamefont {J.~E.}\ \bibnamefont {Debs}}, \bibinfo {author}
  {\bibfnamefont {T.~H.}\ \bibnamefont {Barter}}, \bibinfo {author}
  {\bibfnamefont {J.~D.}\ \bibnamefont {Close}}, \bibinfo {author}
  {\bibfnamefont {N.~P.}\ \bibnamefont {Robins}}, \bibinfo {author}
  {\bibfnamefont {S.~A.}\ \bibnamefont {Haine}}, \bibinfo {author}
  {\bibfnamefont {T.~M.}\ \bibnamefont {Hanna}}, \ and\ \bibinfo {author}
  {\bibfnamefont {R.~P.}\ \bibnamefont {Anderson}},\ }\href
  {http://stacks.iop.org/1367-2630/13/i=11/a=119401} {\bibfield  {journal}
  {\bibinfo  {journal} {New J. Phys.}\ }\textbf {\bibinfo {volume} {13}},\
  \bibinfo {pages} {119401} (\bibinfo {year} {2011}{\natexlab{b}})}\BibitemShut
  {NoStop}%
\bibitem [{\citenamefont {Fewell}(2005)}]{Fewell:2005}%
  \BibitemOpen
  \bibfield  {author} {\bibinfo {author} {\bibfnamefont {M.~P.}\ \bibnamefont
  {Fewell}},\ }\href {\doibase 10.1016/j.optcom.2005.04.049} {\bibfield
  {journal} {\bibinfo  {journal} {Opt. Commun.}\ }\textbf {\bibinfo {volume}
  {253}},\ \bibinfo {pages} {125 } (\bibinfo {year} {2005})}\BibitemShut
  {NoStop}%
\bibitem [{\citenamefont {Kasevich}\ and\ \citenamefont
  {Chu}(1992)}]{Kasevich:1992}%
  \BibitemOpen
  \bibfield  {author} {\bibinfo {author} {\bibfnamefont {M.}~\bibnamefont
  {Kasevich}}\ and\ \bibinfo {author} {\bibfnamefont {S.}~\bibnamefont {Chu}},\
  }\href {http://dx.doi.org/10.1007/BF00325375} {\bibfield  {journal} {\bibinfo
   {journal} {Appl. Phys. B}\ }\textbf {\bibinfo {volume} {54}},\ \bibinfo
  {pages} {321} (\bibinfo {year} {1992})},\ \bibinfo {note}
  {10.1007/BF00325375}\BibitemShut {NoStop}%
\bibitem [{\citenamefont {Giltner}\ \emph {et~al.}(1995)\citenamefont
  {Giltner}, \citenamefont {McGowan},\ and\ \citenamefont
  {Lee}}]{Giltner:1995b}%
  \BibitemOpen
  \bibfield  {author} {\bibinfo {author} {\bibfnamefont {D.~M.}\ \bibnamefont
  {Giltner}}, \bibinfo {author} {\bibfnamefont {R.~W.}\ \bibnamefont
  {McGowan}}, \ and\ \bibinfo {author} {\bibfnamefont {S.~A.}\ \bibnamefont
  {Lee}},\ }\href {\doibase 10.1103/PhysRevA.52.3966} {\bibfield  {journal}
  {\bibinfo  {journal} {Phys. Rev. A}\ }\textbf {\bibinfo {volume} {52}},\
  \bibinfo {pages} {3966} (\bibinfo {year} {1995})}\BibitemShut {NoStop}%
\bibitem [{\citenamefont {M\"uller}\ \emph
  {et~al.}(2008{\natexlab{b}})\citenamefont {M\"uller}, \citenamefont {Chiow},\
  and\ \citenamefont {Chu}}]{Muller:2008}%
  \BibitemOpen
  \bibfield  {author} {\bibinfo {author} {\bibfnamefont {H.}~\bibnamefont
  {M\"uller}}, \bibinfo {author} {\bibfnamefont {S.-w.}\ \bibnamefont {Chiow}},
  \ and\ \bibinfo {author} {\bibfnamefont {S.}~\bibnamefont {Chu}},\ }\href
  {\doibase 10.1103/PhysRevA.77.023609} {\bibfield  {journal} {\bibinfo
  {journal} {Phys. Rev. A}\ }\textbf {\bibinfo {volume} {77}},\ \bibinfo
  {pages} {023609} (\bibinfo {year} {2008}{\natexlab{b}})}\BibitemShut
  {NoStop}%
\bibitem [{\citenamefont {Clad{\'e}}\ \emph {et~al.}(9 01)\citenamefont
  {Clad{\'e}}, \citenamefont {Plisson}, \citenamefont {Guellati-Kh{\'e}lifa},
  \citenamefont {Nez},\ and\ \citenamefont {Biraben}}]{Clade:2010}%
  \BibitemOpen
  \bibfield  {author} {\bibinfo {author} {\bibfnamefont {P.}~\bibnamefont
  {Clad{\'e}}}, \bibinfo {author} {\bibfnamefont {T.}~\bibnamefont {Plisson}},
  \bibinfo {author} {\bibfnamefont {S.}~\bibnamefont {Guellati-Kh{\'e}lifa}},
  \bibinfo {author} {\bibfnamefont {F.}~\bibnamefont {Nez}}, \ and\ \bibinfo
  {author} {\bibfnamefont {F.}~\bibnamefont {Biraben}},\ }\href {\doibase
  10.1140/epjd/e2010-00198-0} {\bibfield  {journal} {\bibinfo  {journal} {Eur.
  Phys. J. D}\ }\textbf {\bibinfo {volume} {59}},\ \bibinfo {pages} {349}
  (\bibinfo {year} {2010-09-01})}\BibitemShut {NoStop}%
\bibitem [{\citenamefont {Chiow}\ \emph {et~al.}(2011)\citenamefont {Chiow},
  \citenamefont {Kovachy}, \citenamefont {Chien},\ and\ \citenamefont
  {Kasevich}}]{Chiow:2011}%
  \BibitemOpen
  \bibfield  {author} {\bibinfo {author} {\bibfnamefont {S.-w.}\ \bibnamefont
  {Chiow}}, \bibinfo {author} {\bibfnamefont {T.}~\bibnamefont {Kovachy}},
  \bibinfo {author} {\bibfnamefont {H.-C.}\ \bibnamefont {Chien}}, \ and\
  \bibinfo {author} {\bibfnamefont {M.~A.}\ \bibnamefont {Kasevich}},\ }\href
  {\doibase 10.1103/PhysRevLett.107.130403} {\bibfield  {journal} {\bibinfo
  {journal} {Phys. Rev. Lett.}\ }\textbf {\bibinfo {volume} {107}},\ \bibinfo
  {pages} {130403} (\bibinfo {year} {2011})}\BibitemShut {NoStop}%
\bibitem [{\citenamefont {Hardman}\ \emph {et~al.}(2014)\citenamefont
  {Hardman}, \citenamefont {Kuhn}, \citenamefont {McDonald}, \citenamefont
  {Debs}, \citenamefont {Bennetts}, \citenamefont {Close},\ and\ \citenamefont
  {Robins}}]{Hardman:2014}%
  \BibitemOpen
  \bibfield  {author} {\bibinfo {author} {\bibfnamefont {K.~S.}\ \bibnamefont
  {Hardman}}, \bibinfo {author} {\bibfnamefont {C.~C.~N.}\ \bibnamefont
  {Kuhn}}, \bibinfo {author} {\bibfnamefont {G.~D.}\ \bibnamefont {McDonald}},
  \bibinfo {author} {\bibfnamefont {J.~E.}\ \bibnamefont {Debs}}, \bibinfo
  {author} {\bibfnamefont {S.}~\bibnamefont {Bennetts}}, \bibinfo {author}
  {\bibfnamefont {J.~D.}\ \bibnamefont {Close}}, \ and\ \bibinfo {author}
  {\bibfnamefont {N.~P.}\ \bibnamefont {Robins}},\ }\href {\doibase
  10.1103/PhysRevA.89.023626} {\bibfield  {journal} {\bibinfo  {journal} {Phys.
  Rev. A}\ }\textbf {\bibinfo {volume} {89}},\ \bibinfo {pages} {023626}
  (\bibinfo {year} {2014})}\BibitemShut {NoStop}%
\bibitem [{\citenamefont {Scully}\ and\ \citenamefont
  {Zubairy}(1997)}]{Scully:1997}%
  \BibitemOpen
  \bibfield  {author} {\bibinfo {author} {\bibfnamefont {M.~O.}\ \bibnamefont
  {Scully}}\ and\ \bibinfo {author} {\bibfnamefont {M.~S.}\ \bibnamefont
  {Zubairy}},\ }\href@noop {} {\emph {\bibinfo {title} {Quantum Optics}}},\
  \bibinfo {edition} {1st}\ ed.\ (\bibinfo  {publisher} {Cambridge University
  Press},\ \bibinfo {year} {1997})\BibitemShut {NoStop}%
\bibitem [{\citenamefont {Pezz\'e}\ and\ \citenamefont
  {Smerzi}(2009)}]{Pezze:2009}%
  \BibitemOpen
  \bibfield  {author} {\bibinfo {author} {\bibfnamefont {L.}~\bibnamefont
  {Pezz\'e}}\ and\ \bibinfo {author} {\bibfnamefont {A.}~\bibnamefont
  {Smerzi}},\ }\href {\doibase 10.1103/PhysRevLett.102.100401} {\bibfield
  {journal} {\bibinfo  {journal} {Phys. Rev. Lett.}\ }\textbf {\bibinfo
  {volume} {102}},\ \bibinfo {pages} {100401} (\bibinfo {year}
  {2009})}\BibitemShut {NoStop}%
\bibitem [{\citenamefont {Stefszky}\ \emph {et~al.}(2012)\citenamefont
  {Stefszky}, \citenamefont {Mow-Lowry}, \citenamefont {Chua}, \citenamefont
  {Shaddock}, \citenamefont {Buchler}, \citenamefont {Vahlbruch}, \citenamefont
  {Khalaidovski}, \citenamefont {Schnabel}, \citenamefont {Lam},\ and\
  \citenamefont {McClelland}}]{Stefszky:2012}%
  \BibitemOpen
  \bibfield  {author} {\bibinfo {author} {\bibfnamefont {M.~S.}\ \bibnamefont
  {Stefszky}}, \bibinfo {author} {\bibfnamefont {C.~M.}\ \bibnamefont
  {Mow-Lowry}}, \bibinfo {author} {\bibfnamefont {S.~S.~Y.}\ \bibnamefont
  {Chua}}, \bibinfo {author} {\bibfnamefont {D.~A.}\ \bibnamefont {Shaddock}},
  \bibinfo {author} {\bibfnamefont {B.~C.}\ \bibnamefont {Buchler}}, \bibinfo
  {author} {\bibfnamefont {H.}~\bibnamefont {Vahlbruch}}, \bibinfo {author}
  {\bibfnamefont {A.}~\bibnamefont {Khalaidovski}}, \bibinfo {author}
  {\bibfnamefont {R.}~\bibnamefont {Schnabel}}, \bibinfo {author}
  {\bibfnamefont {P.~K.}\ \bibnamefont {Lam}}, \ and\ \bibinfo {author}
  {\bibfnamefont {D.~E.}\ \bibnamefont {McClelland}},\ }\href
  {http://stacks.iop.org/0264-9381/29/i=14/a=145015} {\bibfield  {journal}
  {\bibinfo  {journal} {Class. Quantum Grav.}\ }\textbf {\bibinfo {volume}
  {29}},\ \bibinfo {pages} {145015} (\bibinfo {year} {2012})}\BibitemShut
  {NoStop}%
\bibitem [{\citenamefont {Walls}\ and\ \citenamefont
  {Milburn}(2008)}]{Walls:2008}%
  \BibitemOpen
  \bibfield  {author} {\bibinfo {author} {\bibfnamefont {D.~F.}\ \bibnamefont
  {Walls}}\ and\ \bibinfo {author} {\bibfnamefont {G.~J.}\ \bibnamefont
  {Milburn}},\ }\href@noop {} {\emph {\bibinfo {title} {Quantum Optics}}},\
  \bibinfo {edition} {2nd}\ ed.\ (\bibinfo  {publisher} {Springer-Verlag},\
  \bibinfo {address} {Berlin and Heidelberg},\ \bibinfo {year}
  {2008})\BibitemShut {NoStop}%
\bibitem [{Note1()}]{Note1}%
  \BibitemOpen
  \bibinfo {note} {Actually, this phase factor $\protect \qopname \relax
  o{exp}(i q c t)$ is an artefact of writing the pulse shape $u_p(\protect
  \textbf {r}, t) \protect \qopname \relax o{exp}(i \protect \textbf {k}_1
  \cdot \protect \textbf {r})$ as a single frequency pulse with a
  slowly-varying envelope. Strictly, this pulse contains a range of frequencies
  about $k_1$ which exactly cancel $\protect \qopname \relax o{exp}(i q c t)$
  in the above integral.}\BibitemShut {Stop}%
\bibitem [{\citenamefont {Pezz\'e}\ and\ \citenamefont
  {Smerzi}(2008)}]{Pezze:2008}%
  \BibitemOpen
  \bibfield  {author} {\bibinfo {author} {\bibfnamefont {L.}~\bibnamefont
  {Pezz\'e}}\ and\ \bibinfo {author} {\bibfnamefont {A.}~\bibnamefont
  {Smerzi}},\ }\href {\doibase 10.1103/PhysRevLett.100.073601} {\bibfield
  {journal} {\bibinfo  {journal} {Phys. Rev. Lett.}\ }\textbf {\bibinfo
  {volume} {100}},\ \bibinfo {pages} {073601} (\bibinfo {year}
  {2008})}\BibitemShut {NoStop}%
\bibitem [{\citenamefont {Lang}\ and\ \citenamefont {Caves}(2013)}]{Lang:2013}%
  \BibitemOpen
  \bibfield  {author} {\bibinfo {author} {\bibfnamefont {M.~D.}\ \bibnamefont
  {Lang}}\ and\ \bibinfo {author} {\bibfnamefont {C.~M.}\ \bibnamefont
  {Caves}},\ }\href {\doibase 10.1103/PhysRevLett.111.173601} {\bibfield
  {journal} {\bibinfo  {journal} {Phys. Rev. Lett.}\ }\textbf {\bibinfo
  {volume} {111}},\ \bibinfo {pages} {173601} (\bibinfo {year}
  {2013})}\BibitemShut {NoStop}%
\bibitem [{\citenamefont {Steel}\ \emph {et~al.}(1998)\citenamefont {Steel},
  \citenamefont {Olsen}, \citenamefont {Plimak}, \citenamefont {Drummond},
  \citenamefont {Tan}, \citenamefont {Collett}, \citenamefont {Walls},\ and\
  \citenamefont {Graham}}]{Steel:1998}%
  \BibitemOpen
  \bibfield  {author} {\bibinfo {author} {\bibfnamefont {M.~J.}\ \bibnamefont
  {Steel}}, \bibinfo {author} {\bibfnamefont {M.~K.}\ \bibnamefont {Olsen}},
  \bibinfo {author} {\bibfnamefont {L.~I.}\ \bibnamefont {Plimak}}, \bibinfo
  {author} {\bibfnamefont {P.~D.}\ \bibnamefont {Drummond}}, \bibinfo {author}
  {\bibfnamefont {S.~M.}\ \bibnamefont {Tan}}, \bibinfo {author} {\bibfnamefont
  {M.~J.}\ \bibnamefont {Collett}}, \bibinfo {author} {\bibfnamefont {D.~F.}\
  \bibnamefont {Walls}}, \ and\ \bibinfo {author} {\bibfnamefont
  {R.}~\bibnamefont {Graham}},\ }\href {\doibase 10.1103/PhysRevA.58.4824}
  {\bibfield  {journal} {\bibinfo  {journal} {Phys. Rev. A}\ }\textbf {\bibinfo
  {volume} {58}},\ \bibinfo {pages} {4824} (\bibinfo {year}
  {1998})}\BibitemShut {NoStop}%
\bibitem [{\citenamefont {Blakie}\ \emph {et~al.}(2008)\citenamefont {Blakie},
  \citenamefont {Bradley}, \citenamefont {Davis}, \citenamefont {Ballagh},\
  and\ \citenamefont {Gardiner}}]{Blakie:2008}%
  \BibitemOpen
  \bibfield  {author} {\bibinfo {author} {\bibfnamefont {P.~B.}\ \bibnamefont
  {Blakie}}, \bibinfo {author} {\bibfnamefont {A.~S.}\ \bibnamefont {Bradley}},
  \bibinfo {author} {\bibfnamefont {M.~J.}\ \bibnamefont {Davis}}, \bibinfo
  {author} {\bibfnamefont {R.~J.}\ \bibnamefont {Ballagh}}, \ and\ \bibinfo
  {author} {\bibfnamefont {C.~W.}\ \bibnamefont {Gardiner}},\ }\href {\doibase
  10.1080/00018730802564254} {\bibfield  {journal} {\bibinfo  {journal}
  {Advances in Physics}\ }\textbf {\bibinfo {volume} {57}},\ \bibinfo {pages}
  {363} (\bibinfo {year} {2008})}\BibitemShut {NoStop}%
\bibitem [{\citenamefont {Polkovnikov}(2010)}]{Polkovnikov:2010}%
  \BibitemOpen
  \bibfield  {author} {\bibinfo {author} {\bibfnamefont {A.}~\bibnamefont
  {Polkovnikov}},\ }\href {\doibase
  http://dx.doi.org/10.1016/j.aop.2010.02.006} {\bibfield  {journal} {\bibinfo
  {journal} {Ann. Phys. (N.Y.)}\ }\textbf {\bibinfo {volume} {325}},\ \bibinfo
  {pages} {1790} (\bibinfo {year} {2010})}\BibitemShut {NoStop}%
\bibitem [{\citenamefont {Opanchuk}\ and\ \citenamefont
  {Drummond}(2013)}]{Opanchuk:2013}%
  \BibitemOpen
  \bibfield  {author} {\bibinfo {author} {\bibfnamefont {B.}~\bibnamefont
  {Opanchuk}}\ and\ \bibinfo {author} {\bibfnamefont {P.~D.}\ \bibnamefont
  {Drummond}},\ }\href {\doibase 10.1063/1.4801781} {\bibfield  {journal}
  {\bibinfo  {journal} {J. Math. Phys.}\ }\textbf {\bibinfo {volume} {54}},\
  \bibinfo {eid} {042107} (\bibinfo {year} {2013})}\BibitemShut {NoStop}%
\bibitem [{\citenamefont {Gardiner}\ and\ \citenamefont
  {Zoller}(2004)}]{Gardiner:2004b}%
  \BibitemOpen
  \bibfield  {author} {\bibinfo {author} {\bibfnamefont {C.~W.}\ \bibnamefont
  {Gardiner}}\ and\ \bibinfo {author} {\bibfnamefont {P.}~\bibnamefont
  {Zoller}},\ }\href@noop {} {\emph {\bibinfo {title} {Quantum Noise: A
  Handbook of Markovian and Non-Markovian Quantum Stochastic Methods with
  Applications to Quantum Optics}}},\ \bibinfo {edition} {3rd}\ ed.\ (\bibinfo
  {publisher} {Springer},\ \bibinfo {address} {Berlin and Heidelberg},\
  \bibinfo {year} {2004})\BibitemShut {NoStop}%
\bibitem [{\citenamefont {Sinatra}\ \emph {et~al.}(2002)\citenamefont
  {Sinatra}, \citenamefont {Lobo},\ and\ \citenamefont
  {Castin}}]{Sinatra:2002}%
  \BibitemOpen
  \bibfield  {author} {\bibinfo {author} {\bibfnamefont {A.}~\bibnamefont
  {Sinatra}}, \bibinfo {author} {\bibfnamefont {C.}~\bibnamefont {Lobo}}, \
  and\ \bibinfo {author} {\bibfnamefont {Y.}~\bibnamefont {Castin}},\ }\href
  {http://stacks.iop.org/0953-4075/35/i=17/a=301} {\bibfield  {journal}
  {\bibinfo  {journal} {J. Phys. B: At. Mol. Opt. Phys.}\ }\textbf {\bibinfo
  {volume} {35}},\ \bibinfo {pages} {3599} (\bibinfo {year}
  {2002})}\BibitemShut {NoStop}%
\bibitem [{\citenamefont {Johnsson}\ \emph {et~al.}(2013)\citenamefont
  {Johnsson}, \citenamefont {Dennis},\ and\ \citenamefont
  {Hope}}]{Johnsson:2013}%
  \BibitemOpen
  \bibfield  {author} {\bibinfo {author} {\bibfnamefont {M.~T.}\ \bibnamefont
  {Johnsson}}, \bibinfo {author} {\bibfnamefont {G.~R.}\ \bibnamefont
  {Dennis}}, \ and\ \bibinfo {author} {\bibfnamefont {J.~J.}\ \bibnamefont
  {Hope}},\ }\href {http://stacks.iop.org/1367-2630/15/i=12/a=123024}
  {\bibfield  {journal} {\bibinfo  {journal} {New J. Phys.}\ }\textbf {\bibinfo
  {volume} {15}},\ \bibinfo {pages} {123024} (\bibinfo {year}
  {2013})}\BibitemShut {NoStop}%
\bibitem [{\citenamefont {Olsen}\ and\ \citenamefont
  {Bradley}(2009)}]{Olsen:2009}%
  \BibitemOpen
  \bibfield  {author} {\bibinfo {author} {\bibfnamefont {M.}~\bibnamefont
  {Olsen}}\ and\ \bibinfo {author} {\bibfnamefont {A.}~\bibnamefont
  {Bradley}},\ }\href {\doibase 10.1016/j.optcom.2009.06.033} {\bibfield
  {journal} {\bibinfo  {journal} {Opt. Commun.}\ }\textbf {\bibinfo {volume}
  {282}},\ \bibinfo {pages} {3924 } (\bibinfo {year} {2009})}\BibitemShut
  {NoStop}%
\bibitem [{\citenamefont {Haine}\ \emph
  {et~al.}(2014{\natexlab{b}})\citenamefont {Haine}, \citenamefont {Szigeti},
  \citenamefont {Lang},\ and\ \citenamefont {Caves}}]{Haine:2014b}%
  \BibitemOpen
  \bibfield  {author} {\bibinfo {author} {\bibfnamefont {S.~A.}\ \bibnamefont
  {Haine}}, \bibinfo {author} {\bibfnamefont {S.~S.}\ \bibnamefont {Szigeti}},
  \bibinfo {author} {\bibfnamefont {M.~D.}\ \bibnamefont {Lang}}, \ and\
  \bibinfo {author} {\bibfnamefont {C.~M.}\ \bibnamefont {Caves}},\ }\href@noop
  {} {\bibfield  {journal} {\bibinfo  {journal} {arXiv:1411.5111}\ } (\bibinfo
  {year} {2014}{\natexlab{b}})}\BibitemShut {NoStop}%
\bibitem [{\citenamefont {van~der Stam}\ \emph {et~al.}(2007)\citenamefont
  {van~der Stam}, \citenamefont {van Ooijen}, \citenamefont {Meppelink},
  \citenamefont {Vogels},\ and\ \citenamefont {van~der
  Straten}}]{van_der_Stam:2007}%
  \BibitemOpen
  \bibfield  {author} {\bibinfo {author} {\bibfnamefont {K.~M.~R.}\
  \bibnamefont {van~der Stam}}, \bibinfo {author} {\bibfnamefont {E.~D.}\
  \bibnamefont {van Ooijen}}, \bibinfo {author} {\bibfnamefont
  {R.}~\bibnamefont {Meppelink}}, \bibinfo {author} {\bibfnamefont {J.~M.}\
  \bibnamefont {Vogels}}, \ and\ \bibinfo {author} {\bibfnamefont
  {P.}~\bibnamefont {van~der Straten}},\ }\href {\doibase
  DOI:10.1063/1.2424439} {\bibfield  {journal} {\bibinfo  {journal} {Rev. Sci.
  Instrum.}\ }\textbf {\bibinfo {volume} {78}},\ \bibinfo {pages} {013102}
  (\bibinfo {year} {2007})}\BibitemShut {NoStop}%
\bibitem [{\citenamefont {Dennis}\ \emph {et~al.}(2013)\citenamefont {Dennis},
  \citenamefont {Hope},\ and\ \citenamefont {Johnsson}}]{Dennis:2012}%
  \BibitemOpen
  \bibfield  {author} {\bibinfo {author} {\bibfnamefont {G.~R.}\ \bibnamefont
  {Dennis}}, \bibinfo {author} {\bibfnamefont {J.~J.}\ \bibnamefont {Hope}}, \
  and\ \bibinfo {author} {\bibfnamefont {M.~T.}\ \bibnamefont {Johnsson}},\
  }\href {\doibase 10.1016/j.cpc.2012.08.016} {\bibfield  {journal} {\bibinfo
  {journal} {Comput. Phys. Commun.}\ }\textbf {\bibinfo {volume} {184}},\
  \bibinfo {pages} {201} (\bibinfo {year} {2013})}\BibitemShut {NoStop}%
\end{thebibliography}%

\end{document}